\documentclass[a4paper,11pt]{article}
\pdfoutput=1

\usepackage{jheppub}

\usepackage[T1]{fontenc}
\graphicspath{{img/}}

\usepackage{arydshln}
\usepackage{mathrsfs}                   
\usepackage[dvipsnames,table]{xcolor}   
\usepackage{booktabs}                   
\usepackage{multirow}                   
\usepackage{lipsum}                     
\usepackage{longtable}                  
\usepackage{pifont}                     
\usepackage{slashed}
\usepackage{youngtab}
\usepackage{comment}
\usepackage[compat=1.1.0]{tikz-feynhand}
\usepackage{tikz, pgfplots}
\pgfplotsset{compat=1.18}

\usepackage{subcaption}

\allowdisplaybreaks

\newcommand{\eg}{e.g.\ }

\newcommand{\idlr}[1]{i\!\overleftrightarrow{D}_{\!\!#1}}

\newcommand{\midmidrule}{\arrayrulecolor{black!30}\midrule\arrayrulecolor{black}}
\newcommand{\braket}[1]{\ensuremath{\langle#1\rangle}}

\usepackage[detect-all]{siunitx}
\makeatletter
\g@addto@macro\bfseries{\boldmath}
\makeatother

\title{An EFT approach to baryon number violation: lower limits on the new physics scale and correlations between nucleon decay modes}

\author[a,b]{Arnau Bas i Beneito,}
\author[a,b]{John Gargalionis,}
\author[a,b]{Juan Herrero-Garc\'{i}a,}
\author[a,b]{Arcadi Santamaria,}
\author[c]{Michael A. Schmidt}

\affiliation[a]{Departament de Física Teòrica, Universitat de València, 46100 Burjassot, Spain}

\affiliation[b]{Instituto de Física Corpuscular (CSIC-Universitat de València),
Parc Científic UV, C/Catedrático José Beltrán, 2, E-46980 Paterna, Spain}

\preprint{IFIC/23-52, CPPC-2023-12}

\affiliation[c]{Sydney Consortium for Particle Physics and Cosmology,
  School of Physics, University of New South Wales,
  Sydney, New South Wales 2052, Australia}

\emailAdd{arnau.bas@ific.uv.es}
\emailAdd{john.gargalionis@ific.uv.es}
\emailAdd{juan.herrero@ific.uv.es}
\emailAdd{arcadi.santamaria@ific.uv.es}
\emailAdd{m.schmidt@unsw.edu.au}

\abstract{
Baryon number is an accidental symmetry of the Standard Model at the Lagrangian level. Its violation is arguably one of the most compelling phenomena predicted by physics beyond the Standard Model. Furthermore, there is a large experimental effort to search for it including the Hyper-K, DUNE, JUNO, and THEIA experiments. Therefore, an agnostic, model-independent, analysis of baryon number violation using the power of Effective Field Theory is very timely. In particular, in this work we study the contribution of dimension six and seven effective operators to $|\Delta (B-L)|=0, \, 2$ nucleon decays taking into account the effects of Renormalisation Group Evolution. We obtain lower limits on the energy scale of each operator and study the correlations between different decay modes. We find that for some operators the effect of running is very significant.}


\makeatletter
\gdef\@fpheader{}
\makeatother 
\begin{document} 
\maketitle

\flushbottom

\newpage
\section{Introduction}
\label{sec:intro}

In order to search for physics beyond the Standard Model (SM), a good starting point is to look for processes which are absent in the SM due to fortuitous global symmetries, such as the conservation of baryon number ($B$). Indeed, even though proton decay has not been detected so far (current limits stand at $\tau (p \rightarrow \pi^0 e^+ )> 2.4 \cdot 10^{34}$ years at $90\%$ confidence level \cite{Super-Kamiokande:2020wjk,Super-Kamiokande:2014otb}), it is extremely well motivated by Grand Unified Theories (GUTs)~\cite{Georgi:1974sy,Fritzsch:1974nn} and is a prerequisite for the generation of the baryon asymmetry of the Universe~\cite{Sakharov:1967dj}. Indeed, anomaly cancellation relates quarks and leptons, and any new theory that unifies them is expected to inherently generate proton decay. Moreover, when looking at the SM as an Effective Field Theory (EFT), $B$ is violated in one unit by four-fermion operators at dimension equal or greater than six. In summary, the search for baryon number violation (BNV) stands as one of the most well-grounded strategies for probing new physics.

On the theory side there has been a significant body of research dedicated to the investigation of proton decay in recent decades, not only because of its pivotal significance in GUTs\footnote{Although in some GUTs it is possible to cancel proton decay \cite{Dorsner:2004jj,Fornal:2017xcj,Kobakhidze:2001yk,Berezhiani:2001ub}, in general, nucleon decay is one of their main predictions.} and $R$-parity-violating supersymmetry~\cite{Farrar:1978xj}~\footnote{For a similar study on nucleon decay in the context of R-parity-violation we refer the reader to Ref.~\cite{Chamoun:2020aft}.}, but also many other theories beyond the SM (see Refs.~\cite{Nath:2006ut,FileviezPerez:2022ypk,Dev:2022jbf,Ohlsson:2023ddi,Langacker:1980js} for reviews). Additionally, there are also model-independent studies of BNV using EFT~\cite{deGouvea:2014lva,He:2021mrt,He:2021sbl,Antusch:2020ztu}, alongside a wide variety of models with Leptoquarks (LQs) with BNV phenomenology \cite{Dorsner:2012nq,Murgui:2021bdy,Dorsner:2022twk,Kovalenko:2002eh,Arnold:2012sd,Assad:2017iib,Helo:2019yqp,Davighi:2022qgb,Baldes:2011mh}. Apart from exclusive nucleon decay modes, inclusive nucleon decays~\cite{Heeck:2019kgr} provide an interesting avenue to search for BNV, as well as non-canonical nucleon decays into light new particles~\cite{Fridell:2023tpb,Helo:2018bgb}.

The experimental quest to detect BNV nucleon decay is currently also highly active. Upcoming neutrino experiments, such as Hyper-Kamiokande (Hyper-K)~\cite{Hyper-Kamiokande:2018ofw}, Deep Underground Neutrino Experiment (DUNE)~\cite{DUNE:2016evb, DUNE:2020ypp} and Jiangmen Underground Neutrino Observatory (JUNO)~\cite{JUNO:2015zny} as well as the proposed advanced optical neutrino detector THEIA~\cite{Theia:2019non},
are poised to significantly enhance their sensitivity to nucleon decay, with an expected improvement of about an order of magnitude.\footnote{For Hyper-K, see Figure~1 in  Ref.~\cite{Hyper-Kamiokande:2022smq} and Table 7 in Ref.~\cite{Dev:2022jbf}.} The sensitivity of the last three experiments is expected to be comparable to the that of Hyper-K in modes involving kaons thanks to their better reconstruction of the latter. 
The most sensitive modes are generally semi-leptonic nucleon decays to a meson and a single lepton.\footnote{Decay modes with multiple leptons are phase-space suppressed \cite{Wise:1980ch,Hambye:2017qix} and radiative decays are suppressed by the fine-structure
constant~\cite{Fajfer:2023gfi,Silverman:1980ha}.}

In this work, we carry out a model-independent study of BNV nucleon decay.
Since the scale responsible for BNV is presumably very large,
this can be done by using an EFT framework~\cite{Weinberg:1980wa,Coleman:1969sm,Callan:1969sn,Weinberg:1979sa}.
We assume that all new particles responsible for nucleon decay
have masses of order $\Lambda\gg m_{W}$, where $m_W$ is the $W$-boson mass; then, any physics below
the scale $\Lambda$ can be well described by the SM EFT (SMEFT)~\cite{Buchmuller:1985jz,Grzadkowski:2010es,Brivio:2017vri,Henning:2014wua} with Wilson coefficients (WCs) that can be
directly related to the complete theory above $\Lambda$. Nucleon
decays, however, occur at the mass scale of the nucleons, e.g.\ the mass of the proton $m_p$,  much lower than the electroweak scale, i.e., $m_p\ll m_{W}\ll\Lambda$. Therefore, 
connecting the WCs of the SMEFT at the scale $\Lambda$ with the
physical observables requires going through a tower of EFTs connected
by successive matchings at the thresholds of each EFT, and Renormalisation
Group (RG) evolutions between the different scales.

Since there are basically three energy scales, $\Lambda$, $m_{W}$ and $m_{p}$,
we use three EFTs: 
\begin{enumerate}
\item[a)] For an energy scale $\mu$ such that $m_{W}<\mu<\Lambda$, we use the SMEFT,
the EFT based on the SM gauge group $\mathrm{SU}(3)_C\times \mathrm{SU}(2)_L\times \mathrm{U}(1)_Y$
together with the SM fields. 
\item[b)] For $\mu<m_{W}$, we work with the Low Energy EFT
(LEFT), see e.g.~Refs.~\cite{Jenkins:2017dyc,Jenkins:2017jig}, which is governed by the gauge symmetry $\mathrm{SU}(3)_{C}\times \mathrm{U}(1)_{{\rm EM}}$
and its field content consists of the photon, gluons and all leptons
and quarks with the exception of the top quark. 
\item[c)] For $\mu<2$~GeV, we use Baryon Chiral Perturbation
Theory (B$\chi$PT)~\cite{Claudson:1981gh,JLQCD:1999dld}, which is written in
terms of baryons and mesons.
\end{enumerate}

In the SMEFT we consider $\Delta B= -1$ effective operators up to dimension-7, for which nucleon decays occur at tree level. This includes $\Delta (B-L) =0$ operators at dimension 6 and $|\Delta(B-L)|=2$ operators at dimension 7. These operators are then run, at one loop, from $\Lambda$ down to $m_W$, where they are matched at tree level to dimension-6
$\Delta B=-1$ operators in the LEFT. While evolving the WCs down to
the nucleon scale, we subsequently decouple the bottom quark at $\mu=5$ GeV and the $c$
quark and $\tau$ lepton at $\mu=2$~GeV. Finally, the resulting LEFT is matched to
B$\chi$PT, which is used to calculate (sometimes, with the help of
Lattice QCD results) the relevant two-body semi-leptonic nucleon decays.
Although, many pieces of the calculation are already available in the literature (see Refs.~\cite{Abbott:1980zj,Claudson:1981gh,Kaymakcalan:1983uc,Buchmuller:1985jz,JLQCD:1999dld,Nath:2006ut,Grzadkowski:2010es,Jenkins:2017dyc,Jenkins:2017jig,Liao:2016hru,Zhang:2023ndw}), to our best knowledge, the complete analysis is still lacking. We leave  for Ref.~\cite{Gargalionis:2024nij} an analysis of higher-dimensional operators, for which the matching onto the dimension 6 and 7 operators studied in this paper occurs at the loop level most of the time.

\begin{table}[tbp!]
  
  \begin{center}
    \begin{tabular}{cccc}\toprule
      Process &  $|\Delta (B-L)|$  & $ \tau  \hspace{0.2em} (10^{33}\textrm{ years})$ & \\
      \midrule
      $ p \to \pi^0 \ell^{+}$ & 0 & 24 [16] & \cite{Super-Kamiokande:2020wjk}  \\
      $ p \to \eta^0 \ell^{+}$ & 0 & 10 [4.7] &\cite{Super-Kamiokande:2017gev} \\
      $ p \to K^{+} \nu $ & 0, 2 & 6.61  &\cite{Mine:2016mxy} \\
      $ n \to \pi^{-}\ell^{+}$ & 0 & 5.3 [3.5]& \cite{Super-Kamiokande:2017gev} \\
      $ n \to \pi^{0} \nu $ & 0, 2 & 1.1  &\cite{Super-Kamiokande:2013rwg} \\
      $ p \to K^{0} \ell^{+} $ & 0 & 1 [1.6] &\cite{Super-Kamiokande:2005lev,Super-Kamiokande:2012zik}  \\
      $ p \to \pi^{+} \nu $ & 0, 2 & 0.39  &\cite{Super-Kamiokande:2013rwg} \\
      $ n \to \eta^{0} \nu $ & 0, 2 & 0.158 & \cite{McGrew:1999nd} \\
      $ n \to K^{0} \nu $ & 0, 2 & 0.13  &\cite{Super-Kamiokande:2013rwg} \\
      $ n \to K^{+} \ell^{-} $ & 2 & 0.032 [0.057] & \cite{Frejus:1991ben} \\ 
      \bottomrule
    \end{tabular}
  \end{center}
  
  \caption{\label{tab:NucleonDecayChannels}
    Allowed two-body nucleon decays with 90\% Confidence level upper limit on the lifetime $\tau$ for both the $|\Delta(B-L)|=0,2$ nucleon to pseudoscalar meson decay channels. As the neutrino escapes undetected, both possibilities are included in each channel. For charged leptons $\ell$ the first limit is for the electron and the limit for the muon is quoted in brackets. We do not pursue the study of BNV SMEFT operators with energy dimension higher than 7. Therefore, our analysis excludes the decay channels $n\to\pi^{+}e^{-},~ n \to K^{-}l^{+}, ~p \to \bar K^0 l^{+}$, and $n \to \bar K^0 \nu$.}
\end{table}

We focus on two-body nucleon decay processes with a pseudoscalar meson and a SM lepton in the final state, because they generally provide the most stringent limits.\footnote{See Sec.~\ref{sec:rate-calcs} for details.} See Tab.~\ref{tab:NucleonDecayChannels} for a list of the relevant bounds. Within our study, we establish lower limits on the various SMEFT operators, as well as outline the dominant decay processes, provide estimates for the branching ratios, and discuss the possible correlations among them. 

The remaining sections of this paper are organised as follows. 
Section~\ref{sec:EFT} delves into the methodology we use to compute nucleon decay, including a detailed description of the relevant operators of the different EFTs: SMEFT, the LEFT and B$\chi$PT.
Our main model-independent results are provided in Sec.~\ref{sec:results}. In Sec.~\ref{sec:model} we give a UV model as an example intended to illustrate the use of our results. Finally, our conclusions are outlined in Sec.~\ref{sec:conc}. Technical details are provided in the appendices.

\section{Effective Field Theory framework} \label{sec:EFT}

In this section we discuss the different EFTs necessary for our study. In Sec.~\ref{sec:SMEFT}, we define the relevant effective operators in SMEFT and in Sec.~\ref{sec:LEFT} those of the LEFT.
The final step is the calculation of the nucleon decay rates, which differ in their evaluation of the nuclear matrix elements. One broadly distinguishes between a direct calculation of the matrix elements using lattice quantum chromodynamics (QCD) and the indirect method, where lattice calculations are used to determine low-energy constants in B$\chi$PT, which are then used to evaluate the nuclear matrix elements. See Refs.~\cite{Aoki:2017puj,Yoo:2021gql} for a comparison between the two methods. We opt for the indirect method using B$\chi$PT in this work.
The matching of LEFT to B$\chi$PT is discussed in Sec.~\ref{sec:BchiPT}, and in Sec.~\ref{sec:rate-calcs} we provide the nucleon decay rates.

\subsection{Standard Model Effective Field Theory} \label{sec:SMEFT}

For the analysis, we consider BNV SMEFT operators with $\Delta B=-1$ up to dimension 7. For dimension-6 operators, we employ the Warsaw basis~\cite{Grzadkowski:2010es} and the basis of Ref.~\cite{Liao:2020zyx} for the dimension-7 operators: 
\begin{align}
  \mathcal{L} & = \sum_i C_{i,pqrs} \mathcal{O}_{i,pqrs} +{\rm h.c.}\ ,
\end{align}
where $C$ stands for the dimensionful WC. The field content of each operator is listed as first subscript, and the second subscript, $pqrs$, denotes the family indices of the four fermion fields. These operators comprise three quarks and one lepton. We also introduce the notation $c$ for the dimensionless WC,
\begin{equation}
    C_{i,pqrs} = \frac{c_{i,pqrs}}{\Lambda^{n}}\,,
\end{equation}
with $n=2 \, (3)$ for dimension-6 (7) SMEFT WCs. For definiteness, we work in the basis in which the up-type quark and charged-lepton Yukawa matrices are diagonal. As neutrino masses are much smaller compared to the nucleon mass scale, we neglect them in the following. Consequently, the leptonic mixing matrix is unphysical and thus we set $U_{\alpha i}=\delta_{\alpha i}$.

There are four independent operators with $\Delta B=\Delta L=-1$ at dimension 6~\cite{Grzadkowski:2010es}\footnote{Here and throughout this work we employ two-component Weyl spinor notation. We follow the conventions of Ref.~\cite{Dreiner:2008tw}, including those for the SM fermions defined in Appendix J.}
\begin{equation}
  \label{eq:smeft-ops-1}
  \begin{aligned}
    \mathcal{O}_{qqql,pqrs} & = (Q_p^{i} Q_q^{j}) (Q_r^{l}L_s^k )\epsilon_{ik} \epsilon_{jl}\,,
    &
    \mathcal{O}_{qque,pqrs} & = (Q^{i}_{p} Q^{j}_{q}) (\bar u_r^{\dagger}\bar e_s^\dagger) \epsilon_{ij}\,,
    \\
    \mathcal{O}_{duue,pqrs} & =  ( \bar d_p^{\dagger} \bar u_q^{\dagger}  )(\bar u_r^{\dagger}\bar e_s^\dagger )\,,
    &
    \mathcal{O}_{duql,pqrs} & = (\bar d_p^{\dagger}\bar u_q^{\dagger} ) ( Q_r^{i} L_s^j) \epsilon_{ij}\,,
  \end{aligned}
\end{equation}
where $i,j,k,l$ are $\mathrm{SU}(2)_L$ indices. Colour indices are suppressed here and in the following and it is understood that the three quarks in each bilinear are contracted using a Levi-Civita tensor with indices in the order of the quarks in the operators. Moreover, there are six independent operators with $\Delta B = -\Delta L = -1$ at dimension 7~\cite{Liao:2020zyx}:
\begin{equation}
  \label{eq:smeft-ops-2}
  \begin{aligned}
    \mathcal{O}_{\bar l dddH,pqrs} & = (L_p^{\dagger} \bar d_q^{\dagger}) (\bar d^{\dagger}_{r} \bar d^{\dagger}_{s}) H\,,
    &
    \mathcal{O}_{\bar ldqq \tilde H,pqrs} & = (L_p^{\dagger} \bar d_q^{\dagger}) (Q_r Q_s^{i} ) \tilde H^j  \epsilon_{ij}\,,
    \\
    \mathcal{O}_{\bar e qdd\tilde H,pqrs} & = (\bar e_p Q_q^{i})(\bar d_r^{\dagger} \bar d_s^{\dagger}) \tilde H^j \epsilon_{ij}\,,
    &
    \mathcal{O}_{\bar ldud\tilde H,pqrs} & = (L_p^{\dagger} \bar d_q^{\dagger})(\bar u^{\dagger}_r \bar d^{\dagger}_s) \tilde H\,,
    \\
    \mathcal{O}_{\bar lqdDd,pqrs} & = (L^{\dagger}_p \bar \sigma^\mu Q_q) ( \bar d_r^{\dagger} \idlr{\mu} \bar d_s^{\dagger})\,,
    &
    \mathcal{O}_{\bar edddD,pqrs} & = (\bar e_p \sigma^\mu \bar d_q^{\dagger}) ( \bar d_r^{\dagger} \idlr{\mu} \bar d_s^{\dagger})
    \;,
  \end{aligned}
\end{equation}
where $D_\mu$ denotes a covariant derivative and $A\overleftrightarrow{D_\mu} B = A (D_\mu B) - (D_\mu A) B$. 
At low energies, the dimension-7 SMEFT operators with covariant derivatives are suppressed by $m_p/v$ relative to the first four dimension-7 operators without a covariant derivative, where $v$ denotes the electroweak vacuum expectation value. This can also be seen in their contribution to LEFT operators. While the first four dimension-7 SMEFT operators contribute to dimension-6 LEFT operators with $v/\Lambda^3$, the operators with covariant derivatives only contribute to dimension-7 LEFT operators and thus their contribution to nucleon decay is suppressed relative to the first four dimension-7 operators. We thus neglect the two dimension-7 operators with a covariant derivative in the following.

As BNV operators are generally generated at a very high scale and BNV nucleon decay occurs at the nuclear scale, RG evolution has a significant effect~\cite{Abbott:1980zj}. The RG equations (RGEs) of the BNV dimension-6 and dimension-7 SMEFT operators have been computed in Refs.~\cite{Alonso:2014zka} and~\cite{Liao:2016hru}, respectively. While gauge interactions contribute in a flavour universal way, Yukawa interactions result in mixing between different operators. However, as nucleon decay involves first generation quarks with tiny Yukawa interactions, all but the top Yukawa ($y_t$) are irrelevant for our phenomenological analysis. 
The RGEs take the generic form
\begin{equation}
   \Dot{C}_i \equiv 16\pi^2 \mu \frac{d C_i}{d\mu} = \sum_j \gamma_{ij}C_j
\end{equation}
in terms of the WCs $C_i$ and the anomalous dimension matrices $\gamma_{ij}$.
The RGEs for the $d=6$ BNV WCs take a simple form using the mentioned approximations~\cite{Alonso:2014zka}
\begin{align}\label{duuRGE}
\dot C_{duue,prst} =&  \left( -4 g_3^2 -2 g_1^2 \right) C_{duue,prst}
-\frac{20}{3}g_1^2 \,C_{duue,psrt}\,, \\ \label{duqRGE}
\dot C_{duq\ell,prst} =&  \left( -4 g_3^2 - \frac{9}{2} g_2^2 -\frac{11}{6} g_1^2 \right)C_{duq\ell,prst}\,, \\ \label{qquRGE}
\dot C_{qque,prst} =&  \left( -4 g_3^2 - \frac{9}{2} g_2^2 -\frac{23}{6} g_1^2 \right) C_{qque,prst}\,,\\ \label{qqqRGE}
\dot C_{qqq\ell,prst} =&  \left(- 4 g_3^2 - 3 g_2^2 -\frac13 g_1^2 \right) C_{qqq\ell,prst}
- 4 g_2^2 \left( C_{qqq\ell,rpst} + C_{qqq\ell,srpt} + C_{qqq\ell,psrt} \right) \;,
\end{align}
where $g_i$ are the SM gauge couplings.
The RGEs of the $d=7$ BNV WCs are~\cite{Liao:2016hru,Zhang:2023ndw}
\begin{align}
\label{o1}
\dot{C}_{\bar l dud \tilde H,prst} = &
\left(-4g^2_3-\frac{9}{4}g^2_2 -\frac{17}{12}g^2_1+y_t^2\right)C_{\bar ldud \tilde H,prst}
-\frac{10}{3}g^2_1\,C_{\bar l dud \tilde H,ptsr}\,, \\
\label{o2}
\dot{C}_{\bar l dddH,prst}= &
\Big(-4g^2_3- \frac{9}{4}g^2_2-\frac{13}{12}g^2_1 +y_t^2\Big)C_{\bar l dddH,prst}\,, \\
\label{o3}
\dot{C}_{\bar e qdd \tilde H,prst} = &
\Big(-4g^2_3-\frac{9}{4}g^2_2+\frac{11}{12}g^2_1 +y_t^2\Big)C_{\bar e qdd \tilde H,prst}\,, \\
\label{o4}
\dot{C}_{\bar ldqq \tilde H,prst}=&
\Big(-4g^2_3-\frac{15}{4}g^2_2-\frac{19}{12}g^2_1 +y_t^2\Big)C_{\bar ldqq \tilde H,prst} -3g^2_2\, C_{\bar ldqq \tilde H,prts}\;.
\end{align}
While QCD and top-loop contributions are universal, electroweak corrections depend on the operator. Note that the contribution due to gauge interactions is negative while the one from fermion loops is positive and thus top-loops reduce the RG correction when running from high scale to low scale. The RG evolution of the dimension-6 SMEFT WCs have been evaluated using the {\rm Mathematica} package DSixTools~\cite{Fuentes-Martin:2020zaz,Celis:2017hod}.

\subsection{Low-energy Effective Field Theory}  \label{sec:LEFT}

In the LEFT, the lowest-dimensional operators which violate baryon and lepton numbers appear at dimension 6
\begin{align}
  \mathcal{L} & = \sum_i L_{i,pqrs} \mathcal{O}_{i,pqrs} +{\rm h.c.}\, ,
\end{align}
where $L_i$ denotes the dimensionful WC. In the following, the Lorentz structure of each operator is indicated as a superscript and quark content as subscript followed by the flavour indices $pqrs$.

\begin{table}[tp!]
    \begin{tabular}[t]{lccc}
      \toprule
      Name~\cite{Jenkins:2017jig} (\cite{Nath:2006ut}) & Operator &  Flavour \\
      \midrule
      \([\mathcal{O}_{udd}^{S,LL}]_{111r}\)\; (\(O^{\nu}_{LL}\)) & \((u d)(d \nu_{r})\)  & \((\mathbf{8}, \mathbf{1})\) \\
      \([\mathcal{O}_{udd}^{S,LL}]_{121r}\)\; (\(\tilde{O}^{\nu}_{LL1}\)) & \((u s)(d \nu_{r})\)  & \((\mathbf{8}, \mathbf{1})\) \\
      \([\mathcal{O}_{udd}^{S,LL}]_{112r}\)\; (\(\tilde{O}^{\nu}_{LL2}\)) & \((u d)(s \nu_{r})\)  & \((\mathbf{8}, \mathbf{1})\) \\
      \midmidrule
      \([\mathcal{O}_{duu}^{S,LL}]_{111r}\)\; (\(O^{e}_{LL}\)) &\((d u)(u e_{r})\)  & \((\mathbf{8}, \mathbf{1})\)\\
      \([\mathcal{O}_{duu}^{S,LL}]_{211r}\)\; (\(\tilde{O}^{e}_{LL}\))& \((s u)(u e_{r})\)  & \((\mathbf{8}, \mathbf{1})\) \\
      \midmidrule
      \([\mathcal{O}_{duu}^{S,LR}]_{111r}\)\; (\(O^{e}_{LR}\)) & \((d u)(\bar{u}^{\dagger} \bar{e}^{\dagger}_{r})\)  & \((\bar{\mathbf{3}}, \mathbf{3})\)
      \\
      \([\mathcal{O}_{duu}^{S,LR}]_{211r}\)\; (\(\tilde{O}^{e}_{LR}\)) & \((s u)(\bar{u}^{\dagger} \bar{e}^{\dagger}_{r})\)  &\((\bar{\mathbf{3}}, \mathbf{3})\)\\
      \midmidrule
      \([\mathcal{O}_{duu}^{S,RL}]_{111r}\)\; (\(O^{e}_{RL}\)) & \((\bar{d}^{\dagger} \bar{u}^{\dagger})(u e_{r})\) & \((\mathbf{3}, \bar{\mathbf{3}})\) \\
      \([\mathcal{O}_{duu}^{S,RL}]_{211r}\)\; (\(\tilde{O}^{e}_{RL}\)) & \((\bar{s}^{\dagger} \bar{u}^{\dagger})(u e_{r})\)  & \((\mathbf{3}, \bar{\mathbf{3}})\) \\
      \midmidrule
      \([\mathcal{O}_{dud}^{S,RL}]_{111r}\)\; (\(O^{\nu}_{RL}\)) & \((\bar{d}^{\dagger} \bar{u}^{\dagger})(d \nu_{r})\)  & \((\mathbf{3}, \bar{\mathbf{3}})\) \\
      \([\mathcal{O}_{dud}^{S,RL}]_{211r}\)\; (\(\tilde{O}^{\nu}_{RL1}\)) & \((\bar{s}^{\dagger} \bar{u}^{\dagger})(d \nu_{r})\)  & \((\mathbf{3}, \bar{\mathbf{3}})\) \\
      \([\mathcal{O}_{dud}^{S,RL}]_{112r}\)\; (\(\tilde{O}^{\nu}_{RL2}\)) & \((\bar{d}^{\dagger} \bar{u}^{\dagger})(s \nu_{r})\)  & \((\mathbf{3}, \bar{\mathbf{3}})\)\\
      \midmidrule
      \([\mathcal{O}_{ddu}^{S,RL}]_{[12]1r}\)  & \((\bar{d}^{\dagger} \bar{s}^{\dagger})(u \nu_{r})\)  & \((\mathbf{3}, \bar{\mathbf{3}})\)\\
      \midmidrule
      \([\mathcal{O}_{duu}^{S,RR}]_{111r}\)\; (\(O^{e}_{RR}\)) & \((\bar{d}^{\dagger} \bar{u}^{\dagger})(\bar{u}^{\dagger} \bar{e}^{\dagger}_{r})\)  & \((\mathbf{1}, \mathbf{8})\) \\
      \([\mathcal{O}_{duu}^{S,RR}]_{211r}\)\; (\(\tilde{O}^{e}_{RR}\)) & \((\bar{s}^{\dagger} \bar{u}^{\dagger})(\bar{u}^{\dagger} \bar{e}^{\dagger}_{r})\)  & \((\mathbf{1}, \mathbf{8})\) \\
      \bottomrule
    \end{tabular}
    \hfill
    \begin{tabular}[t]{lcc}
      \toprule
      Name &  Operator  & Flavour \\
      \midrule
      \([\mathcal{O}_{ddd}^{S,LL}]_{[12]r1}\) & \((d s)(\bar{e}_{r} d)\) & \((\mathbf{8}, \mathbf{1})\) \\
      \midmidrule
      \([\mathcal{O}_{udd}^{S,LR}]_{11r1}\) & \((u d)(\nu^{\dagger}_{r} \bar{d}^{\dagger})\)  & \((\bar{\mathbf{3}}, \mathbf{3})\) \\
      \([\mathcal{O}_{udd}^{S,LR}]_{12r1}\) &  \((u s)(\nu^{\dagger}_{r} \bar{d}^{\dagger})\)  & \((\bar{\mathbf{3}}, \mathbf{3})\) \\
      \([\mathcal{O}_{udd}^{S,LR}]_{11r2}\) &  \((u d)(\nu^{\dagger}_{r} \bar{s}^{\dagger})\)  & \((\bar{\mathbf{3}}, \mathbf{3})\) \\
      \midmidrule
      \([\mathcal{O}_{ddu}^{S,LR}]_{[12]r1}\) &  \((d s)(\nu^{\dagger}_{r} \bar{u}^{\dagger})\)  & \((\bar{\mathbf{3}}, \mathbf{3})\) \\
      \midmidrule
      \([\mathcal{O}_{ddd}^{S,LR}]_{[12]r1}\) &  \((d s)(e^{\dagger}_{r} \bar{d}^{\dagger})\)  & \((\bar{\mathbf{3}}, \mathbf{3})\) \\
      \midmidrule
      \([\mathcal{O}_{ddd}^{S,RL}]_{[12]r1}\) &  \((\bar{d}^{\dagger} \bar{s}^{\dagger})(\bar e_r d)
      \)  & \((\mathbf{3},\bar{\mathbf{3}})\)\\
      \midmidrule
      \([\mathcal{O}_{udd}^{S,RR}]_{11r1}\) &  \((\bar{u}^{\dagger} \bar{d}^{\dagger})(\nu^{\dagger}_{r} \bar{d}^{\dagger})\)  & \((\mathbf{1}, \mathbf{8})\) \\
      \([\mathcal{O}_{udd}^{S,RR}]_{12r1}\) &  \((\bar{u}^{\dagger} \bar{s}^{\dagger})(\nu^{\dagger}_{r} \bar{d}^{\dagger})\)  & \((\mathbf{1}, \mathbf{8})\) \\
      \([\mathcal{O}_{udd}^{S,RR}]_{11r2}\) &  \((\bar{u}^{\dagger} \bar{d}^{\dagger})(\nu^{\dagger}_{r} \bar{s}^{\dagger})\)  & \((\mathbf{1}, \mathbf{8})\) \\
      \midmidrule
      \([\mathcal{O}_{ddd}^{S,RR}]_{[12]r1}\) &  \((\bar{d}^{\dagger} \bar{s}^{\dagger})(e^{\dagger}_{r} \bar{d}^{\dagger})\)  & \((\mathbf{1}, \mathbf{8})\) \\
      \bottomrule
    \end{tabular}
  \caption{\label{tab:wet-ops}
    The table shows the $\Delta(B-L)=0$ operators (left) and the $|\Delta(B-L)|=2$ operators (right) in the LEFT basis of Ref.~\cite{Jenkins:2017jig} that give rise to nucleon decays at tree level. In parentheses we also indicate the correspondence to the nomenclature of Ref.~\cite{Nath:2006ut} for the $\Delta (B-L)=0$ operators. The operator $\mathcal{O}_{ddu}^{S,RL}$ is not studied there. Square brackets enclose pairs of indices that are antisymmetric under permutation. The operators are presented in Weyl-spinor notation in the second column, along with their transformation properties under the flavour group $\mathrm{SU}(3)_L \times \mathrm{SU}(3)_R$ in the third column.
  }
\end{table}

\begin{table}[tp!]
  \begin{center}
    \begin{tabular}{lc}
      \toprule
      Name~\cite{Jenkins:2017jig} & SMEFT matching\\ 
      \midrule
      \([\mathcal{O}_{udd}^{S,LL}]_{pqrs}\) & $V_{q'q}V_{r'r}(C_{qqql,r'q'ps}-C_{qqql,q'r'ps}+C_{qqql,q'pr's})$ \\ 
      \([\mathcal{O}_{duu}^{S,LL}]_{pqrs}\)     &  $V_{p'p} (C_{qqql,rqp's}-C_{qqql,qrp's}+C_{qqql,qp'rs})$ \\ 
      \([\mathcal{O}_{duu}^{S,LR}]_{pqrs}\) & $-V_{p'p} (C_{qque,p'qrs}+C_{qque,qp'rs})$\\
      \([\mathcal{O}_{duu}^{S,RL}]_{pqrs}\) & $C_{duql,pqrs}$\\
      \([\mathcal{O}_{dud}^{S,RL}]_{pqrs}\) & $-V_{r'r}C_{duql,pqr's}$\\ 
      \([\mathcal{O}_{duu}^{S,RR}]_{pqrs}\) & $C_{duue,pqrs}$\\
      \midrule
      \midrule
      \([\mathcal{O}_{udd}^{S,LR}]_{pqrs}\) &  $-V_{q'q}C_{\bar ldqq\tilde H,rspq'} \frac{v}{\sqrt{2}} \ $\\ 
      \([\mathcal{O}_{ddd}^{S,LR}]_{pqrs}\) & $V_{p'p}V_{q'q} (C_{\bar l dqq\tilde H,rsq'p'} -C_{\bar l dqq\tilde H,rsp'q'} )\frac{v}{2\sqrt{2}}$\\ 
      \([\mathcal{O}_{ddd}^{S,RL}]_{pqrs}\) & $V_{s's} ( C_{\bar eqdd\tilde H,rs'qp}-C_{\bar eqdd\tilde H,rs'pq})  \frac{v}{2\sqrt{2}}$ \\
      \([\mathcal{O}_{udd}^{S,RR}]_{pqrs}\) & $C_{\bar l dud\tilde H,rspq}\frac{v}{\sqrt{2}} \  $\\
      \([\mathcal{O}_{ddd}^{S,RR}]_{pqrs}\) & $C_{\bar ldddH,rspq} \frac{v}{\sqrt{2}}$ \\
      \bottomrule
    \end{tabular}

  \end{center}
  \caption{Tree-level matching of the dimension-6 and dimension-7 SMEFT operators onto the LEFT (see Refs.~\cite{Jenkins:2017jig,Dekens:2019ept,Liao:2016hru,Liao:2019tep,Liao:2020zyx}). There is an implicit sum over primed indices in the expressions.}
  \label{tab:SMEFT-LEFT}
\end{table}

Disregarding the flavour structure, there are nine independent operators with $\Delta (B-L)=0$ and seven independent dimension-6 operators with $|\Delta (B-L)|=2$, see \eg  Ref.~\cite{Jenkins:2017jig}.\footnote{Note that for the case $|\Delta (B-L)|=2$ dimension-7 LEFT operators neither their RGEs nor the relevant nuclear matrix elements are currently available.} The pertinent\footnote{Two of the BNV dimension-6 operators, $\mathcal{O}_{uud}^{S,XY}$ with $X\neq Y$, $X,Y\in \{L,R\}$, are antisymmetric in the up-type quark flavour indices and do not directly contribute to proton decay at tree-level, but only to charmed $\Lambda_c^+$ decays and, thus, are omitted.}  operators with their flavour structure are listed in Tab.~\ref{tab:wet-ops}. We focus on LEFT operators relevant for nucleon decay and thus only include light quarks $u,d,s$, neutrinos $\nu$ and charged leptons $e$ and $\mu$.
The first column lists all operators following the convention in Ref.~\cite{Jenkins:2017jig} and their relation to Ref.~\cite{Nath:2006ut}. The explicit forms of the operators are presented in the second column using Weyl-spinor notation~\cite{Dreiner:2008tw}. The transformation properties of the LEFT operators with respect to the
flavour group $\mathrm{SU}(3)_L\times \mathrm{SU}(3)_R$, where the three light quarks form a $\mathrm{SU}(3)$ triplet $(u,d,s)\sim \mathbf{3}$, are listed in the third column, which is relevant for matching the LEFT operators to B$\chi$PT.
The matching of the SMEFT operators to the dimension-6 LEFT operators is presented in Tab.~\ref{tab:SMEFT-LEFT}. Note that some of the dimension-6 LEFT operators are only generated at higher-order in the SMEFT; examples are $\mathcal{O}^{S,RL}_{ddu}$, $\mathcal{O}^{S,LL}_{ddd}$, and $\mathcal{O}^{S,LR}_{ddu}$.

The RGEs are dominated by QCD interactions, resulting in a universal rescaling of all $d=6$ BNV WCs~\cite{Jenkins:2017dyc}, 
\begin{align}
    \dot L_{i,prst} = -4 g_3^2\, L_{i,prst}\,, 
\end{align}
to leading order. For the actual calculation of the running, from $m_W$  to $\mu=5$ GeV we use the {\rm Mathematica} package DSixTools~\cite{Fuentes-Martin:2020zaz,Celis:2017hod}, which also includes QED corrections. At $\mu=5$ GeV we decouple the bottom quark, and from there we solve numerically the RGE of the LEFT with $N_q=4$ down to $2$ GeV, finding $L_{i,prst}(2~\mathrm{GeV}) \simeq 1.26\, L_{i,prst}(m_W)$. Finally, at $\mu=2$ GeV, the matching to B$\chi$PT is performed.

\subsection{Baryon Chiral Perturbation Theory} \label{sec:BchiPT}

Nucleon decay rates are well described using B$\chi$PT, which describes the interaction of baryons with light mesons. We provide a brief introduction to set the notation. The discussion follows the conventions in Ref.~\cite{Claudson:1981gh}. 

\subsubsection{Baryon-number-conserving interactions}
 In B$\chi$PT the relevant fields are the baryon $B$  and meson $M$ fields\footnote{In this section we will use four-component Dirac notation for the field $B$.} with the respective transformation properties under the flavour group $\mathrm{SU}(3)_L\times \mathrm{SU}(3)_R$ 
\begin{align}
    B&=\sum_{a=1}^8 B_a \frac{\lambda_a}{\sqrt{2}} =\begin{pmatrix}
    \frac{\Sigma^0}{\sqrt{2}}+\frac{\Lambda^0}{\sqrt{6}} & \Sigma^+ & p\\
    \Sigma^- & -\frac{\Sigma^0}{\sqrt{2}} + \frac{\Lambda^0}{\sqrt{6}} & n\\
    \Xi^- & \Xi^0 & -\sqrt{\frac{2}{3}}\Lambda^0
    \end{pmatrix}
    &
    B&\to UBU^\dagger &
    \end{align}
    \begin{align}
    M & =\sum_{a=1}^8 M_a\frac{\lambda_a}{\sqrt{2}} = \begin{pmatrix}
    \frac{\pi^0}{\sqrt{2}}+\frac{\eta}{\sqrt{6}} & \pi^+ & K^+\\
    \pi^- & -\frac{\pi^0}{\sqrt{2}} + \frac{\eta}{\sqrt{6}} & K^0\\
    K^- & \bar K^0 & -\sqrt{\frac{2}{3}}\eta
    \end{pmatrix}
    \end{align}
    \begin{align}
    \Sigma \equiv e^{2iM/f_\pi}& \to L\Sigma R^\dagger &
    \xi\equiv e^{iM/f_\pi} &\to L\xi U^\dagger = U\xi R^\dagger
\end{align}
with $\xi^2=\Sigma$ and the pion decay constant $f_\pi=130.41(20)\, \mathrm{MeV}$~\cite{ParticleDataGroup:2022pth}. Here $L \in \mathrm{SU}(3)_L$, $R\in \mathrm{SU}(3)_R$, and $U$ is an element of the diagonal subgroup $\mathrm{SU}(3)_V \supset \mathrm{SU}(3)_L\times \mathrm{SU}(3)_R$. The meson fields $M$ are pseudo-Goldstone bosons of the spontaneously broken flavour group.
The combinations $\xi B \xi$, $\xi^\dagger B \xi^\dagger$, $\xi B \xi^\dagger$, $\xi^\dagger B \xi$ have well-defined transformation properties with respect to the flavour group $\mathrm{SU}(3)_L\times \mathrm{SU}(3)_R$
\begin{align}
    \xi B \xi & \to L \xi B \xi R^\dagger 
    &
    \xi^\dagger B \xi^\dagger & \to R \xi^\dagger B \xi^\dagger L^\dagger 
    \\
    \xi B \xi^\dagger & \to L \xi B \xi^\dagger L^\dagger 
    &
    \xi^\dagger B \xi & \to R \xi^\dagger B \xi R^\dagger 
\end{align}
and thus can be identified with the representations $\xi B\xi \sim(\mathbf{3},\mathbf{\bar 3})$, $\xi^\dagger B \xi^\dagger \sim(\mathbf{\bar 3},\mathbf{3})$, $\xi B \xi^\dagger \sim (\mathbf{8},\mathbf{1})$ and $\xi^\dagger B \xi \sim (\mathbf{1},\mathbf{8})$.

The lowest order chiral Lagrangian for the interactions of mesons and baryons is~\cite{Claudson:1981gh}
\begin{align}\label{eq:L0}
    \mathcal{L}_0 & = \frac18 f_\pi^2 \mathrm{tr}(\partial_\mu \Sigma \partial^\mu \Sigma^\dagger) + \mathrm{Tr}(\bar B (i \slashed{D} -M_B) B) - \frac{D}{2} \mathrm{Tr}(\bar B \gamma^\mu \gamma_5\{\xi_\mu,B\}) -\frac{F}{2} \mathrm{Tr}(\bar B \gamma^\mu \gamma_5 [\xi_\mu,B])
\end{align}
with the covariant derivatives $D_\mu B = \partial_\mu B + [\Gamma_\mu ,B]$ and
\begin{align}
    \Gamma_\mu &= \frac12\left[ \xi^\dagger \partial_\mu \xi + \xi \partial_\mu \xi^\dagger\right]
    &
\xi_\mu &= i\left[ \xi^\dagger \partial_\mu \xi - \xi \partial_\mu  \xi^\dagger\right]
\;.
\end{align}
The low-energy constants $D$ and $F$ can be obtained from a phenomenological analysis, see e.g.~Ref.~\cite{Aoki:2008ku},
\begin{align}
    D&=0.80(1)\;, & 
    F& =0.47(1) \;,
\end{align}
or calculated on the lattice~\cite{Bali:2022qja},
\begin{align}
D&=0.730(11)\;,
& 
F&=0.447^{(6)}_{(7)}\;.
\end{align}

For the following discussion, it is sufficient to include operators with at most one strange quark, because there are no couplings of a nucleon to a meson and a baryon with two strange quarks as part of the baryon-number conserving $D$ and $F$ terms. In fact, the operators parameterised by $D$ and $F$ in Eq.~\eqref{eq:L0} lead to the following couplings of a meson with two baryons, where at least one of the baryons is a proton or neutron~\cite{Nath:2006ut}
\begin{equation}\label{eq:LagDF}
\begin{aligned}
    \mathcal{L}_0  \supset &   \left(
     \frac{D-F}{f_\pi}\,\overline{\Sigma^+} \gamma^\mu \gamma_5 p
    -\frac{D+3F}{\sqrt{6}f_\pi}\, \overline{\Lambda^0}  \gamma^\mu \gamma_5 n 
    -\frac{D-F}{\sqrt{2}f_\pi} \,\overline{\Sigma^0} \gamma^\mu \gamma_5n
    \right)\, \partial_\mu \bar K^0 
     \\ &
    +\left(
      \frac{D-F}{\sqrt{2}f_\pi}\,\overline{\Sigma^0} \gamma^\mu \gamma_5 p
     - \frac{D+3F}{\sqrt{6}f_\pi} \, \overline{\Lambda^0} \gamma^\mu \gamma_5 p 
     + \frac{D-F}{f_\pi}\, \overline{\Sigma^-} \gamma^\mu \gamma_5n 
     \right)\,\partial_\mu K^-
    \\ &
     + \frac{3F-D}{2\sqrt{6}f_\pi}\,\left(\overline{p} \gamma^\mu \gamma_5p \, + \,\overline{n} \gamma^\mu \gamma_5n \,
    \right)\partial_\mu \eta
    \\ &
    + \frac{D+F}{f_\pi}\,\overline{p} \gamma^\mu \gamma_5n \,\partial_\mu \pi^+ 
    \\  &
    + \frac{D+F}{2\sqrt{2}f_\pi}\,\left(\overline{p} \gamma^\mu \gamma_5p \, -\,\overline{n} \gamma^\mu \gamma_5n \,
    \right)\partial_\mu \pi^0 +\mathrm{h.c.} \;.
\end{aligned}
\end{equation}
Note that the terms with $\eta$ and $\pi^0$ are hermitian and thus the hermitian conjugation results in a factor $2$. We corrected the last term of the first line of equation (527) in Ref.~\cite{Nath:2006ut}. For brevity, we neglect $\mathrm{SU}(3)_V$ breaking effects such as quark masses.

\subsubsection{Baryon-number-violating interactions}

Due to parity conservation of strong interactions and isospin symmetry, there are only two independent nuclear matrix elements for proton decay parameterised in terms of two real numbers $\alpha$ and $\beta$~\cite{JLQCD:1999dld}
\begin{align}
\braket{0|\epsilon^{abc} (\bar u_a^\dagger \bar d_b^\dagger) u_c|p^{(s)}} &= \alpha P_L u_p^{(s)}\,, 
&
\braket{0|\epsilon^{abc} (u_a d_b) u_c|p^{(s)}} &= \beta P_L u_p^{(s)} \,,
\end{align}
where $a,b,c$ denote the colour indices, $s$ the proton spin, $u_p^{(s)}$ the $4$-component proton spinor and $P_{L,R}$ the chirality projectors.  
The matrix elements with an initial state neutron with $4$-component spinor $u_n^{(s)}$ are obtained from imposing isospin symmetry
\begin{align}
\braket{0|\epsilon^{abc} (\bar d_a^\dagger \bar u_b^\dagger) d_c|n^{(s)}} &= \alpha P_L u_n^{(s)} \;,
&
\braket{0|\epsilon^{abc} (d_a u_b) d_c|n^{(s)}} &= \beta P_L u_n^{(s)} \,,
\end{align}
and the ones for right-handed spinors can be related by parity conservation 
\begin{equation} 
\begin{aligned}
\braket{0|\epsilon^{abc} (u_a d_b) \bar u_c^\dagger |p^{(s)}} &= -\alpha P_R u_p^{(s)} \;,
&
\braket{0|\epsilon^{abc} (\bar u_a^\dagger \bar d_b^\dagger) \bar u_c^\dagger |p^{(s)}} &= -\beta P_R u_p^{(s)} \;,
\\
\braket{0|\epsilon^{abc} (d_a  u_b) \bar d_c^\dagger|n^{(s)}} &= -\alpha P_R u_n^{(s)} \;,
&
\braket{0|\epsilon^{abc} (\bar d_a^\dagger  \bar u_b^\dagger ) \bar d_c^\dagger|n^{(s)}} &= -\beta P_R u_n^{(s)}\;, 
\end{aligned}
\end{equation}
where the values of the $\alpha$ and $\beta$ constants are~\cite{Yoo:2021gql} 
\begin{align}
    \beta & = 0.01269(107)~\rm{GeV}^3 \; , & 
    \alpha & = -0.01257(111)~\rm{GeV}^3\;.
\end{align}
The matching of the BNV dimension-6 LEFT operators to B$\chi$PT is governed by the flavour symmetry listed in the last column of Tab.~\ref{tab:wet-ops}. Demanding the same flavour symmetry properties in LEFT and B$\chi$PT uniquely identifies the B$\chi$PT operator formed out of fields $\xi$, $\xi^\dagger$ and $B$. For the matching we impose the full SU(3)$_V$ flavour symmetry to relate LEFT operators, which annihilate $\Sigma$, $\Lambda$ and $\Xi$ baryons, to $\alpha$ and $\beta$. A projection operator $P_{ij}$ is used to project out the correct flavour indices. The projection operator $P_{ij}$ is defined by setting the $(i,j)$ entry to $1$ and all other entries $0$. We similarly define $\tilde P_{ij}\equiv -P_{ij}$, e.g.\ the $(1,2)$ entry of $P_{12}$ ($\tilde P_{12}$) equals $1$ ($-1$), while all other entries vanish. We provide the relevant matching conditions for nucleon decay in Tabs.~\ref{tab:MatchingBchiPT0} and \ref{tab:MatchingBchiPT2} in App.~\ref{sec:BchiPT-matching}.
The explicit negative signs are from the relation between nuclear matrix elements, while the minus signs due to group theory factors are indicated by a tilde on the projection operator.

\subsection{Nucleon decay rates} \label{sec:rate-calcs}

The semi-leptonic two-body nucleon decay rates are straightforwardly obtained in B$\chi$PT, see App.~\ref{sec:rateformulae}. Here we provide the expressions of the different $|\Delta (B-L)|=2$ decays of a nucleon into a lepton and a pseudoscalar meson,\footnote{We do not consider vector mesons because 
computations within B$\chi$PT rely on unspecified low-energy constants~\cite{Kaymakcalan:1983uc}.} which, to the best of our knowledge, are missing in the literature.
The three strangeness-conserving BNV nucleon decays are
\begin{align}
    \Gamma \left(p \to \pi^{+} \nu_r \right) &= (32\pi f_{\pi}^2m_p^3)^{-1} (m_p^2-m_{\pi}^2)^2 \left| \alpha \left[L_{udd}^{S,LR}\right]_{11r1} + \beta \left[L_{udd}^{S,RR }\right]_{11r1}\right|^2\left(1+D+F\right)^2
    \;,
    \\
\Gamma \left(n \to \pi^{0} \nu_r \right) &= (32\pi f_{\pi}^2m_n^3)^{-1} (m_n^2-m_{\pi}^2)^2 \,\frac{1}{2}\,\left|\alpha \left[L_{udd}^{S,LR}\right]_{11r1} + \beta \left[L_{udd}^{S,RR }\right]_{11r1}\right|^2(1+D+F)^2
\;,
\\
\Gamma \left(n \to \eta^{0} \nu_r \right) & = (32\pi f_{\pi}^2m_n^3)^{-1} (m_n^2-m_{\eta}^2)^2 \, \frac{3}{2} \, \times  \nonumber \\ &\biggl| \alpha \left[L_{udd}^{S,LR}\right]_{11r1}\left(-\frac{1}{3}-\frac{D}{3}+F\right) + \beta \left[L_{udd}^{S,RR}\right]_{11r1}\left(1-\frac{D}{3}+F\right) \biggr|^2
\;,
\end{align}
and the three strangeness-changing BNV nucleon decays are given by
\begin{align}
\Gamma& \left(p \to K^{+} \nu_r \right)  = (32\pi f_{\pi}^2m_p^3)^{-1} (m_p^2-m_{K}^2)^2 \times  \nonumber \\ & \biggl| \beta \left[L_{udd}^{S,RR}\right]_{12r1} + \alpha \left[L_{udd}^{S,LR}\right]_{12r1}  +
    \frac{m_p}{2m_{\Sigma}}\left(\beta \left[L_{udd}^{S,RR}\right]_{11r2} + \alpha \left[L_{udd}^{S,LR}\right]_{11r2}\right)\left(D-F\right) \nonumber \\
    & + \frac{m_p}{6m_{\Lambda}}\left( 
    \beta \left[L_{udd}^{S,RR}\right]_{11r2} + \alpha \left[L_{udd}^{S,LR}\right]_{11r2} + 2\beta \left[L_{udd}^{S,RR}\right]_{12r1} + 2\alpha \left[L_{udd}^{S,LR}\right]_{12r1}\right)\left( D+3F\right)
    \biggr|^2
    \;, \label{eq:pnuk}
\\
    \Gamma& \left(n \to K^{0} \nu_r\right)  = (32\pi f_{\pi}^2m_n^3)^{-1} (m_n^2-m_{K}^2)^2 \times  \nonumber \\ & 
    \biggl| - \alpha \left[L_{udd}^{S,LR}\right]_{12r1} + \beta \left[L_{udd}^{S,RR}\right]_{12r1} + \alpha \left[L_{udd}^{S,LR}\right]_{11r2} + \beta \left[L_{udd}^{S,RR}\right]_{11r2} \nonumber \\ & 
    - \frac{m_n}{2m_{\Sigma}}\left( \alpha \left[L_{udd}^{S,LR}\right]_{12r1} + \beta \left[L_{udd}^{S,RR}\right]_{12r1} \right) \left(D-F\right) \nonumber \\ &
    + \frac{m_n}{6m_{\Lambda}}\left( \alpha \left[L_{udd}^{S,LR}\right]_{12r1} + \beta \left[L_{udd}^{S,RR}\right]_{12r1} + 2\alpha \left[L_{udd}^{S,LR}\right]_{11r2} + 2\beta \left[L_{udd}^{S,RR}\right]_{11r2}\right)\left( D+3F\right)
    \biggr|^2
    \;, \label{eq:nnuk}
\\
    \Gamma& \left(n \to K^{+} e_r^{-} \right)  = (32\pi f_{\pi}^2m_n^3)^{-1} (m_n^2-m_{K}^2)^2 \times \nonumber \\ & \biggl\{ \biggl|\beta \left[L_{ddd}^{S,LL}\right]_{12r1} - \alpha \left[L_{ddd}^{S,RL}\right]_{12r1}+ \frac{m_n}{m_{\Sigma}}\left( \alpha \left[L_{ddd}^{S,RL}\right]_{12r1} + \beta \left[L_{ddd}^{S,LL}\right]_{12r1} \right)\left( D-F \right)\biggr|^2  \nonumber \\ & 
    + \biggl| \beta \left[L_{ddd}^{S,RR}\right]_{12r1} - \alpha \left[L_{ddd}^{S,LR}\right]_{12r1} + \frac{m_n}{m_{\Sigma}}\left( \alpha \left[L_{ddd}^{S,LR}\right]_{12r1} + \beta \left[L_{ddd}^{S,RR}\right]_{12r1} \right)\left( D-F \right)
    \biggr|^2\biggr\} \label{eq:nek}
    \;,
\end{align}
where we neglect the final state lepton mass and follow the notation in Tab.~\ref{tab:MatchingBchiPT2}. Here $L$ denotes the WC that comes with each $|\Delta (B-L)| =2 $ LEFT operator and we have set $m_N = m_n$ or $m_N = m_p$ in the expressions of App.~\ref{sec:rateformulae} depending on the nucleon present in each channel. Note that in the last decay width $[L_{ddd}^{S,LL}]_{12r1}$ is not generated by $d=7$ SMEFT operators at tree level. The first bilinear of the operator $\mathcal{O}_{ddd}^{S,XY}$ with $X,Y\in \{L,R\}$ always contains a strange quark due to antisymmetry. In this work we do not study the $|\Delta (B-L)|=2$ nucleon decay $n \to \pi^{+}e^{-}$ because it is not generated at dimension-6 in the LEFT. It may be generated by a dimension-9 SMEFT operator or by the combination of a dimension-7 SMEFT operator and a $W$-boson exchange.

Furthermore, we found a discrepancy with respect to the $\Delta (B-L)=0$ decay width of $p\to K^0 e^{+}_r$ presented in Ref.~\cite{Nath:2006ut}. We obtained the following expression
\begin{align}
     \Gamma&(p \to K^0 e^{+}_r)  = (32\pi f_{\pi}^2m_p^3)^{-1} (m_p^2-m_{K}^2)^2 \times \nonumber \\& \biggl\{\left| \beta \left[L^{S,LL}_{duu}\right]_{211r} - \alpha \left[L^{S,RL}_{duu}\right]_{211r} + \frac{m_p}{m_{\Sigma^0}}(D-F)\left(\beta \left[L^{S,LL}_{duu}\right]_{211r} + \alpha \left[L^{S,RL}_{duu}\right]_{211r}\right )\right|^2 \nonumber \\
    &+\left| \beta \left[L^{S,RR}_{duu}\right]_{211r} - \alpha \left[L^{S,LR}_{duu}\right]_{211r} + \frac{m_p}{m_{\Sigma^0}}(D-F)\left( \beta \left[L^{S,RR}_{duu}\right]_{211r} + \alpha \left[L^{S,LR}_{duu}\right]_{211r}\right )\right|^2  \biggr\},
\end{align}
where we follow the notation in Tab.~\ref{tab:MatchingBchiPT0} and $L_i$ denotes the WC that comes with each $\Delta (B-L) =0 $ LEFT operator. The expressions for the other two-body $\Delta (B-L) =0 $ decay channels are listed in Appendix F of Ref.~\cite{Nath:2006ut}.
Note that an electron in the final state can always be replaced by a muon. Nucleon decays with tau leptons are kinematically forbidden. BNV processes with tau leptons are indirectly bounded via proton-decay searches with a virtual $\tau$ lepton~\cite{Marciano:1994bg,Crivellin:2023ter} as well as those mediated by higher generation quarks~\cite{Hou:2005iu}.

Assuming isospin conservation,
the following straightforward relations hold:
\begin{equation}
    \Gamma(p\to \pi^{+}  \nu) = 2 \Gamma(n\to \pi^{0} \nu)\,, \label{eq:1SR}
\end{equation}
\begin{equation}
    \Gamma(n\to \pi^{-} e^{+} ) = 3 \Gamma(p\to \pi^0 e^{+})\,. \label{eq:2SR}
\end{equation}
We have checked that, in general, it is not possible to find relations between the decay widths of different processes using the expressions of Ref.~\cite{Nath:2006ut}, or other sum rules, unless two LEFT WCs are set to zero.

In order to translate a possible nucleon decay signature into the list of SMEFT WCs that contribute to that particular channel, it is useful to rewrite the decay rates of the nine $\Delta (B-L)=0$ processes using the following structure 
\begin{equation} \label{eq:kd6}
    \Gamma_{(i)}^{d=6} \equiv \:  10^{-4} \: c_j^{*} \: \boldsymbol{\kappa}_{(i)}^{jk}\:c_k \: \frac{m_p^5}{\Lambda^4}, 
\end{equation}
for dimension-6 operators, and similarly for the six $|\Delta (B-L)|=2$ processes using
\begin{equation} \label{eq:kd7}
     \Gamma_{(i)}^{d=7} \equiv \: c_j^{*} \: \boldsymbol{\kappa}_{(i)}^{jk}\:c_k \: \frac{m_p^7}{\Lambda^6}\, 
\end{equation}
for dimension-7 operators. Here $(i)$ labels the 9 (6) $|\Delta (B-L)|=0$ (2) processes, the indices $j,k$ run over the 10 (9) $|\Delta (B-L)|=0$ (2) SMEFT WCs, and $\boldsymbol{\kappa}_{(i)}^{jk}$ are dimensionless numbers corresponding to process $(i)$ that can be computed using the analytic formulae derived in Appendix~\ref{sec:rateformulae} for $|\Delta (B-L)| =2$ processes and those in Ref.~\cite{Nath:2006ut} for $\Delta (B-L) =0$ processes. This results in 15 hermitian matrices $\boldsymbol{\kappa}_{(i)}$, whose $(j,k)$-entry represents the numerical coefficient of the product of WCs $c_j^* c_k$ in the decay-rate expression. Note that these matrices are scale-dependent, as they also include the RG effects from $\Lambda = 10^{16}$ GeV down to $2$ GeV. Therefore, they serve as a dictionary to relate BNV processes and SMEFT WCs at the UV scale, and they are useful to easily visualise which combination of SMEFT WCs contribute more significantly. The numerical values of the matrices are provided in Appendix~\ref{sec:appendixMatrices}, and we also make the these available online in CSV format~\cite{kappa-matrices-zenodo}.

\section{Results}
\label{sec:results}

In this section we provide several results of our analysis, including the limits on the UV scale for each operator in Sec. \ref{sec:limits}, correlations between the decay channels of interest listed in Tab. \ref{tab:NucleonDecayChannels} and SMEFT operators, and a discussion of the possible flat directions in Sec.~\ref{sec:correlations}.

\subsection{Lower limits on the scale of operators}
\label{sec:limits}

\begin{figure}[tbp!]
    \centering
    \includegraphics[width = 16.2cm]{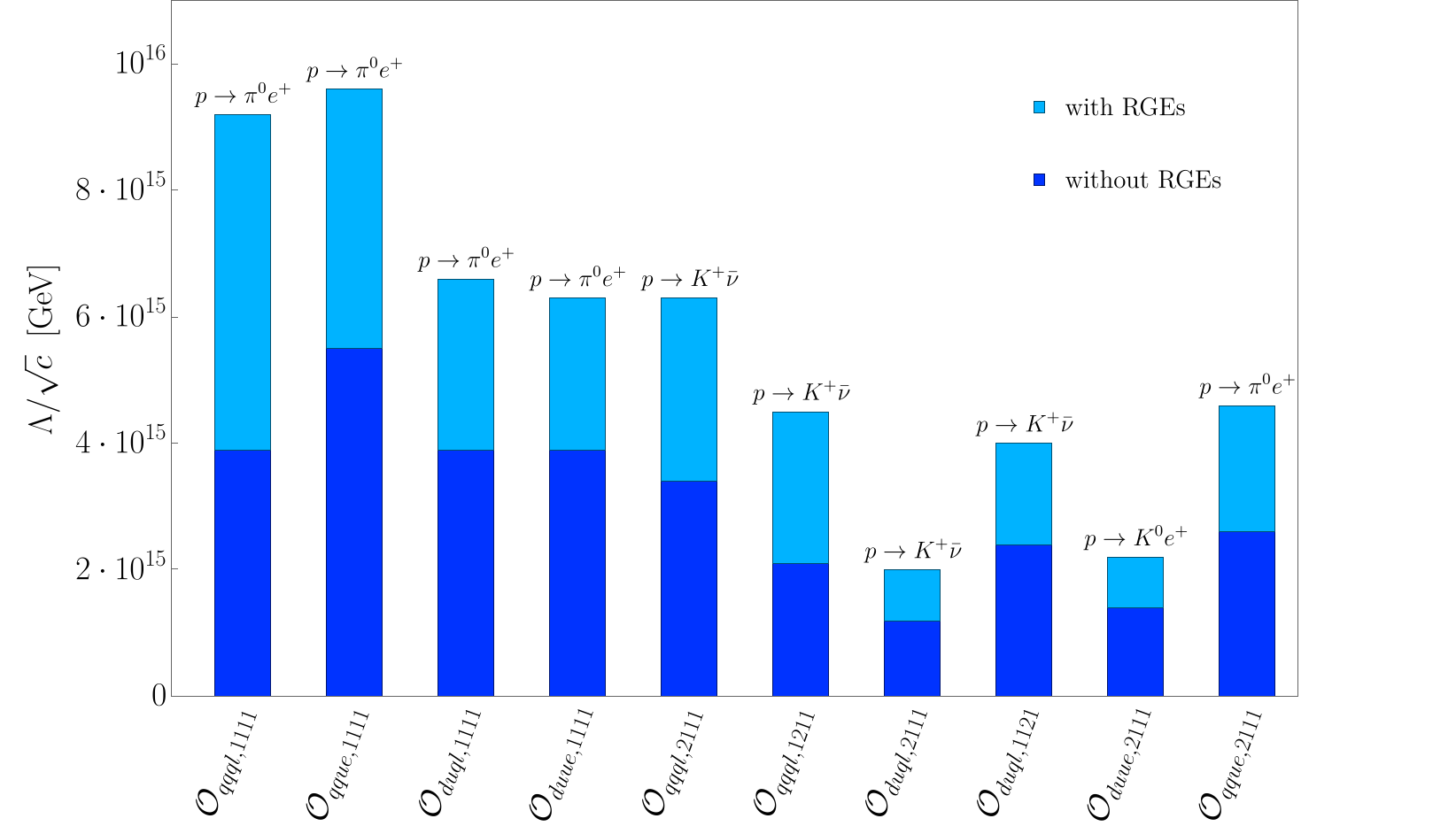}
    \caption{
    Limits on the UV scale $\Lambda/\sqrt{c}$ for each of the dimension-6 $\Delta (B-L)=0$ SMEFT operators without (with) RG effects in dark (light) blue. The numerical 
 are given in Tab.~\ref{tab:general-limits-d6}.
    }
    \label{fig:stacked6}
\end{figure}

\begin{figure}[tbp!]
    \centering
    \includegraphics[width = 16.2cm]{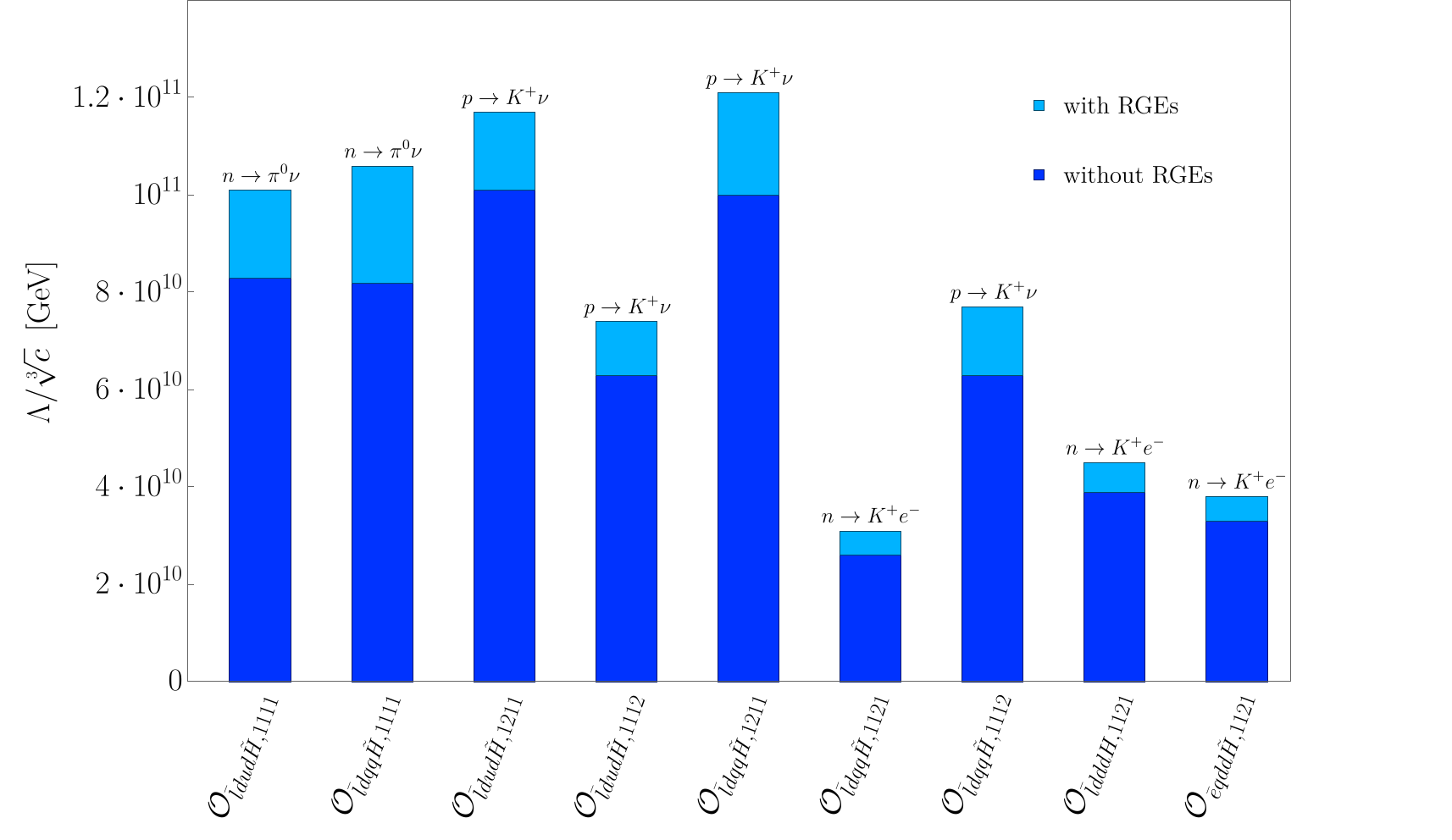}
    \caption{The figure is analogous to Fig.~\ref{fig:stacked6} but for the dimension-7 $|\Delta (B-L)|=2$ SMEFT operators, with the values provided in Tab.~\ref{tab:general-limits-d7}. The expression bounded in this case is $\Lambda / \sqrt[3]{c}$.}
    \label{fig:stacked7}
\end{figure}

Here we provide lower limits on the UV scale $\Lambda$ underlying each of the operators  in Tab.~\ref{tab:SMEFT-LEFT}. To obtain those we have set all SMEFT WCs to zero except for the one we are interested in and computed which process constrains it the most using the experimental data of Tab.~\ref{tab:NucleonDecayChannels}. The numerical values for the limits of the dimension-6 (dimension-7) operators can be found in Tab.~\ref{tab:general-limits-d6} (Tab.~\ref{tab:general-limits-d7}) and in Fig.~\ref{fig:stacked6} (Fig.~\ref{fig:stacked7}). We show the results obtained both with and without incorporating the RG effects.


\begin{table}[tbp!]
    \centering\renewcommand{\arraystretch}{1.5}
    \makebox[\textwidth][c]{
    \begin{tabular}{c c c c l  l }\toprule
  Operator & $\Lambda^{\textrm{w/o}}/\sqrt{c}$ [$10^{15}$ GeV] & $\Lambda^{\textrm{w}}/\sqrt{c}$ [$10^{15}$ GeV] & $\Lambda^{\textrm{w}}/\Lambda^{\textrm{w/o}}$ & Process & LEFT WC \\\midrule 
    $\mathcal{O}_{qqql, 1 1 1 1}$&$3.9 \; (4.0) $ & $9.2 \; (9.4) $ &2.3&$p \to \pi^{0} e^{+}$ & $[L_{duu}^{S,LL}]_{1111}$ \\
    $ \mathcal{O}_{qque, 1 1 1 1}$&$5.5 \; (5.6) $ & $9.6 \; (9.8) $ &1.8& $p \to \pi^{0} e^{+}$ & $[L_{duu}^{S,LR}]_{1111}$ \\
    $\mathcal{O}_{duql, 1 1 1 1}$&$3.9 \; (4.0) $ & $6.6 \; (6.8)  $ &1.7&$p \to \pi^{0} e^{+}$ & $[L_{duu}^{S,RL}]_{1111}$ \\
    $\mathcal{O}_{duue, 1 1 1 1}$ &$3.9\; (4.0) $ & $6.3 \; (6.5)  $ &1.6&$p \to \pi^{0} e^{+}$ & $[L_{duu}^{S,RR}]_{1111}$ \\ \cdashline{1-6}
    $\mathcal{O}_{qqql, 2 1 1 1}$ &$3.4 \; (3.4) $ & $  6.3 \;(6.3)   $ &1.9& $p \to K^{+} \bar \nu$ & $[L_{udd}^{S,LL}]_{1121}$ \\
    $\mathcal{O}_{qqql, 1 2 1 1}$ &$2.1 \; (2.1) $ & $4.5 \; (4.5) $ &2.1& $p \to K^{+}\bar \nu $ & $[L_{dud}^{S,LL}]_{2111}$ \\
    $\mathcal{O}_{duql, 2 1 1 1}$ &$1.2 \;(1.3) $ & $2.0 \; (2.1)  $ &1.6& $p \to K^{+} \bar \nu    $ & $[L_{dud}^{S,RL}]_{2111}$ \\
    $\mathcal{O}_{duql, 1 1 2 1}$ &$2.4 \;(2.4) $ & $4.0 \;(4.1) $ &1.7&$p \to K^{+} \bar \nu  $ & $[L_{dud}^{S,RL}]_{1121}$ \\
    $\mathcal{O}_{duue, 2 1 1 1}$ &$1.4 \;(1.4) $ & $2.2 \; (2.2)  $ &1.6& $p \to K^0 e^{+}$ & $[L_{duu}^{S,RR}]_{2111}$ \\
    $ \mathcal{O}_{qque, 2 1 1 1}$ &$2.6 \;(2.7) $ & $4.6 \;(4.7) $ &1.7&$ p \to \pi^0 e^{+} $ & $[L_{duu}^{S,LR}]_{1111}$ \\ \bottomrule
    \end{tabular}
    }
    \caption{
    Lower limits on the UV scale $\Lambda/\sqrt{c}$ for each $d=6$ SMEFT operator without RG effects (second column, $\Lambda^{\textrm{w/o}}$) and including them (third column, $\Lambda^{\textrm{w}}$) using the values of the B$\chi$PT low-energy constants $F=0.447$ and $D=0.730$ of Ref.~\cite{Bali:2022qja}, and in parenthesis, $F=0.47$ and $D=0.80$ of Ref.~\cite{Aoki:2008ku}. The ratio $\Lambda^{\textrm{w}}/\Lambda^{\textrm{w/o}}$ has been computed using the limits outside of the parenthesis. 
    The dashed line separates $\Delta s =0$ and $\Delta s =1$ SMEFT operators.
    In the fifth column we show the most constraining process for each operator.
    In the last column we show the LEFT WC which gives the strongest bound.
    }
    \label{tab:general-limits-d6}
\end{table}
\begin{table}[tbp!]
    \centering\renewcommand{\arraystretch}{1.3}
    \makebox[\textwidth][c]{
    \begin{tabular}{c c c c l  l }\toprule
    Operator & $\Lambda^{\textrm{w/o}}/\sqrt[3]{c}$ [$10^{10}$ GeV] & $\Lambda^{\textrm{w}}/\sqrt[3]{c}$ [$10^{10}$ GeV] & $\Lambda^{\textrm{w}}/\Lambda^{\textrm{w/o}}$ & Process & LEFT WC \\\midrule 
    $\mathcal{O}_{\bar l dud \tilde H, 1 1 1 1}$ &$8.3 \; (8.4)  $ & $10.1 \;  (10.2) $ &1.2& $ n \to \pi^0 \nu$ & $[L_{udd}^{S,RR}]_{1111}$ \\
    $\mathcal{O}_{\bar l dqq \tilde H, 1 1 1 1}$ &$8.2 \; (8.4) $ & $ 10.6 \;  (10.7)  $ &1.3& $ n \to \pi^0 \nu$  & $[L_{udd}^{S,LR}]_{1111}$ \\\cdashline{1-6}
    $\mathcal{O}_{\bar l dud \tilde H, 1 2 1 1}$ &$10.1 \; (10.2)  $ & $11.7 \; (11.9) $ &1.2& $ p \to K^{+} \nu $ & $[L_{udd}^{S,RR}]_{1112}$ \\
    $\mathcal{O}_{\bar l dud \tilde H, 1 1 1 2}$ &$6.3 \; (6.5) $ & $ 7.4 \;  (7.8)  $ &1.2& $ p \to K^{+} \nu $  & $[L_{udd}^{S,RR}]_{1211}$ \\
    $\mathcal{O}_{\bar l dqq \tilde H, 1 2 1 1}$ &$10.0 \; (10.1) $ & $ 12.1 \; (12.2)$ &1.2& $ p \to K^{+} \nu $ & $[L_{udd}^{S,LR}]_{1112}$ \\
    $\mathcal{O}_{\bar l dqq \tilde H, 1 1 2 1}$ &$2.6 \; (2.6) $ & $ 3.1 \;  (3.1) $ &1.2& $ n \to K^{+} e^{-} $ & $[L_{ddd}^{S,LR}]_{2111}$ \\
    $\mathcal{O}_{\bar l dqq \tilde H, 1 1 1 2}$ & $6.3 \; (6.5) $ & $ 7.7 \;  (7.9)  $ &1.2& $ p \to K^{+} \nu $ & $[L_{udd}^{S,LR}]_{1211}$ \\
    $\mathcal{O}_{\bar l dddH, 1 1 2 1}$ &$3.9 \;(3.9) $ & $ 4.5 \;  (4.6) $ &1.2& $ n \to K^{+} e^{-} $ & $[L_{ddd}^{S,RR}]_{1211}$ \\
    $\mathcal{O}_{\bar e qdd \tilde H, 1 1 2 1}$ &$3.3 \; (3.2) $ & $ 3.8 \;  (3.8) $ &1.2& $ n \to K^{+} e^{-} $ & $[L_{ddd}^{S,RL}]_{2111}$ \\\bottomrule
    \end{tabular}
    }
    \caption{Same as Tab.~\ref{tab:general-limits-d6} for $d=7$ SMEFT operators. Lower limits are now on $\Lambda / \sqrt[3]{c}$.}
    \label{tab:general-limits-d7}
\end{table}

The first column in the tables denotes the operator. In Tab.~\ref{tab:general-limits-d6} (Tab.~\ref{tab:general-limits-d7}) the second (third) column shows the lower limits on the scale over the square root (cubic root) of the WC for $d=6$ ($d=7$) including (excluding) RG corrections, which have been obtained using the most-constraining process listed in the fifth column. The fourth column quantifies the RG correction using the low-energy constants in Ref.~\cite{Bali:2022qja}. The other choice of low-energy constants~\cite{Aoki:2008ku} result in the same values to the level of accuracy shown. The LEFT WCs responsible for the most constraining BNV nucleon decay is listed in last column. For each of the limits, we provide two numerical values, one based on the low-energy constants $F=0.447$ and $D=0.730$~\cite{Bali:2022qja} and the second one in parentheses based on $F=0.47$ and $D=0.8$~\cite{Aoki:2008ku}.  

In Ref.~\cite{Gargalionis:2024nij}, some of us have computed the limits using the nuclear matrix elements directly computed on the lattice as mentioned in Sec.~\ref{sec:EFT} and found agreement with the channels giving the strongest bound on the UV scale, except for the operator $\mathcal{O}_{qqql,1211}$, for which the most relevant decay channel is $p\to \pi^0 e^{+}$ rather than $p\to K^{+}\nu$. Additionally, the lifetimes computed using the direct method are a factor 2-3 times larger, as it was pointed out in Ref.~\cite{Aoki:2017puj}, which translates to a 40-70\% (26-44)\% weaker constraints on the scale of the dimension-6 (dimension-7) operators.

The limits incorporating the RG effects have been computed following the procedure explained in Sec.~\ref{sec:EFT}: We take into account the several EFTs defined between the UV scale $\Lambda$ and $2$ GeV, where the matching onto B$\chi$PT has been performed~\cite{Aoki:2017puj}. RG corrections always result in a more stringent limit on the UV scale because of the universal sign of the QCD contribution to the RG evolution.
One can also appreciate that the parameter space of $\Delta (B-L) =0$ WCs gets more constrained by the RG effects than that of the $|\Delta (B-L)|=2$ WCs, c.f. Tab.~\ref{tab:general-limits-d6} and Tab.~\ref{tab:general-limits-d7}.

Notice that the strongest bound on the UV scale of operator $\mathcal{O}_{qque,2111}$ results from a strangeness-conserving channel, as can be seen in the last row of Tab.~\ref{tab:general-limits-d6}. This channel is generated when we match the SMEFT onto the LEFT and non-diagonal elements of the CKM matrix induce strangeness-conserving LEFT operators (see Tab.~\ref{tab:SMEFT-LEFT}). Although they are suppressed by the Cabbibo angle, it is possible that when combined with the experimental limits of Tab.~\ref{tab:NucleonDecayChannels} such channels can give rise to more stringent bounds on the UV scale compared to the limit computed in a strangeness-changing nucleon channel.

Nucleon decay searches constrain the scale of the dimension-6 (dimension-7) operators to be $\sim(1-10)\times 10^{15}$ GeV ($\sim(2-12)\times 10^{10}$ GeV). RG corrections strengthen the limits by 60-130\% for dimension-6 operators compared to a tree-level analysis and by 20-30\% for dimension-7 operators. Most of the enhancement of the limits in Tabs.~\ref{tab:general-limits-d6} and \ref{tab:general-limits-d7} is induced by the RGEs of the SMEFT WCs described in Sec.~\ref{sec:SMEFT}. All anomalous dimensions of dimension-6 and 7 WCs have the same contribution from the $\mathrm{SU}(3)_C$ gauge couplings, which in most cases is the dominant contribution to the growth. However, it can be seen in Tab.~\ref{tab:general-limits-d6} that the growth of $\mathcal{O}_{qqql,1111}$ is significantly larger than the rest of the WCs, which can be understood from Eq.~\eqref{qqqRGE} since the contribution from the $\mathrm{SU}(2)_L$ gauge couplings is in this case more significant. In a similar vein, we see that RG corrections are smaller for dimension-7 operators due to top-quark contributions, which enter with the opposite sign (see Eqs.~\eqref{o1} - \eqref{o4}), and the smaller energy scale range. All in all, the RG effects always lead to a stronger limit for both dimension-6 and 7 operators, as can be appreciated in the fourth columns of Tabs.~\ref{tab:general-limits-d6} and \ref{tab:general-limits-d7}.

We do not show the limits on the SMEFT operators involving the muon since they can be obtained by rescaling the numerical values using the experimental limits shown in Tab.~\ref{tab:NucleonDecayChannels}. Similarly, we have not included the limits of all components of the SMEFT operators. For operators with left-handed quarks, they are related by the non-diagonal CKM matrix elements and can be obtained by a simple rescaling.

\begin{figure}[t!]
    \centering
    \includegraphics[width = 1\linewidth]
    {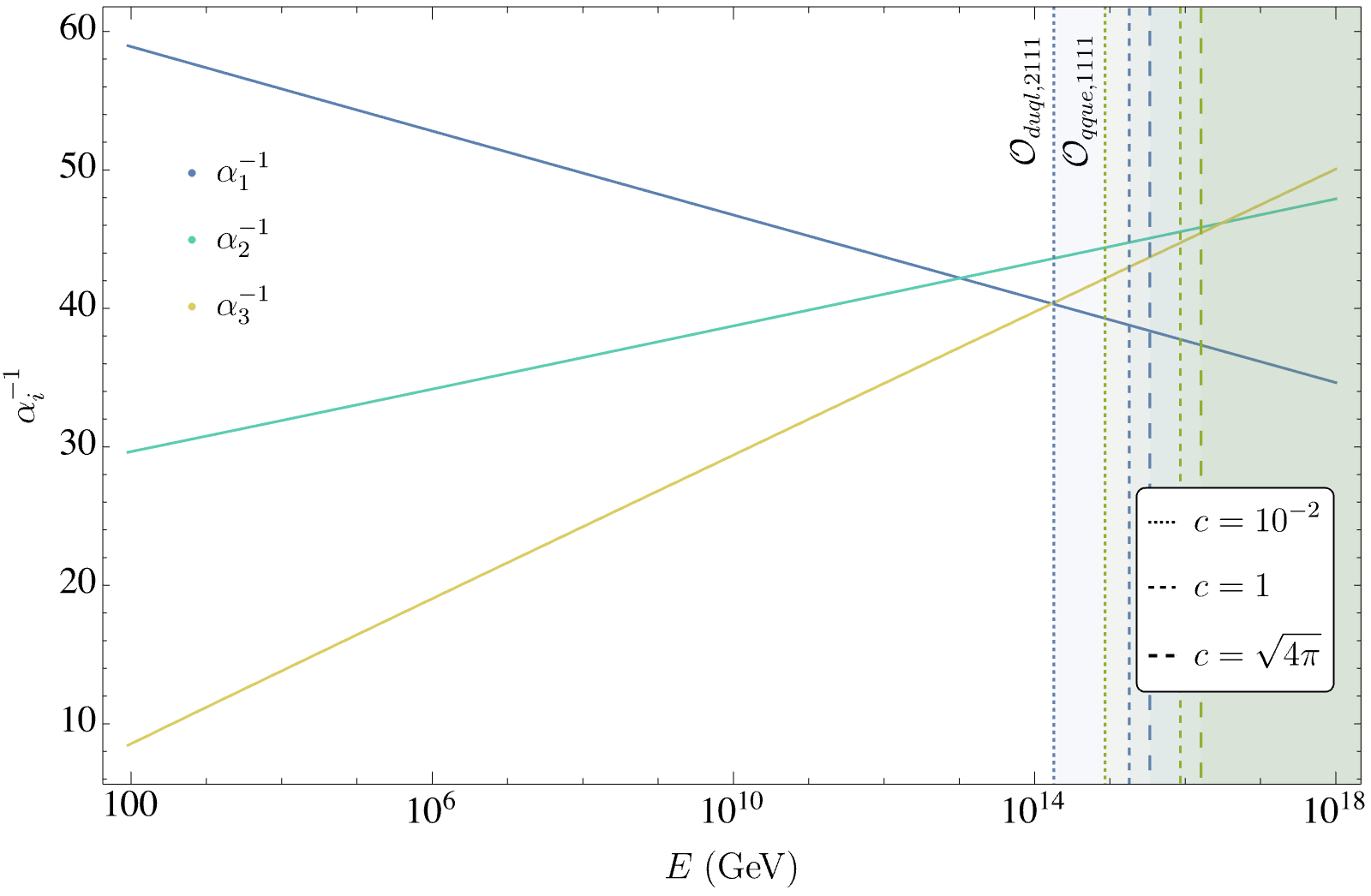}
    \caption{Running of the three SM gauge couplings alongside the lower limits on the UV scale for three different values of the WCs $c$ in blue (green) for the least (most) constraining operators, i.e. $\mathcal{O}_{duql,2111}$ ($\mathcal{O}_{qque,1111}$).}
    \label{fig:unification}
\end{figure}

We can use the limits on the UV scale, as derived in Tab.~\ref{tab:general-limits-d6}, to make a qualitative prediction about the GUT scale, under the assumption that there are no light particles between the electroweak scale and the GUT scale. Fig.~\ref{fig:unification} illustrates the conventional plot depicting the running of the three gauge couplings,\footnote{We take into account the usual factor 3/5 for the normalization of the $U(1)_Y$ hypercharge} alongside the various limits on the scales underlying each dimension-6 operator for three different values of the dimensionless WCs $c$. We have selected the most and least constraining limits on these operators, obtained from Tab.~\ref{tab:general-limits-d6}. We observe compatibility of gauge coupling unification and the proton decay lower limits if the unification scale is provided by the intersection of the $\alpha_2$ and $\alpha_3$ running couplings. The $\beta-$functions of the gauge couplings are computed at 3-loops in the electroweak sector and at 5-loops for the QCD coupling and their values at the electroweak scale are those provided in DSixTools 2.1~\cite{Fuentes-Martin:2020zaz,Celis:2017hod}.

\subsection{Correlations between decay modes and flat directions} \label{sec:correlations}

\begin{figure}[h]
    \centering
    \includegraphics[width = \linewidth]
    {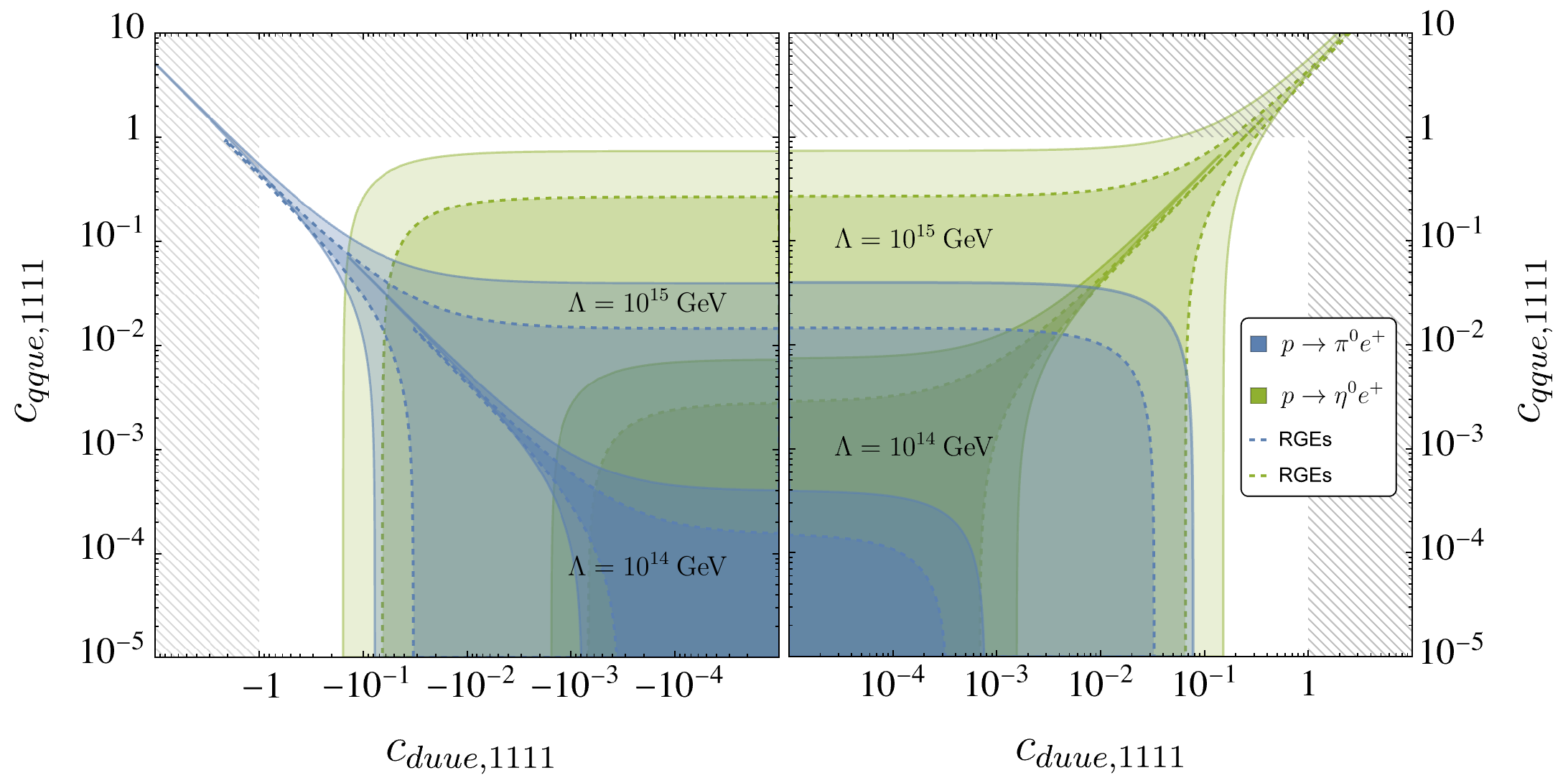}
    \caption{Allowed parameter space of the dimensionless WCs $c_{duue,1111}$ and $c_{qque,1111}$ at the UV scale using the current limits on $p \to \pi^0 e^{+}$~\cite{Super-Kamiokande:2020wjk} and $p \to \eta^0 e^{+}$~\cite{Super-Kamiokande:2017gev}. Inner (outer) regions are obtained by setting $\Lambda=10^{14} \, (10^{15})$ GeV. Dashed lines indicate the effect of including the RGEs, leading always to a shrinking in the available parameter space. The hatched region corresponds to the non-perturbative regime for the WCs.}
    \label{fig:flatdirections6}
\end{figure}

\begin{figure}[h]
    \centering
    \includegraphics[width = \linewidth]
    {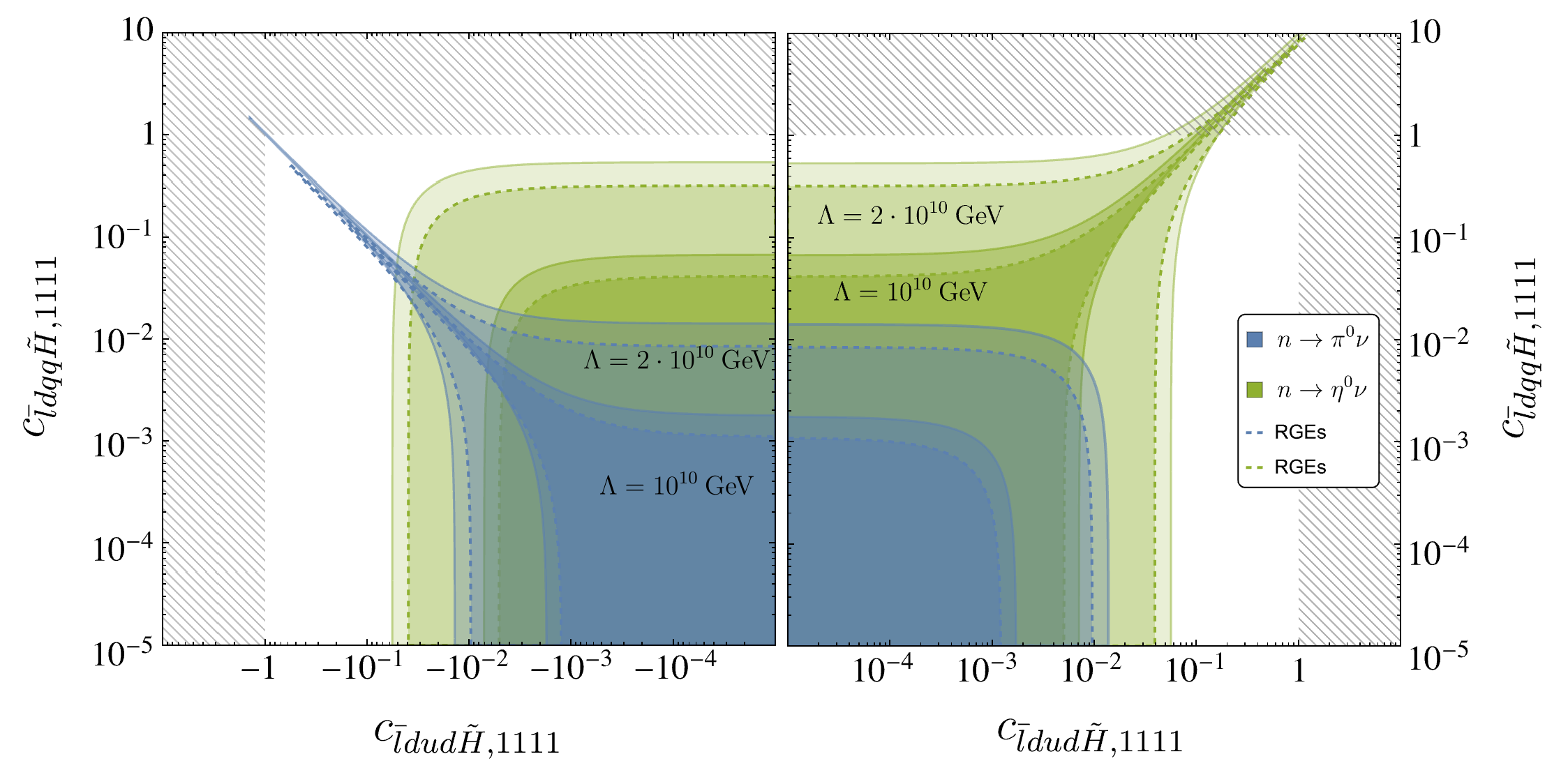}
    \caption{The figure is analogous to Fig.~\ref{fig:flatdirections6} but for $c_{\bar l dud \tilde H,1111}$ and $c_{\bar ld qq \tilde H,1111}$, where the relevant current limits are in this case on $n \to \pi^0\nu$ \cite{Super-Kamiokande:2013rwg} and $n \to \eta^0 \nu $ \cite{McGrew:1999nd}. Inner (outer) regions are obtained by setting $\Lambda=10^{10} \, (2 \cdot 10^{10})$ GeV.}
    \label{fig:flatdirections7}
\end{figure}

We now turn to the discussion of BNV nucleon decay in the presence of two WCs. In Fig.~\ref{fig:flatdirections6} (Fig.~\ref{fig:flatdirections7}) we show the allowed regions in parameter space of the WCs of two dimension-$6$ (dimension-$7$) operators, $\mathcal{O}_{duue,1111}$ and $\mathcal{O}_{qque,1111}$ ($\mathcal{O}_{\bar l dud \tilde H,1111}$ and $\mathcal{O}_{\bar ld qq \tilde H,1111}$). We use two different the energy scales in each plot. We also show the non-perturbative regime for the WCs as hatched regions. The dimension-6 operators induce proton decay via $p\to \pi^0 e^+$ and $p\to\eta^0 e^+$ and the dimension-7 operators neutron decay via $n\to \pi^0 \nu$ and $n\to\eta^0 \nu$. One can observe that the allowed parameter space of the dimensionless WCs always gets reduced once running effects are incorporated.


The figures also clearly illustrate the complementarity of the different nucleon decay searches. While $p\to \pi^0 e^+$ is not sensitive to $c_{qque,1111} = -0.44\; c_{duue,1111}$ and $p\to \eta^0 e^+$ is not sensitive to $c_{qque,1111}=4.1\; c_{duue,1111}$, as illustrated in Fig.~\ref{fig:flatdirections6}, the combination of the two searches does not allow for unconstrained flat directions in parameter space. A similar observation can be made for Fig.~\ref{fig:flatdirections7}. 
Studying the parameter space of pairs of WCs through channels involving the pseudoscalar mesons $\pi^0$ and $\eta^0$ gives us insight into the possibility of flat directions. More generally, the number of decay widths and SMEFT WCs that enter them allow us to determine the possible existence of flat directions. For the strangeness-conserving decay channels, the number of SMEFT WCs saturate the number of observables and no flat directions can be found. However, the reduced number of strangeness-changing processes compared to the many WCs entering the analytic expressions allows us to find flat directions in this sector.

\begin{figure}[tbp!]
    \centering
    \includegraphics[width=1\textwidth]{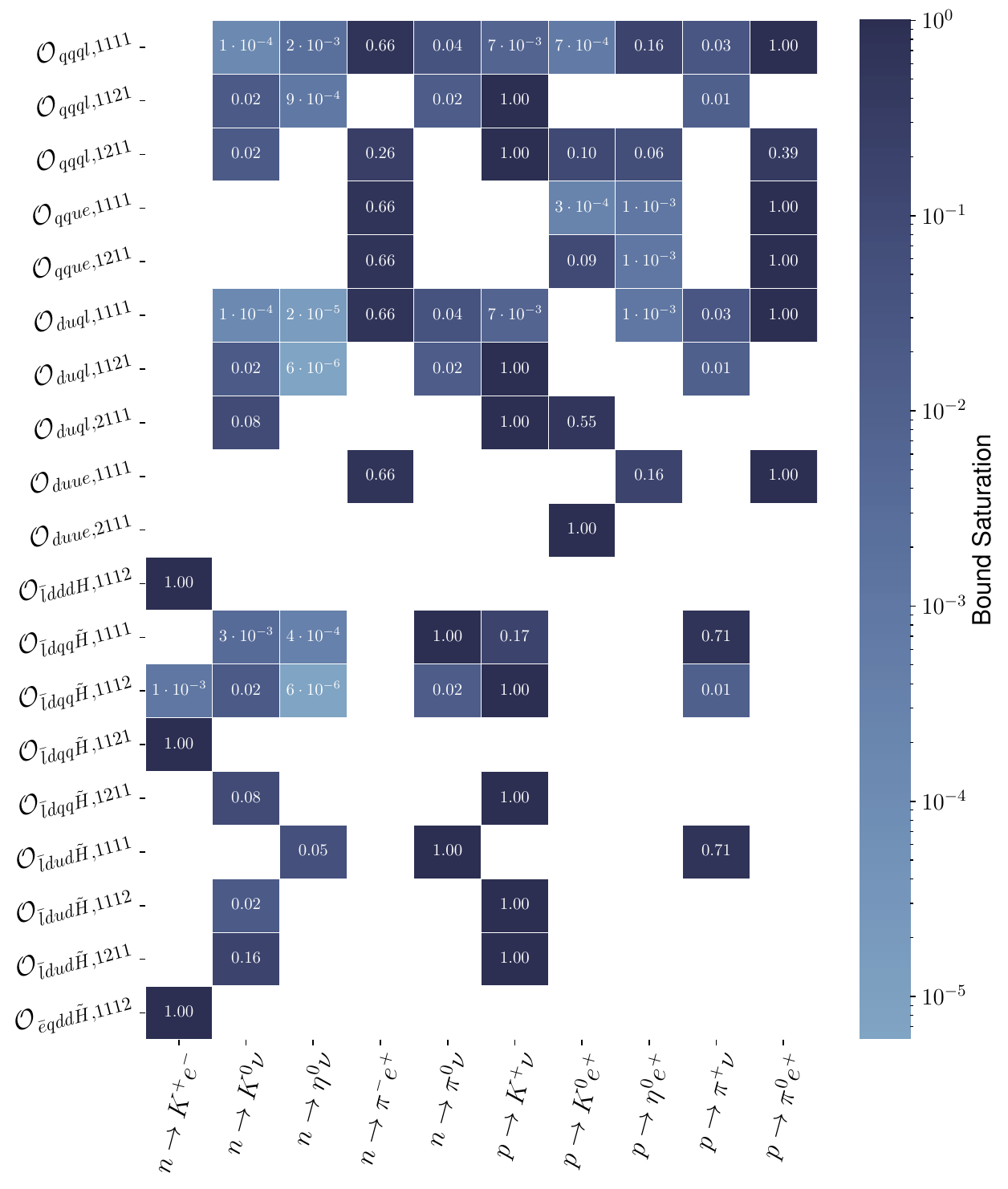}
    \caption{The figure shows correlations among different BNV nucleon decays for one active SMEFT operator at a time. The colour and corresponding numerical value shown in each entry represents the \emph{Bound Saturation} defined in Eq.~\eqref{eq:bound-saturation}, 
    $(\Gamma_{i,\rm th} /\Gamma_{\rm max,th}) /(\Gamma_{i,\rm exp}/\Gamma_{\rm max,exp})$.
    Final states with neutrinos are not distinguished from those with antineutrinos in the figure. Decays mediated by dimension-6 operators feature antineutrinos, while those from dimension-7 operators imply neutrino final states.}
    \label{fig:correlations}
\end{figure}

Before concluding this section, we provide in Fig.~\ref{fig:correlations} the correlations between each SMEFT WC that enters the different $|\Delta (B-L)| = 0, \; 2$ decay widths. We define \emph{Bound\,Saturation} as the ratio between each observable $\Gamma_i$ and the most constrained decay rate $\Gamma_{\rm max}$, both normalised to their current experimental limits,
\begin{align}\label{eq:bound-saturation}
    \text{Bound\,Saturation}\equiv \frac{\Gamma_{i,\rm th} / \Gamma_{\rm max,th}}{\Gamma_{i,\rm exp}/\Gamma_{\rm max,exp}}\,.
\end{align}
Therefore, the closer this quantity is to 1 (darker colour), the more relevant the process is when the SMEFT WC is present. For each operator (row), we have incorporated the RG effects induced by the running. Notice that the channels $p\to \pi^0 e^{+}$ and $p\to K^{+}\nu$ are responsible for the saturation of most of the SMEFT WCs, as can be seen in Tabs.~\ref{tab:general-limits-d6} and \ref{tab:general-limits-d7}. These correlations complement the numerical values for the $\boldsymbol{\kappa}_{(i)}$ matrices, given in App.~\ref{sec:appendixMatrices}, and allow us to determine, for each operator and given the experimental limits of Tab.~\ref{tab:NucleonDecayChannels}, which channel/s are the most promising ones to search for BNV.

\begin{figure}[h]
    \centering
    \includegraphics[width=0.6\textwidth]{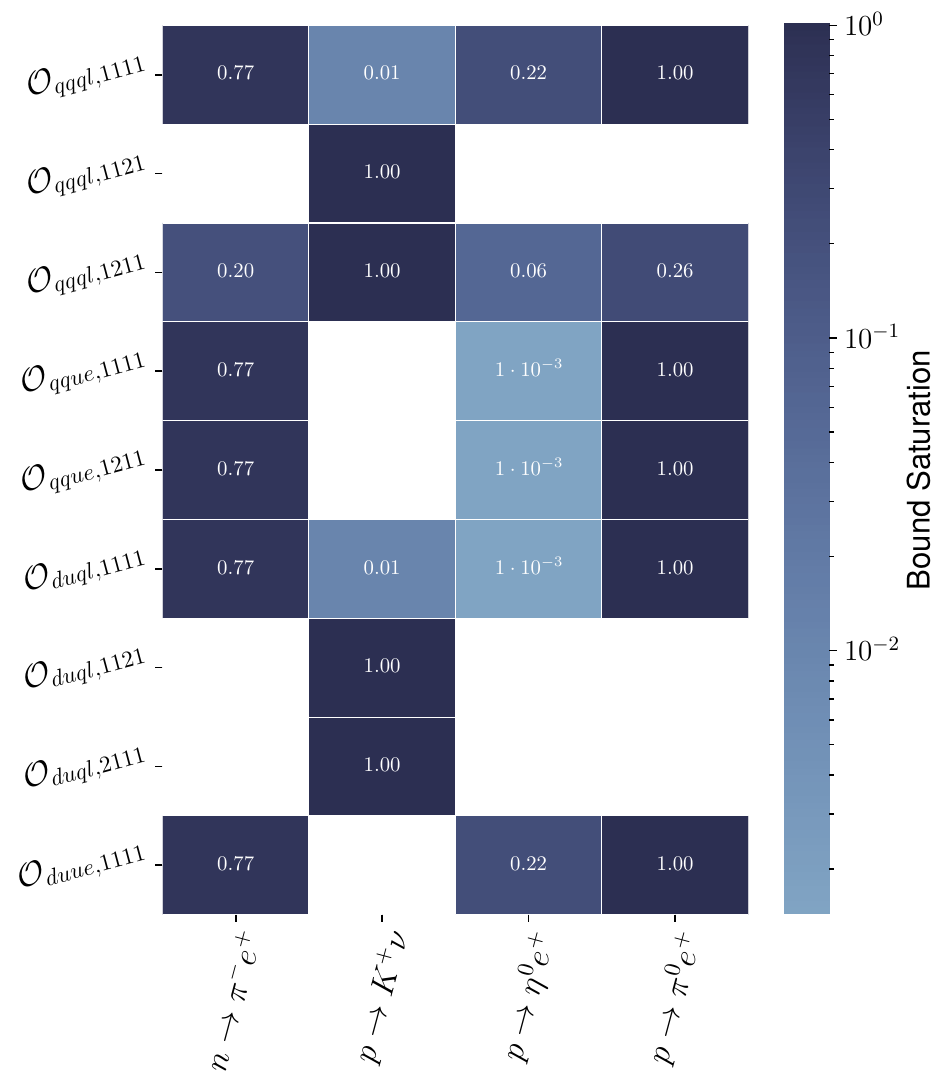}
    \caption{Same as Fig.~\ref{fig:correlations} for correlations between future sensitivities of $\Delta (B-L)=0$ nucleon decay channels and operators; see the main text for more details.}
    \label{fig:futurecorrelations}
\end{figure}

In Fig.~\ref{fig:futurecorrelations}, we illustrate the correlations between distinct decay modes using the future sensitivities of Hyper-K~\cite{Hyper-Kamiokande:2018ofw}. Note that we only provide the correlations for the neutrino and electronic decay channels whose future sensitivities are available in the literature, as summarised in Ref.~\cite{Gargalionis:2024nij}.
Specifically, we concentrate on correlations among $\Delta (B-L) = 0$ SMEFT operators and processes. This decision stems from the limited number of $|\Delta (B-L)| = 2$ channels with explicit sensitivity estimates provided, which are $p\to K^{+}\nu$ and $n \to K^{+} e^{-}$. Notably, the operator $\mathcal{O}_{\bar l dqq\tilde H,1112}$ is of particular interest here as it could potentially induce both processes. However, the absence of additional expected sensitivities for neutrino and muonic $|\Delta (B-L)| = 2$ channels complicates the disentanglement of the action of various SMEFT operators.

This plot provides insights into the operators governing nucleon decays in the event of a future BNV signal at HK. For instance, if detection of the channel $p \to K^{+} \nu$ is closely followed by the channel $p \to \pi^0 e^{+}$, it would suggest that the operator responsible for these decays could be $\mathcal{O}_{qqql,1211}$.

\section{Example: A simplified model} \label{sec:model}

Even though the results derived in this paper are completely general and model-independent, for a particular UV model, a specific flavour structure in the SMEFT WCs is inherited, leading to different nucleon decay channels and relations between SMEFT WCs. To illustrate this we provide a simplified model that induces nucleon decays via $|\Delta (B-L)| = 2$ processes. This model enlarges the SM content with one scalar leptoquark $\omega_2$ which transforms as $(3,1,2/3)$ with respect to the SM gauge group $\left(\mathrm{SU}(3)_C,\mathrm{SU}(2)_L,\mathrm{U}(1)_Y\right)$, and two 2-component Weyl spinors $Q_1,\bar Q_1$ transforming as $Q_1 \sim(3,2,1/6)$ and $\bar Q_1^{\dagger} \sim(3,2,1/6)$, which together form a vector-like fermion, where we have followed the notation of Ref.~\cite{deBlas:2017xtg}. The SM Lagrangian is enlarged with 
\begin{equation}
    \Delta \mathcal{L} = \mathcal{L}_{\omega_2} + \mathcal{L}_{Q_1} + \mathcal{L}_{\rm int}
\end{equation}
where
\begin{equation}
    \mathcal{L}_{\omega_2} =- \omega_2^{\dagger}\left(M_{\omega}^2 + D^2 \right)\omega_2,
\end{equation}
\begin{equation}
    \mathcal{L}_{Q_1} =  i Q_1^{\dagger} \bar \sigma_{\mu}D^{\mu}  Q_1 + i \bar Q_1^{\dagger} \bar \sigma_{\mu}D^{\mu} \bar Q_1 - M_{Q} \Bar{Q}_1 Q_1 - M_{Q} \Bar{Q}_1^{\dagger} Q_1^{\dagger}
\end{equation}
\begin{equation}
    \mathcal{L}_{\rm int} = y_{1,ij} \;\omega_2\Bar{d}^{\dagger i}\Bar{d}^{\dagger j} + y_{2,k}  H^{\dagger}Q_1 \bar d^k + y_{3,k} Q_1 \epsilon H \Bar{u}^k + y_{4,l} \omega_2 \bar Q_1 L^l + \mathrm{h.c.} \,,\label{eq:Lint}
\end{equation}
and the colour indices are implicitly contracted as $\epsilon_{abc}\omega_2^a \bar d^{\dagger b} \bar d^{\dagger c}$. We do not show the $\mathrm{SU}(2)_L$ indices but an anti-symmetric symbol $\epsilon$ is explicitly introduced in expression~\eqref{eq:Lint} to make clear the term is a singlet. The Yukawa coupling $y_1$ is antisymmetric, $y_{1,ij} = - y_{1,ji}$.

\begin{figure}[tb!]
	\centering
	\begin{subfigure}[H]{0.45\textwidth}
		\centering
        \includegraphics[width = 6cm]{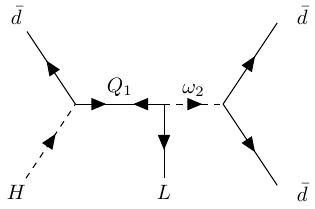}
	\end{subfigure} $\quad \quad$
	\begin{subfigure}[H]{0.45\textwidth}
		\centering
        \includegraphics[width = 6cm]{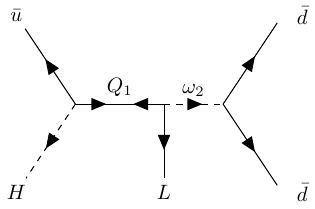}
	\end{subfigure}
    \caption{Feynman diagrams of two $|\Delta (B-L)|=2$ processes mediated by the UV particles $\omega_2$ and $Q_1$. We follow Ref.~\cite{Dreiner:2008tw} for the arrow convention in the diagrams.}
    \label{fig:2models}
\end{figure}

Once we integrate out $\omega_2$ and $Q_1$, we are left with two $d=7$ $|\Delta (B-L)|=2$ SMEFT operators\footnote{A complete analysis of the tree-level matching of UV models and SMEFT operators up to dimension 6 and 7 can be found in Refs.~\cite{deBlas:2017xtg} and~\cite{Li:2023cwy}, respectively.}
\begin{equation}
    \mathcal{L}_{\rm eff} \supset \frac{ y_{1,ij} y_{2,k}^* y_{4,l}^*}{M^2_{\omega}M_Q} (L^{\dagger l} \bar d^{\dagger k}) (\bar d^{\dagger i} \bar d^{\dagger j})H + \frac{ y_{1,ij} y_{3,k}^* y_{4,l}^*}{M^2_{\omega}M_Q}
     (L^{\dagger l}\bar u^{\dagger k})(\bar d^{\dagger i}\bar d^{\dagger j})\tilde H + \mathrm{h.c.} 
     \;.
     \label{eq:effL}
\end{equation}
These are depicted in Fig.~\ref{fig:2models}.

To use the basis for dimension-7 SMEFT operators of Ref.~\cite{Liao:2020zyx} and the BNV operators defined in Sec.~\ref{sec:SMEFT} we need to apply a Fierz transformation to the second operator to get
\begin{equation}
     \frac{ y_{1,ij} y_{3,k}^* y_{4,l}^*}{M^2_{\omega}M_Q}
     (L^{\dagger l}\bar u^{\dagger k})(\bar d^{\dagger i}\bar d^{\dagger j})\tilde H   = 2 \; \frac{ y_{1,ij} y_{3,k}^* y_{4,l}^*}{M^2_{\omega}M_Q}  (L^{\dagger l}\bar d^{\dagger i})(\bar u^{\dagger k}\bar d^{\dagger j})\tilde H\,.
\end{equation}
Therefore we can rewrite our effective operators relevant to nucleon decay as
\begin{equation}
    \mathcal{L}_{\rm eff} \supset C_{\bar l dddH,pqrs} \mathcal{O}_{\bar l dddH,pqrs} + C_{\bar l dud \tilde H,pqrs} \mathcal{O}_{\bar l dud \tilde H,pqrs} + \mathrm{h.c.} 
     \;,
\end{equation}
where the WCs are
\begin{align}
    C_{\bar l dddH,pqrs} &=  \frac{y_{1,rs} y_{2,q}^* y_{4,p}^*}{M_{\omega}^2M_Q} \,,\label{eq:effC1}
\\
C_{\bar l dud \tilde H,pqrs} &= 2 \; \frac{y_{1,qs} y_{3,r}^* y_{4,p}^*}{M_{\omega}^2M_Q}\,. \label{eq:effC2}
\end{align}
In this model, due to the anti-symmetry of the coupling $y_1$, two conclusions can be extracted:
\begin{itemize}
 \item All nucleon decays must be strangeness-changing, such as $p \to K^{+} \nu$, $n \to K^0 \nu$, and $n \to K^{+} e^{-}$. Then only the SMEFT WCs $C_{\bar l dddH,1121}$, $C_{\bar l dud \tilde H,1211}$, and $C_{\bar l dud \tilde H,1112}$ contribute to nucleon decay.
 \item From Eq.~\eqref{eq:effC2}, we can see that $C_{\bar l dud \tilde H,1211} = - C_{\bar l dud \tilde H,1112}$. 
\end{itemize}
Whereas the first two decay channels mentioned above are induced by the LEFT operators $\mathcal{O}^{S,RR}_{udd}$, the channel involving $e^{-}$ in the final state is triggered by the LEFT operator $\mathcal{O}^{S,RR}_{ddd}$ through the matching conditions specified in Tab.~\ref{tab:SMEFT-LEFT}. 
As there are trivially no interference terms between the LEFT WCs $L^{S,RR}_{udd}$ and $L^{S,RR}_{ddd}$ in the decay-width formulae of Sec.~\ref{sec:rate-calcs}, we can use the results of Tab.~\ref{tab:general-limits-d7} for the LEFT WC $L^{S,RR}_{ddd}$ to derive a direct bound on the quantity in Eq.~\eqref{eq:effC1}.
On the other hand, the SMEFT operators $\mathcal{O}_{\bar l dud \tilde H,1211}$ and $\mathcal{O}_{\bar l dud \tilde H,1112}$ match onto 2 different LEFT operators, $\left[\mathcal{O}^{S,RR}_{udd}\right]_{1112}$ and $\left[\mathcal{O}^{S,RR}_{udd}\right]_{1211}$, respectively. Therefore, to derive a bound on the quantity  in Eq.~\eqref{eq:effC2} we use the decay rates computed in Sec.~\ref{sec:rate-calcs} along with the RGEs of Secs.~\ref{sec:SMEFT} and \ref{sec:LEFT} to conclude that the channel $p \to K^{+}\nu$  leads to a more stringent limit on the WC.

\begin{figure}[tb!]
	\centering
        \includegraphics[width = 10cm]{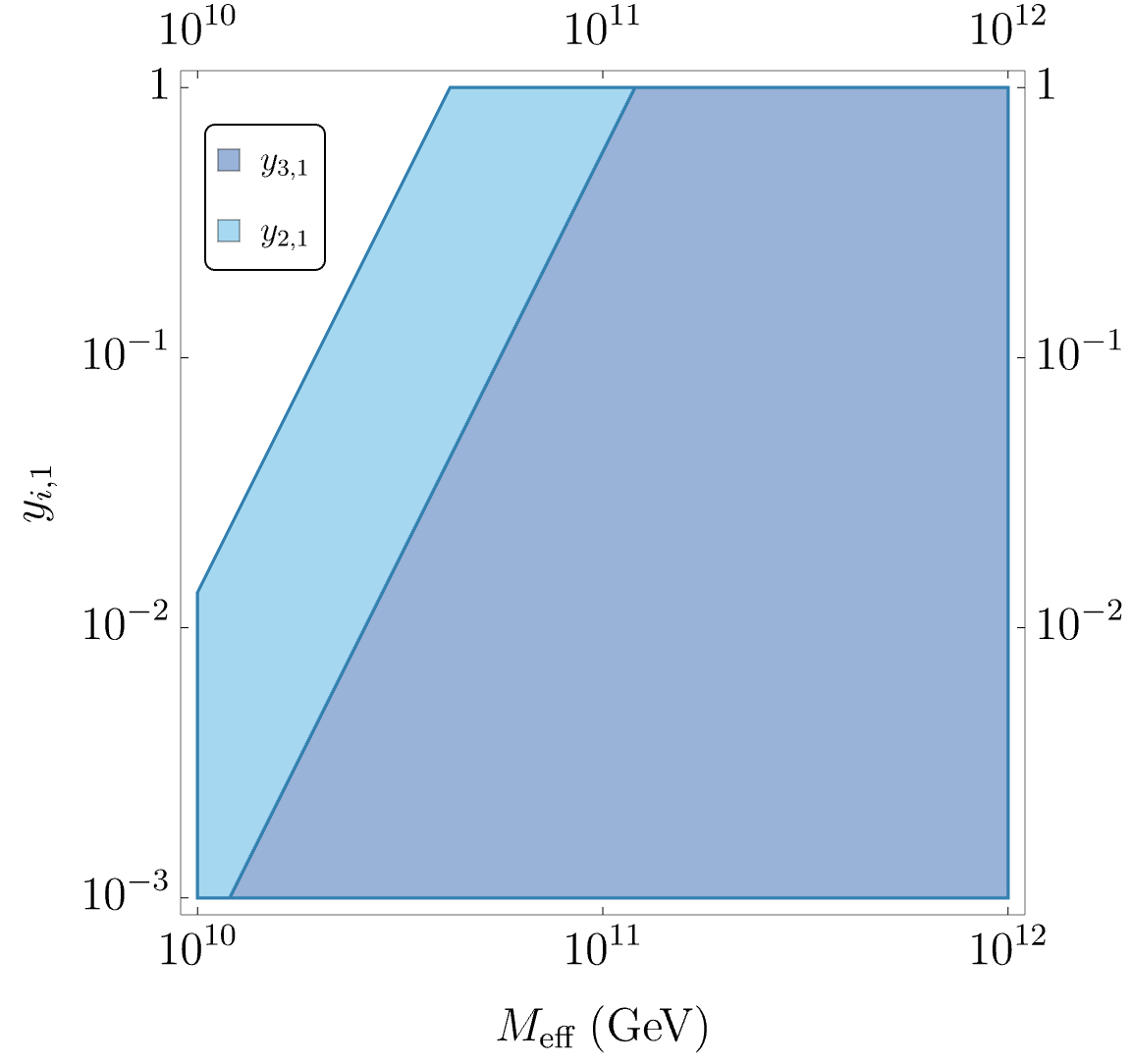}
    \caption{Allowed parameter space (blue coloured region) of the UV model involving one scalar $\omega_2 \sim (3,1,2/3)$ and two 2-component Weyl spinors $Q_1 \sim(3,2,1/6)$ and $\bar Q_1^{\dagger} \sim(3,2,1/6)$.}
    \label{fig:UVmodel}
\end{figure}

In order to present the results, it is useful to define an effective mass,
\begin{equation} \label{eq:meff}
    M_{\rm eff}^3 \equiv \frac{M_{\omega}^2M_Q}{y_{1,12}\,y_{4,1}^*}\,.
\end{equation}
To simplify the analysis, we assume that $M_{\omega}$ and $M_Q$ are of the same order and that we decouple them from the UV theory at the same energy scale. In Fig.~\ref{fig:UVmodel} we plot the allowed parameter space in the plane $y_{2,1}\, (y_{3,1})$ versus $M_{\rm eff}$ in light (dark) blue. We have used the experimental bounds of Tab.~\ref{tab:NucleonDecayChannels}.

We have incorporated the RG effects using the UV scale $1.2 \cdot  10^{11}$ GeV for the parameters $y_{2,1}$ and $y_{3,1}$. However, the choice for the UV scale is rather arbitrary due to the several parameters entering expression~\eqref{eq:meff}, which also depends on the mass hierarchies of the particles $\omega_2$ and $Q_1$. In the fourth column of Tab.~\ref{tab:general-limits-d7} we have shown that the RG corrections for the dimension-7 SMEFT WCs are small. They only lead to an order $\sim$ 0.1 difference on the bound of the WC in Eq.~\eqref{eq:effC2} when doing the running from $\Lambda = 10^{12}$ GeV or from $\Lambda = 10^{10}$ GeV.

\section{Conclusions} \label{sec:conc}

Proton decay is one of the few available probes to test the largest energy scales present in some of the best motivated theories beyond the SM, including 
 Grand Unified Theories and Quantum-Gravity schemes. Any observation of proton decay implies the violation of baryon and lepton numbers, two accidental symmetries of the SM. Depending on the proton decay mode, different combinations of baryon and lepton numbers are broken. The lowest dimensional SMEFT operators contributing to proton decay are $\Delta (B-L) =0$ operators at dimension-6 and $|\Delta(B-L)|=2$ operators at dimension-7. With the increased sensitivity in the upcoming experiments searching for proton decay like Hyper-K, DUNE, JUNO and THEIA, it is timely to revisit baryon number violation from a bottom-up perspective.

In this work, we employed the power of EFT to derive model-independent lower limits on the BNV scale of dimension-6 $\Delta (B-L) = 0$ and dimension-7 $|\Delta (B-L)| = 2$ SMEFT operators following the conventional approach: We matched the SMEFT onto the LEFT and used B$\chi$PT, including RG effects in both the SMEFT and the LEFT. We also computed the $|\Delta (B-L)|=2$ decay rates in the B$\chi$PT formalism, which were absent in the literature to the best of our knowledge.  
The main uncertainty originates from the nuclear matrix elements, which we evaluated using B$\chi$PT, where the low-energy constants are determined using lattice QCD. This method is sometimes also denoted indirect method~\cite{Aoki:2017puj} in contrast to the evaluation of all nuclear matrix elements on the lattice, the so-called direct method.

Running effects consistently result in more stringent lower limits on the scale of the effective operator. It should be noted that for some dimension-6 operators, the RG effects are significant, and the next-order corrections of the anomalous dimensions could be relevant. Naively, our results show that proton decay is consistent with grand unification scales. 
In Appendix \ref{sec:appendixMatrices} we provide the numerical values of the matrices entering together with SMEFT WCs in the decay widths, computed at the largest allowed scale, which can be easily used for a particular model. 
For a given SMEFT operator, the ratios of the decay rates of different modes are independent of the scale and of some of the uncertainties. Positive signals in 2-3 channels may point to particular operators, and therefore set the path to possible UV completions, \eg GUT theories.
Our analysis may also be used to classify and bound possible new particles that couple to SM fields, in the line of Ref.~\cite{Herrero-Garcia:2019czj}. This study will be pursued in an upcoming work.

Finally, we studied possible correlations among various processes. Specifically, we investigated the potential existence of flat directions. The answer turns out to be negative, with the combination of all possible decay modes ensuring that the WCs are always strongly constrained. Similarly, our investigation did not reveal any novel \emph{sum rule} between distinct channels, apart from the already established correlations due to isospin symmetry.

A positive signal in the next decade for long-sought proton decay would open up a new era in the realm of particle physics. If this was the case, pinning down the operator underlying the observed decay mode and looking for the other processes predicted by it would be the first step. If that day were to come, the necessary ingredients for its analysis may be found here.

\acknowledgments

We are grateful to Avelino Vicente, Mart\'in Gonz\'alez-Alonso, Luiz Vale Silva, Alberto Ramos, Tong Li, and Chang-Yuan Yao for useful discussions. We thank Julian Heeck for pointing out a wrong sign in Eq.~\eqref{eq:LagDF}.
The Feynman diagrams were generated using the Ti\textit{k}Z-Feynman and Ti\textit{k}Z-FeynHand packages for \LaTeX~\cite{Ellis:2016jkw,Dohse:2018vqo}. MS acknowledges support by the Australian Research Council through the ARC Discovery Project DP200101470.
JHG, AS, JG, and ABB are partially funded by the Spanish ``Agencia Estatal de Investigación'', MICIN/AEI/10.13039/501100011033, through the grants PID2020-113334GB-I00, and in addition, JHG by PID2020-113644GB-I00 and by the ``Consolidación Investigadora'' Grant CNS2022-135592 and JG by the ``Juan de la Cierva'' programme reference FJC2021-048111-I, both funded also by ``European Union NextGenerationEU/PRTR''. This research is also financed by the ``Generalitat Valenciana'': JHG is also supported through the GenT Excellence Program (CIDEGENT\slash 2020\slash 020), AS is supported by the ``PROMETEO'' programme under grant CIPROM/2022/66, while ABB is funded by grant CIACIF/2021/061.

\appendix
\newpage
\section{Matching to \texorpdfstring{B$\chi$PT}{BchiPT} of BNV dimension-6 LEFT operators}
\label{sec:BchiPT-matching}
\begin{table}[bth!]
  \begin{center}
    \resizebox{\linewidth}{!}{%
 \renewcommand{\arraystretch}{1.3}%
    \begin{tabular}{lcc}
      \toprule
      Name & LEFT & Flavour/B$\chi$PT  \\
      \midrule
      \rowcolor{lightgray}
      \([\mathcal{O}_{udd}^{S,LL}]_{rstu}\) & \((u_{r} d_{s})(d_{t} \nu_{u})\) & \((\mathbf{8}, \mathbf{1})\)     \\ 
       \([\mathcal{O}_{udd}^{S,LL}]_{111r}\) & \((ud)(d\nu_r)\)    & \(-\beta \overline{\nu_{Lr}^c} \mathrm{tr}(\xi B \xi^\dagger P_{32})  \supset -\beta \overline{\nu_{Lr}^c} n - \frac{i \beta}{f_\pi} \overline{\nu_{Lr}^c} \left( \sqrt{\frac32} n \eta - \frac{1}{\sqrt{2}} n\pi^0 + p\pi^-\right)   \)    \\ 
       \([\mathcal{O}_{udd}^{S,LL}]_{121r}\) & \((us)(d\nu_r)\)    & \(-\beta \overline{\nu_{Lr}^c} \mathrm{tr}(\xi B \xi^\dagger \tilde P_{22})  \supset-\beta \overline{\nu_{Lr}^c} \left(-\frac{\Lambda^0}{\sqrt{6}}+\frac{\Sigma^0}{\sqrt{2}}\right) - \frac{i\beta}{f_\pi} \overline{\nu_{Lr}^c} n \bar K^0 \)  \\ 
       \([\mathcal{O}_{udd}^{S,LL}]_{112r}\) &  \((ud)(s\nu_r)\)     & \(-\beta \overline{\nu_{Lr}^c} \mathrm{tr}(\xi B \xi^\dagger P_{33}) \supset  \beta \sqrt{\frac{2}{3}} \overline{\nu_{Lr}^c} \Lambda^0 - \frac{i\beta}{f_\pi} \overline{\nu_{Lr}^c} \left(n \bar K^0 + p K^-\right) \)  \\ 
       \midrule
      \rowcolor{lightgray}
      \([\mathcal{O}_{duu}^{S,LL}]_{rstu}\) & \((d_{r} u_{s})(u_{t} e_{u})\)  & \((\mathbf{8}, \mathbf{1})\)    \\ 
      \([\mathcal{O}_{duu}^{S,LL}]_{111r}\) & \((du)(ue_r)\)    & \(-\beta \overline{e_{Lr}^c}\mathrm{tr}(\xi B \xi^\dagger \tilde P_{31})  \supset\beta \overline{e_{Lr}^c} p +\frac{i\beta}{f_\pi} \overline{e_{Lr}^c}\left(\sqrt{\frac32} p \eta + \frac{1}{\sqrt{2}} p\pi^0 + n\pi^+\right)  \)  \\ 
      \([\mathcal{O}_{duu}^{S,LL}]_{211r}\) & \((su)(ue_r)\)    & \(-\beta \overline{e_{Lr}^c}\mathrm{tr}(\xi B \xi^\dagger  P_{21})   \supset -\beta \overline{e_{Lr}^c}\Sigma^+ +\frac{i\beta}{f_\pi} \overline{e_{Lr}^c} p \bar K^0  \)  \\ 
      \midrule
      \rowcolor{lightgray}
      \([\mathcal{O}_{uud}^{S,LR}]_{[rs]tu}\)  & \((u_{r} u_{s})(\bar{d}^{\dagger}_{t} \bar{e}^{\dagger}_{u})\)  & ---     \\ 
      \midrule
      \rowcolor{lightgray}
      \([\mathcal{O}_{duu}^{S,LR}]_{rstu}\) & \((d_{r} u_{s})(\bar{u}^{\dagger}_{t} \bar{e}^{\dagger}_{u})\)  & \((\bar{\mathbf{3}}, \mathbf{3})\)   \\ 
      \([\mathcal{O}_{duu}^{S,LR}]_{111r}\) & \((d u)(\bar{u}^{\dagger} \bar{e}^{\dagger}_{r})\) &     $\alpha \overline{e_{Rr}^c} \mathrm{tr}(\xi^\dagger B \xi^\dagger \tilde P_{31})  \supset-\alpha \overline{e_{Rr}^c}p 
      +\frac{i\alpha}{f_\pi}\overline{e_{Rr}^c}\left(-\frac{1}{\sqrt{6}} p\eta + \frac{1}{\sqrt{2}} p\pi^0 + n\pi^+\right)  $  \\ 
      \([\mathcal{O}_{duu}^{S,LR}]_{211r}\) & \((s u)(\bar{u}^{\dagger} \bar{e}^{\dagger}_{r})\) &     $\alpha \overline{e_{Rr}^c}\mathrm{tr}(\xi^\dagger B \xi^\dagger P_{21}) \supset\alpha\overline{e_{Rr}^c} \Sigma^+ -\frac{i\alpha}{f_\pi} \overline{e_{Rr}^c} p \bar K^0 $ \\ 
      \midrule
      \rowcolor{lightgray}
      \([\mathcal{O}_{uud}^{S,RL}]_{[rs]tu}\)  & \((\bar{u}^{\dagger}_{r} \bar{u}^{\dagger}_{s})(d_{t} e_{u})\)  & ---    \\
      \midrule
      \rowcolor{lightgray}
      \([\mathcal{O}_{duu}^{S,RL}]_{rstu}\) & \((\bar{d}^{\dagger}_{r} \bar{u}^{\dagger}_{s})(u_{t} e_{u})\)  & \((\mathbf{3}, \bar{\mathbf{3}})\)  \\
      \([\mathcal{O}_{duu}^{S,RL}]_{111r}\) & \((\bar{d}^{\dagger} \bar{u}^{\dagger})(u e_{r})\) &    $-\alpha \overline{e_{Lr}^c} \mathrm{tr}(\xi B \xi \tilde P_{31})  \supset\alpha\overline{e_{Lr}^c}p +\frac{i\alpha}{f_\pi}\overline{e_{Lr}^c} \left(-\frac{1}{\sqrt{6}}p\eta + \frac{1}{\sqrt{2}} p\pi^0 + n\pi^+\right)  $  \\ 
      \([\mathcal{O}_{duu}^{S,RL}]_{211r}\) & \((\bar{s}^{\dagger} \bar{u}^{\dagger})(u e_{r})\) &    $-\alpha \overline{e_{Lr}^c}\mathrm{tr}(\xi B \xi P_{21}) \supset-\alpha\overline{e_{Lr}^c}\Sigma^+ -\frac{i\alpha}{f_\pi} \overline{e_{Lr}^c} p \bar K^0  $  \\ 
      \midrule
      \rowcolor{lightgray}
      \([\mathcal{O}_{dud}^{S,RL}]_{rstu}\)  & \((\bar{d}^{\dagger}_{r} \bar{u}^{\dagger}_{s})(d_{t} \nu_{u})\)  & \((\mathbf{3}, \bar{\mathbf{3}})\)    \\
      \([\mathcal{O}_{dud}^{S,RL}]_{111r}\)  & \((\bar{d}^{\dagger} \bar{u}^{\dagger})(d \nu_{r})\)    & $\alpha\overline{\nu_{Lr}^c}\mathrm{tr}(\xi B \xi \tilde P_{32})     \supset-\alpha\overline{\nu_{Lr}^c}n
      +\frac{i\alpha}{f_\pi} \overline{\nu_{Lr}^c} \left(\frac{1}{\sqrt{6}}n\eta + \frac{1}{\sqrt{2}}n\pi^0 - p\pi^-\right)  $  \\ 
      \([\mathcal{O}_{dud}^{S,RL}]_{211r}\)  & \((\bar{s}^{\dagger} \bar{u}^{\dagger})(d \nu_{r})\)     & $\alpha\overline{\nu_{Lr}^c}\mathrm{tr}(\xi B \xi  P_{22})  \supset\alpha\overline{\nu_{Lr}^c} \left(\frac{\Lambda^0}{\sqrt{6}}-\frac{\Sigma^0}{\sqrt{2}}\right) +\frac{i\alpha}{f_\pi} \overline{\nu_{Lr}^c} n \bar K^0   $  \\ 
      \([\mathcal{O}_{dud}^{S,RL}]_{112r}\)  & \((\bar{d}^{\dagger} \bar{u}^{\dagger})(s \nu_{r})\)     & $\alpha\overline{\nu_{Lr}^c}\mathrm{tr}(\xi B \xi  \tilde P_{33})  \supset\alpha\overline{\nu_{Lr}^c} \sqrt{\frac{2}{3}} \Lambda^0  -\frac{i\alpha}{f_\pi} \overline{\nu_{Lr}^c} \left(n\bar K^0 + p K^-\right)   $ \\ 
      \([\mathcal{O}_{dud}^{S,RL}]_{212r}\)  & \((\bar{s}^{\dagger} \bar{u}^{\dagger})(s \nu_{r})\)   & $\alpha\overline{\nu_{Lr}^c}\mathrm{tr}(\xi B \xi  P_{23}) \supset\alpha\overline{\nu_{Lr}^c}\Xi^0    $  \\ 
      \midrule
      \rowcolor{lightgray}
      \([\mathcal{O}_{ddu}^{S,RL}]_{[rs]tu}\)  & \((\bar{d}^{\dagger}_{r} \bar{d}^{\dagger}_{s})(u_{t} \nu_{u})\)  & \((\mathbf{3}, \bar{\mathbf{3}})\)  \\
      \([\mathcal{O}_{ddu}^{S,RL}]_{[12]1r}\)  & \((\bar{d}^{\dagger} \bar{s}^{\dagger})(u \nu_{r})\) &    $ -\alpha \overline{\nu_{Lr}^c}  \mathrm{tr}(\xi B \xi P_{11})   \supset\alpha \overline{\nu_{Lr}^c} \left(\frac{\Lambda^0}{\sqrt{6}} + \frac{\Sigma^0}{\sqrt{2}}\right) -\frac{i\alpha}{f_\pi} \overline{\nu_{Lr}^c} p K^-$  \\ 
      \midrule
      \rowcolor{lightgray}
      \([\mathcal{O}_{duu}^{S,RR}]_{rstu}\)  & \((\bar{d}^{\dagger}_{r}     \bar{u}^{\dagger}_{s})(\bar{u}^{\dagger}_{t} \bar{e}^{\dagger}_{u})\)  & \((\mathbf{1}, \mathbf{8})\)   \\
      \([\mathcal{O}_{duu}^{S,RR}]_{111r}\)  & \((\bar{d}^{\dagger}     \bar{u}^{\dagger})(\bar{u}^{\dagger} \bar{e}^{\dagger}_{r})\) &    $\beta \overline{e_{Rr}^c} \mathrm{tr}(\xi^\dagger B \xi \tilde P_{31})    \supset -\beta \overline{e_{Rr}^c} p +\frac{i\beta}{f_\pi} \overline{e_{Rr}^c} \left(\sqrt{\frac32} p \eta + \frac{1}{\sqrt{2}} p \pi^0 + n \pi^+\right)  $ \\ 
      \([\mathcal{O}_{duu}^{S,RR}]_{211r}\)  & \((\bar{s}^{\dagger}     \bar{u}^{\dagger})(\bar{u}^{\dagger} \bar{e}^{\dagger}_{r})\) &    $\beta  \overline{e_{Rr}^c}\mathrm{tr}(\xi^\dagger B \xi P_{21})   \supset \beta \overline{e_{Rr}^c} \Sigma^+ +\frac{i\beta}{f_\pi} \overline{e_{Rr}^c} p \bar K^0 $   \\ 
      \bottomrule
      \end{tabular}}
\end{center}

  \caption{\label{tab:MatchingBchiPT0} Matching of LEFT to B$\chi$PT for $\Delta (B-L)=0$ operators. The colour indices are contracted as $\epsilon_{abc}(q^a q^b)(q^c \ell)$ and minus signs originate from the nuclear matrix elements.}
\end{table}      

 \begin{table}[tbp!]
  \begin{center}
    \resizebox{\linewidth}{!}{%
\renewcommand{\arraystretch}{1.3}%
    \begin{tabular}{lcc}
      \toprule
      Name & LEFT  & Flavour/B$\chi$PT  \\
      \midrule
      \rowcolor{lightgray}
      \([\mathcal{O}_{ddd}^{S,LL}]_{[rs]tu}\) & \((d_{r} d_{s})(\bar{e}_{t} d_{u})\)  & \((\mathbf{8}, \mathbf{1})\)   \\
      \([\mathcal{O}_{ddd}^{S,LL}]_{[12]r1}\) & \((d s)(\bar{e}_{r} d)\) &    \(-\beta\overline{e_{Rr}}  \mathrm{tr}(\xi B \xi^\dagger  P_{12})  \supset-\beta\overline{e_{Rr}} \Sigma^- +\frac{i\beta}{f_\pi} \overline{e_{Rr}} n K^-  \)  \\ 
      \([\mathcal{O}_{ddd}^{S,LL}]_{[12]r2}\) & \((d s)(\bar{e}_{r} s)\) &   \(-\beta \overline{e_{Rr}} \mathrm{tr}(\xi B \xi^\dagger  P_{13}) \supset -\beta\overline{e_{Rr}}\Xi^-  \) \\ 
      \midrule
      \rowcolor{lightgray}
      \([\mathcal{O}_{udd}^{S,LR}]_{rstu}\) & \((u_{r} d_{s})(\nu^{\dagger}_{t} \bar{d}^{\dagger}_{u})\)  & \((\bar{\mathbf{3}}, \mathbf{3})\) \\
      \([\mathcal{O}_{udd}^{S,LR}]_{11r1}\) & \((u d)(\nu^{\dagger}_{r} \bar{d}^{\dagger})\) &   $\alpha\overline{\nu_{Lr}}\mathrm{tr}(\xi^\dagger B \xi^\dagger P_{32}) \supset \alpha\overline{\nu_{Lr}} n +\frac{i\alpha}{f_\pi} \overline{\nu_{Lr}} \left(\frac{1}{\sqrt{6}} n \eta + \frac{1}{\sqrt{2}} n \pi^0 -p\pi^-\right)  $ \\ 
      \([\mathcal{O}_{udd}^{S,LR}]_{12r1}\) & \((u s)(\nu^{\dagger}_{r} \bar{d}^{\dagger})\) &  $\alpha\overline{\nu_{Lr}}\mathrm{tr}(\xi^\dagger B \xi^\dagger \tilde P_{22})  \supset \alpha \overline{\nu_{Lr}} \left(\frac{\Sigma^0}{\sqrt{2}}-\frac{\Lambda^0}{\sqrt{6}}\right)  +\frac{i\alpha}{f_\pi} \overline{\nu_{Lr}}n\bar K^0 $ \\ 
      \([\mathcal{O}_{udd}^{S,LR}]_{11r2}\) & \((u d)(\nu^{\dagger}_{r} \bar{s}^{\dagger})\) &  $\alpha\overline{\nu_{Lr}}\mathrm{tr}(\xi^\dagger B \xi^\dagger  P_{33})  \supset -\alpha \sqrt{\frac{2}{3}} \overline{\nu_{Lr}}\Lambda^0 -\frac{i\alpha}{f_\pi} \overline{\nu_{Lr}} \left(n\bar K^0+pK^-\right) $ \\ 
     \([\mathcal{O}_{udd}^{S,LR}]_{12r2}\) & \((u s)(\nu^{\dagger}_{r} \bar{s}^{\dagger})\) & $\alpha\overline{\nu_{Lr}}\mathrm{tr}(\xi^\dagger B \xi^\dagger \tilde P_{23}) \supset -\alpha \overline{\nu_{Lr}} \Xi^0 $ \\ 
      \midrule
      \rowcolor{lightgray}
      \([\mathcal{O}_{ddu}^{S,LR}]_{[rs]tu}\)  & \((d_{r} d_{s})(\nu^{\dagger}_{t} \bar{u}^{\dagger}_{u})\)  & \((\bar{\mathbf{3}}, \mathbf{3})\)   \\ 
      \([\mathcal{O}_{ddu}^{S,LR}]_{[12]r1}\)  & \((d s)(\nu^{\dagger}_{r} \bar{u}^{\dagger})\) &     $\alpha \overline{\nu_{Lr}}\mathrm{tr}(\xi^\dagger B \xi^\dagger P_{11}) \supset \alpha\overline{\nu_{Lr}} \left(\frac{\Lambda^0}{\sqrt{6}}+\frac{\Sigma^0}{\sqrt{2}}\right) -\frac{i\alpha}{f_\pi} \overline{\nu_{Lr}}p K^- $\\ 
      \midrule
      \rowcolor{lightgray}
      \([\mathcal{O}_{ddd}^{S,LR}]_{[rs]tu}\) & \((d_{r} d_{s})(e^{\dagger}_{t} \bar{d}^{\dagger}_{u})\)  & \((\bar{\mathbf{3}}, \mathbf{3})\)  \\
      \([\mathcal{O}_{ddd}^{S,LR}]_{[12]r1}\) & \((d s)(e^{\dagger}_{r} \bar{d}^{\dagger})\) &   $\alpha\overline{e_{Lr}} \mathrm{tr}(\xi^\dagger B \xi^\dagger P_{12})  \supset\alpha\overline{e_{Lr}} \Sigma^- -\frac{i\alpha}{f_\pi} \overline{e_{Lr}}n K^-   $\\ 
      \([\mathcal{O}_{ddd}^{S,LR}]_{[12]r2}\) & \((d s)(e^{\dagger}_{r} \bar{s}^{\dagger})\) &   $\alpha \overline{e_{Lr}}\mathrm{tr}(\xi^\dagger B \xi^\dagger P_{13})    \supset\alpha\overline{e_{Lr}} \Xi^-  $ \\ 
      \midrule
      \rowcolor{lightgray}
      \([\mathcal{O}_{ddd}^{S,RL}]_{[rs]tu}\)  & \((\bar{d}^{\dagger}_{r} \bar{d}^{\dagger}_{s})(\bar e_{t} d_{u})\)  & \((\mathbf{3}, \bar{\mathbf{3}})\)   \\
      \([\mathcal{O}_{ddd}^{S,RL}]_{[12]r1}\)  & \((\bar{d}^{\dagger} \bar{s}^{\dagger})(\bar e_{r} d)\) &     $\alpha \overline{e_{Rr}}\mathrm{tr}(\xi B \xi P_{12}) \supset \alpha \overline{e_{Rr}} \Sigma^- + \frac{i\alpha}{f_\pi} \overline{e_{Rr}}n K^-$\\ 
      \([\mathcal{O}_{ddd}^{S,RL}]_{[12]r2}\)  & \((\bar{d}^{\dagger} \bar{s}^{\dagger})(\bar e_{r} s)\) &    $\alpha  \overline{e_{Rr}}\mathrm{tr}(\xi B \xi P_{13}) \supset \alpha\overline{e_{Rr}} \Xi^-$   \\ 
      \midrule
      \rowcolor{lightgray}
      \([\mathcal{O}_{udd}^{S,RR}]_{rstu}\) & \((\bar{u}^{\dagger}_{r} \bar{d}^{\dagger}_{s})(\nu^{\dagger}_{t} \bar{d}^{\dagger}_{u})\)  & \((\mathbf{1}, \mathbf{8})\)  \\
      \([\mathcal{O}_{udd}^{S,RR}]_{11r1}\) & \((\bar{u}^{\dagger} \bar{d}^{\dagger})(\nu^{\dagger}_{r} \bar{d}^{\dagger})\) &     $\beta  \overline{\nu_{Lr}}\mathrm{tr}(\xi^\dagger B \xi P_{32}) \supset \beta \overline{\nu_{Lr}} n -\frac{i\beta}{f_\pi} \overline{\nu_{Lr}} \left(\sqrt{\frac32} n\eta - \frac{1}{\sqrt{2}}n\pi^0 + p\pi^-\right)  $\\ 
      \([\mathcal{O}_{udd}^{S,RR}]_{12r1}\) & \((\bar{u}^{\dagger} \bar{s}^{\dagger})(\nu^{\dagger}_{r} \bar{d}^{\dagger})\) &     $\beta  \overline{\nu_{Lr}}\mathrm{tr}(\xi^\dagger B \xi \tilde P_{22}) \supset \beta \overline{\nu_{Lr}} \left(\frac{\Sigma^0}{\sqrt{2}} - \frac{\Lambda^0}{\sqrt{6}}\right) -\frac{i\beta}{f_\pi} \overline{\nu_{Lr}}n \bar K^0
 $\\ 
       \([\mathcal{O}_{udd}^{S,RR}]_{11r2}\) & \((\bar{u}^{\dagger} \bar{d}^{\dagger})(\nu^{\dagger}_{r} \bar{s}^{\dagger})\) &     $\beta  \overline{\nu_{Lr}}\mathrm{tr}(\xi^\dagger B \xi P_{33})  \supset-\beta \sqrt{\frac23} \overline{\nu_{Lr}}\Lambda^0 -\frac{i\beta}{f_\pi} \overline{\nu_{Lr}} \left(n\bar K^0 + p K^-\right)  $\\ 
      \([\mathcal{O}_{udd}^{S,RR}]_{12r2}\) & \((\bar{u}^{\dagger} \bar{s}^{\dagger})(\nu^{\dagger}_{r} \bar{s}^{\dagger})\) &   $\beta \overline{\nu_{Lr}} \mathrm{tr}(\xi^\dagger B \xi \tilde P_{23}) \supset -\beta \overline{\nu_{Lr}}\Xi^0 $  \\ 
      \midrule
      \rowcolor{lightgray}
      \([\mathcal{O}_{ddd}^{S,RR}]_{[rs]tu}\)  & \((\bar{d}^{\dagger}_{r} \bar{d}^{\dagger}_{s})(e^{\dagger}_{t} \bar{d}^{\dagger}_{u})\)  & \((\mathbf{1}, \mathbf{8})\)   \\
      \([\mathcal{O}_{ddd}^{S,RR}]_{[12]r1}\)  & \((\bar{d}^{\dagger} \bar{s}^{\dagger})(e^{\dagger}_{r} \bar{d}^{\dagger})\) &    $\beta\overline{e_{Lr}} \mathrm{tr}(\xi^\dagger B \xi P_{12})  \supset \beta \overline{e_{Lr}} \Sigma^- + \frac{i\beta}{f_\pi} \overline{e_{Lr}}nK^- $\\
     \([\mathcal{O}_{ddd}^{S,RR}]_{[12]r2}\)  & \((\bar{d}^{\dagger} \bar{s}^{\dagger})(e^{\dagger}_{r} \bar{s}^{\dagger})\) &   $\beta\overline{e_{Lr}} \mathrm{tr}(\xi^\dagger B \xi P_{13}) \supset \beta \overline{e_{Lr}} \Xi^-  $ \\
      \bottomrule
    \end{tabular}}
  \end{center}
  \caption{\label{tab:MatchingBchiPT2} Matching of LEFT to B$\chi$PT for $|\Delta (B-L)|=2$ operators. The colour indices are contracted as $\epsilon_{abc}(q^a q^b)(q^c \ell)$ and minus signs originate from the nuclear matrix elements.}
\end{table}

\clearpage

\section{Nucleon decay rates based on the effective Lagrangian}
\label{sec:rateformulae}

The relevant interactions for nucleon decay, can be mapped onto the effective Lagrangian
\begin{align}\label{eq:LagBL2}
    \mathcal{L} & = g_{MB}^N \overline{B} \gamma^\mu \gamma_5 N \partial_\mu M
    + m_{B\alpha,X} \overline{\ell_{\alpha}} P_{\bar X} B + i y_{M\alpha,X}^N \overline{\ell_{\alpha}} P_{\bar X} N M + \mathrm{h.c.}
\end{align}
where $\ell_\alpha=\nu_\alpha,e_\alpha$ and $X=L,R$ denotes the lepton chirality, and $\bar X$ denotes the opposite chirality, i.e.~$\bar L = R$ and $\bar R = L$. 

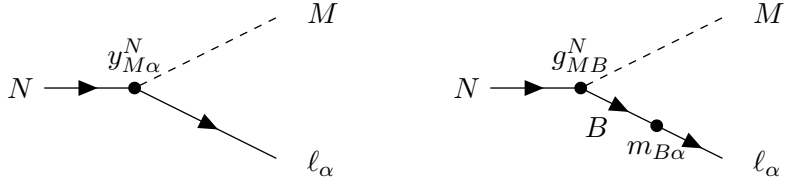
\begin{figure}[h!]
    \centering
   \begin{tikzpicture}
    \begin{feynhand}
    \vertex (i) at (-1.5,0) {$N$};
    \vertex[dot,label=above:$y_{M\alpha}^N$] (v) at (0,0) {};
    \vertex[label={right:$M$}] (m) at (2,1) {};
    \vertex[label={right:$\ell_\alpha$}] (l) at (2,-1) {};
    \propag [fermion] (i) to (v);
    \propag [fermion] (v) to (l);
    \propag [scalar] (v) to (m);
    \end{feynhand}
    \end{tikzpicture}
    \hspace{1cm}
    \begin{tikzpicture}
    \begin{feynhand}
    \vertex (i) at (-1.5,0) {$N$};
    \vertex[dot,label=above:$g_{MB}^N$] (v) at (0,0) {};
    \vertex[dot,label=below:$m_{B\alpha}$] (w) at (1,-0.5) {};
    \vertex[label={right:$M$}] (m) at (2,1) {};
    \vertex[label={right:$\ell_\alpha$}] (l) at (2,-1) {};
    \propag [fermion] (i) to (v);
    \propag [fermion] (v) to[edge label'=$B$] (w);
    \propag [fermion] (w) to (l);
    \propag [scalar] (v) to (m);
    \end{feynhand}
    \end{tikzpicture}
    
    \caption{Nucleon decay to a meson and a lepton induced by $|\Delta (B-L)|=2$ interactions.}
    \label{fig:Ndecay}
\end{figure}

At leading order there are two contributions to nucleon decay to a meson and a charged lepton, pole and non-pole contributions which are illustrated in Fig.~\ref{fig:Ndecay}. 
Assuming the fermion flow in the diagrams goes from left to right, we thus find for the leading-order amplitude of nucleon decay for $|\Delta (B-L)|=2$ violating processes
\begin{align}\label{eq:ampBL2}
    i\mathcal{M} =  \bar u_\alpha \sum_{X=L,R} P_{\bar X}\left(-y_{M\alpha,X}^N   + \sum_B m_{B\alpha,X} \frac{\slashed{k}+m_B}{k^2-m_B^2} g_{MB}^N \slashed{p}\gamma_5\right)  u_N  
\end{align}
with the lepton momentum $k$ and meson momentum $p$, where depending on the operator the lepton chirality is fixed. The summation over the baryons includes any of $p,n,\Lambda^0,\Sigma^{0,\pm}$.
The spin-averaged squared matrix element thus becomes
\begin{equation} 
\begin{aligned}\label{eq:MsquaredBL2}
    \frac{\overline{|M|^2}}{m_N^2} 
    &= 
    \frac12\sum_{X=L,R} |y_{M\alpha,X}^N|^2  \left(1+x_\alpha^2-x_M^2\right) 
    + x_\alpha \mathrm{Re}(y_{M\alpha,L}^N y_{M\alpha,R}^{N*})
    \\
    &+\sum_{B,B^\prime} g_{MB}^N  g_{MB^\prime}^{N*} \left( \frac12\sum_{X=L,R}   m_{B\alpha,X}m_{B^\prime\alpha,X}^* g(x_B,x_{B^\prime})
    + \mathrm{Re}(m_{B\alpha,L} m_{B^\prime\alpha,R}^*)  g_{LR}(x_B,x_{B^\prime})
    \right)
    \\&
    + \sum_{B} \mathrm{Re}(
    y_{M\alpha,R} m_{B\alpha,R}^* g_{MB}^{N*} 
    -y_{M\alpha,L} m_{B\alpha,L}^* g_{MB}^{N*}
    ) h(x_B)  
    \\
    &
    + \sum_{B} \mathrm{Re}(
    y_{M\alpha,R} m_{B\alpha,L}^* g_{MB}^{N*} 
    -y_{M\alpha,L} m_{B\alpha,R}^* g_{MB}^{N*} 
    ) h_{LR}(x_B)  
\end{aligned}
\end{equation}
where $m_N$ is the mass of the nucleon $N=p,n$ and we define the mass ratios $x_M=m_M/m_N$, $x_\alpha=m_\alpha/m_N$, $x_B=m_B/m_N$ together with the kinematic functions
\begin{equation}
\begin{aligned}
     g(x_1,x_2)&=\frac{(x_\alpha^2+x_1x_2)(1-x_M^2 -x_\alpha^2(2+x_M^2) +x_\alpha^4) -2 x_\alpha^2 x_M^2 (x_1+ x_2)}{(x_1^2-x_\alpha^2)(x_2^2-x_\alpha^2)}\;,
     \\
     g_{LR}(x_1,x_2)&=x_\alpha\frac{-2x_M^2(x_\alpha^2+x_1x_2) +(1-x_M^2-x_\alpha^2(x_M^2+2)+x_\alpha^4)(x_1+x_2)}{(x_1^2-x_\alpha^2)(x_2^2-x_\alpha^2)} \;,
     \\
     h(x_B) & = \frac{x_B(1-x_M^2-x_\alpha^2) -x_\alpha^2(1+x_M^2-x_\alpha^2)}{x_B^2-x_\alpha^2}\;,
     \\
     h_{LR}(x_B) &=x_\alpha \frac{1-x_M^2-x_\alpha^2-x_B(1+x_M^2-x_\alpha^2)}{x_B^2-x_\alpha^2} \;.
\end{aligned}
\end{equation}
The branching ratio is then given by
\begin{align}
    \Gamma(N\to M \ell_\alpha) & = m_N\, \frac{\lambda^{1/2}(1,x_M^2,x_\alpha^2)}{16\pi} \frac{\overline{|M|^2}}{m_N^2}
\end{align}
in terms of the K\"all\'en function $\lambda(x,y,z)=x^2+y^2+z^2-2xy-2xz-2yz$. 

Neglecting the final state lepton mass there is no interference between the different chiralities and the decay width reduces to
\begin{equation}
\begin{aligned}
    \Gamma(N\to M \ell_\alpha) & = 
    \frac{m_N}{32\pi}\sum_{X=L,R} |y_{M\alpha,X}^N|^2  \left(1-x_M^2\right)^2 
    \\
    &+\frac{m_N}{32\pi}\sum_{B,B^\prime} g_{MB}^N  g_{MB^\prime}^{N*} 
     \sum_{X=L,R}   m_{B\alpha,X}m_{B^\prime\alpha,X}^*  \frac{(1-x_M)^2}{x_Bx_{B^\prime}}
    \\&
    + \frac{m_N}{16\pi}\sum_{B} \mathrm{Re}(
    y_{M\alpha,R} m_{B\alpha,R}^* g_{MB}^{N*} 
    -y_{M\alpha,L} m_{B\alpha,L}^* g_{MB}^{N*}
    ) \frac{(1-x_M^2)^2}{x_B}\,.
\end{aligned}
\end{equation}
The effective $\Delta(B-L)=0$ Lagrangian is obtained by replacing $\ell \to \ell^c$ and the projection operator $P_{\bar X}$ by $P_X$ in Eq.~\eqref{eq:LagBL2}. Similarly, the amplitude for $\Delta(B-L)=0$ is obtained by replacing $X\to\bar X$ in Eq.~\eqref{eq:ampBL2} and the squared matrix element for $\Delta(B-L)=0$ violating nucleon decays is obtained from Eq.~\eqref{eq:MsquaredBL2} by reversing the overall sign in the last two lines.
Note that there is no interference between $\Delta(B-L)=0$ and $|\Delta(B-L)|=2$ decays for massless neutrinos. Explicit expressions for the $|\Delta(B-L)|=2$ nucleon decay rates are provided in Sec.~\ref{sec:rate-calcs}.

\section{Numerical values of the $\boldsymbol{\kappa}_{(i)}$ matrices entering the decay rates} \label{sec:appendixMatrices}

Here we provide the numerical matrices  $\boldsymbol{\kappa}_{(i)}$ introduced in Sec.~\ref{sec:rate-calcs} for both $|\Delta (B-L)| = 0, \: 2$ decays. We have used 
$D=0.730$, $F=0.447$~\cite{Bali:2022qja} and $\beta = -\alpha=0.0126$~\cite{Yoo:2021gql} for the low-energy constants, and we have incorporated the running effects from the UV scale $10^{16} \; (1.5 \cdot 10^{11})$ GeV down to $2$ GeV for the $|\Delta (B-L)| = 0\; (2)$ decays into a scalar meson and anti-lepton (lepton). From Eqs.~\eqref{eq:kd6} and \eqref{eq:kd7} one sees that these matrices are hermitian. However, the presence of non-real components stems solely from first-to-third quark transitions, which are suppressed in the CKM matrix by a factor of $\mathcal{O}(10^{-4})$. Consequently, we may confidently presume the $\kappa$-matrix elements to be real. Finally, note that these matrices provide the weight of each SMEFT operator obtained after matching the UV theory to the SMEFT, and may be used to obtain the contributions of each operator to the different nucleon decay modes. We have made the matrices available online for ease of use by the community~\cite{kappa-matrices-zenodo}.

\begin{figure}[tbp!]
        \begin{subfigure}[H]{0.55\textwidth}
		\centering
        \includegraphics[width = 1.01\textwidth]{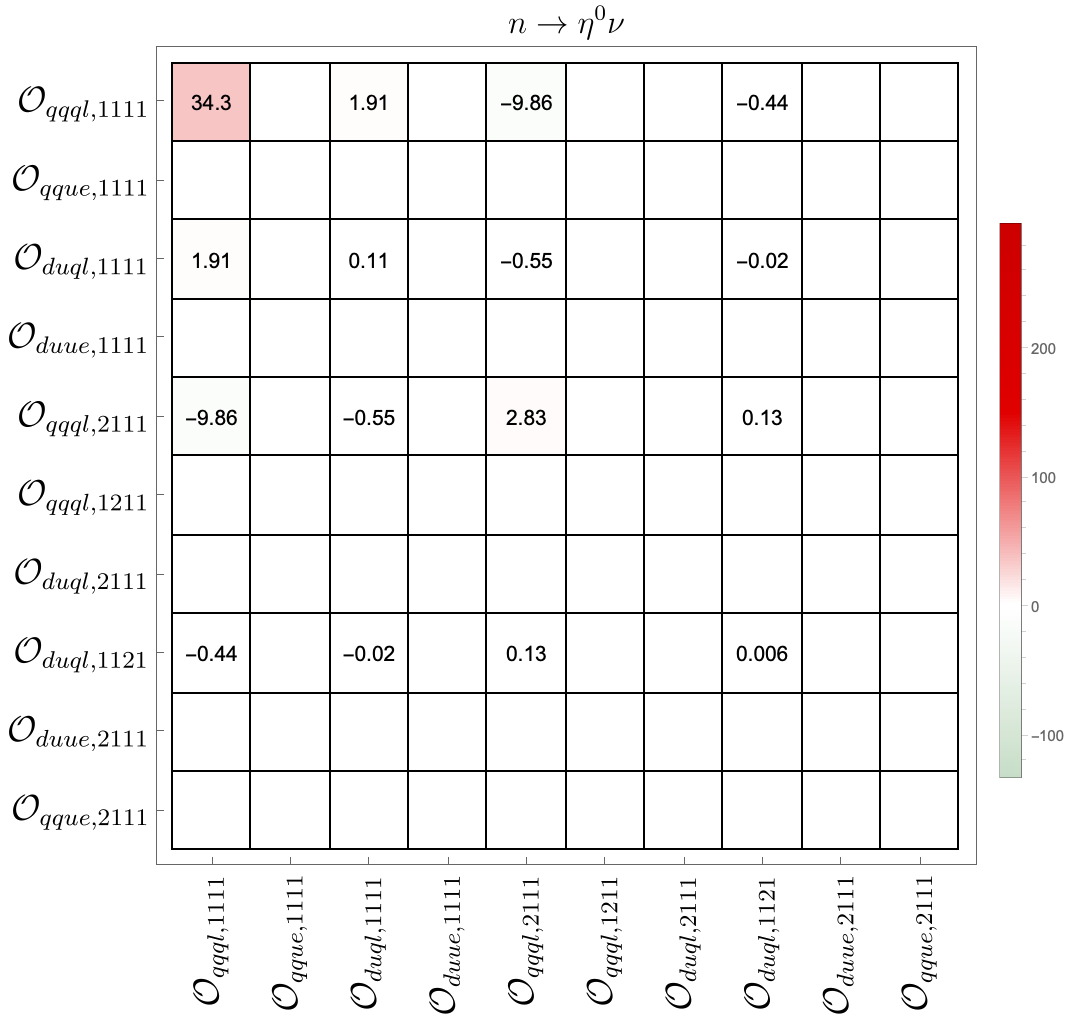}
	\end{subfigure}
	\begin{subfigure}[H]{0.55\textwidth}
		\centering
        \includegraphics[width = 1.01\textwidth]{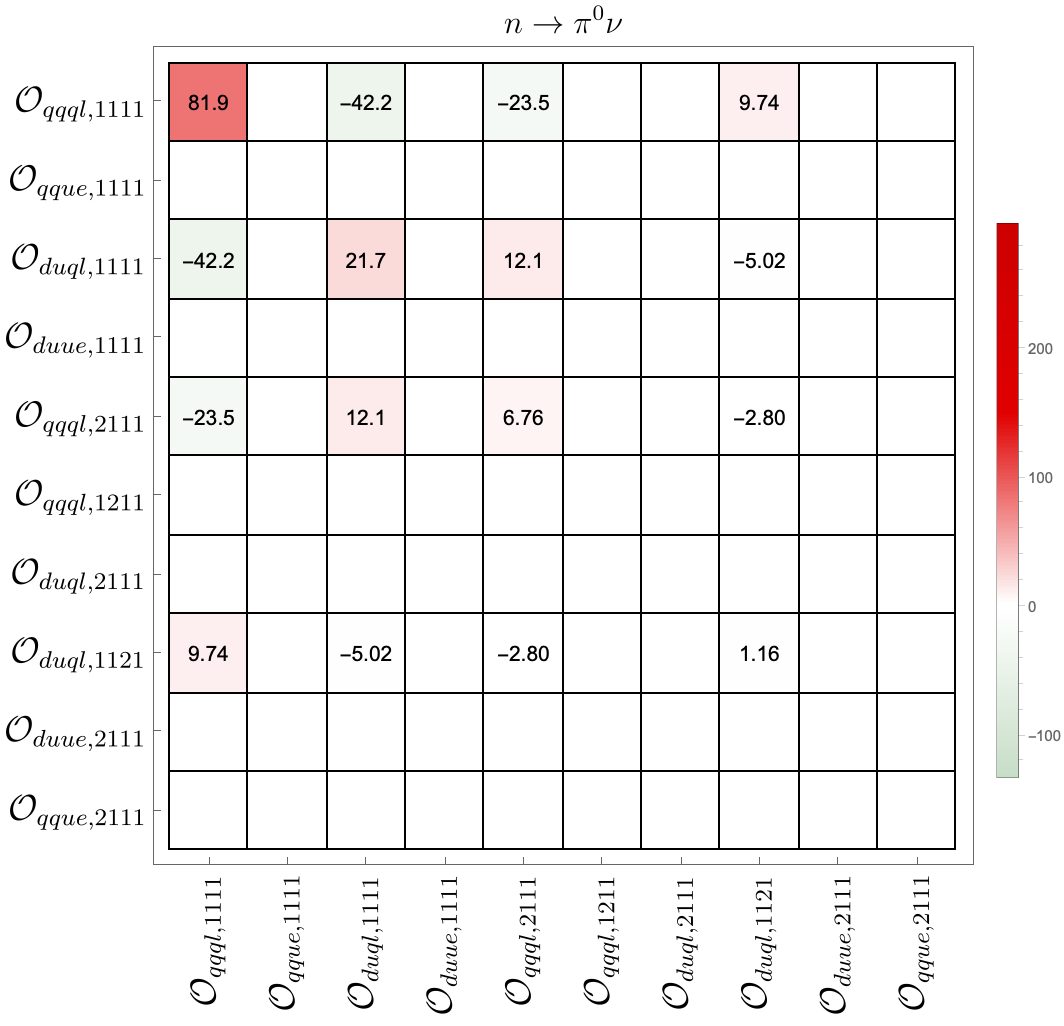}
	\end{subfigure}
 	\begin{subfigure}[H]{0.55\textwidth}
		\centering
        \includegraphics[width = 1.01\textwidth]{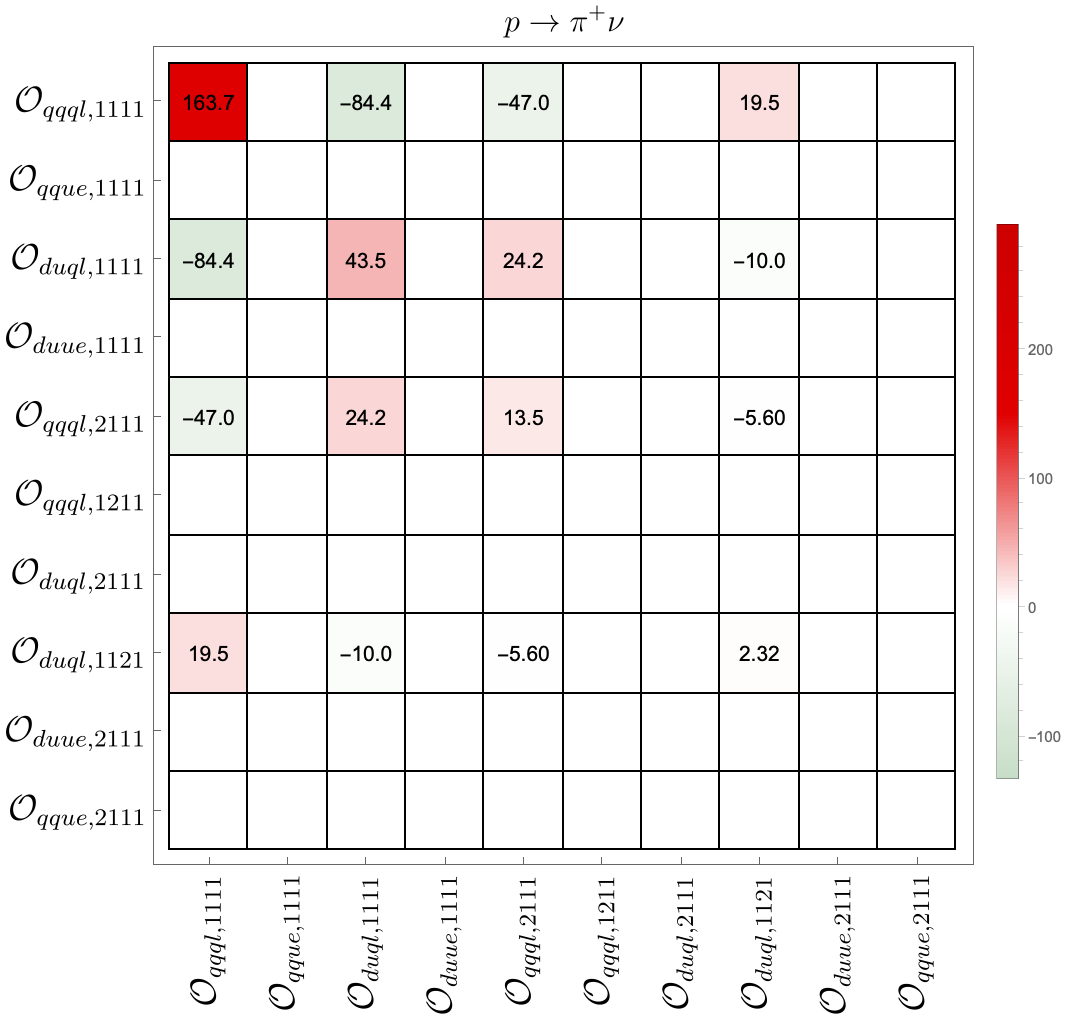}
	\end{subfigure}
        \begin{subfigure}[H]{0.55\textwidth}
		\centering
        \includegraphics[width = 1.01\textwidth]{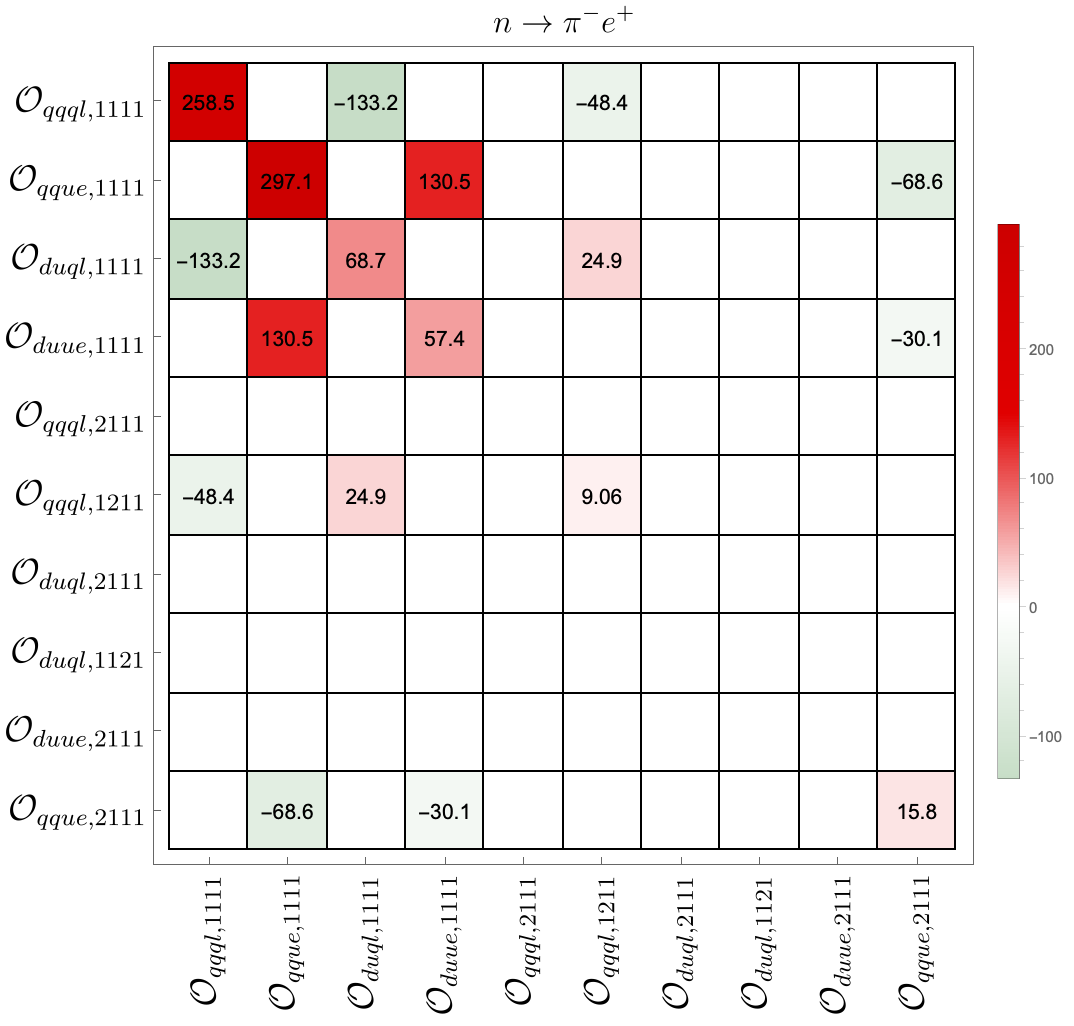}
	\end{subfigure}
        \begin{subfigure}[H]{0.55\textwidth}
		\centering
        \includegraphics[width = 1.01\textwidth]{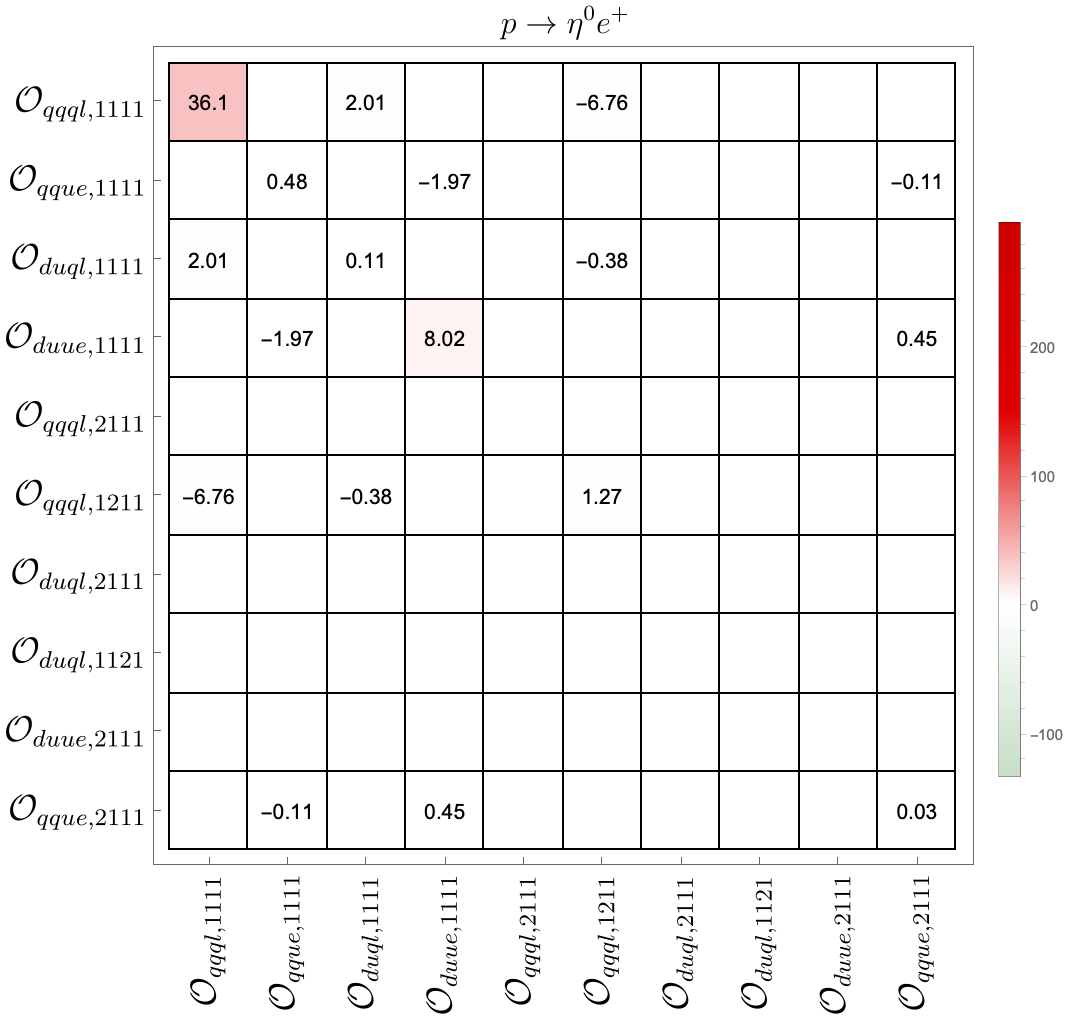}
	\end{subfigure}
        \begin{subfigure}[H]{0.55\textwidth}
		\centering
        \includegraphics[width = 1.01\textwidth]{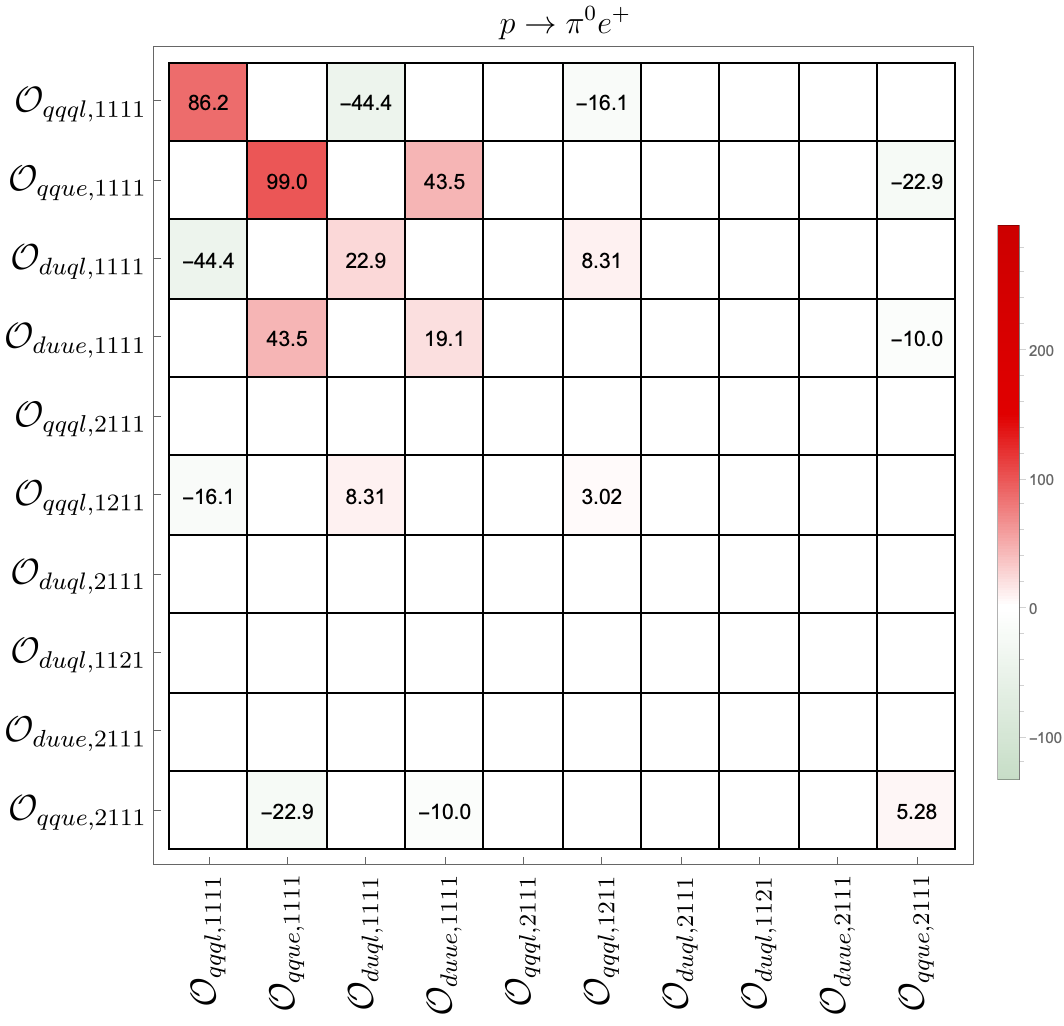}
	\end{subfigure}
\end{figure}
\begin{figure}[tbp!]
        \begin{subfigure}[H]{0.55\textwidth}
		\centering
        \includegraphics[width = 1.01\textwidth]{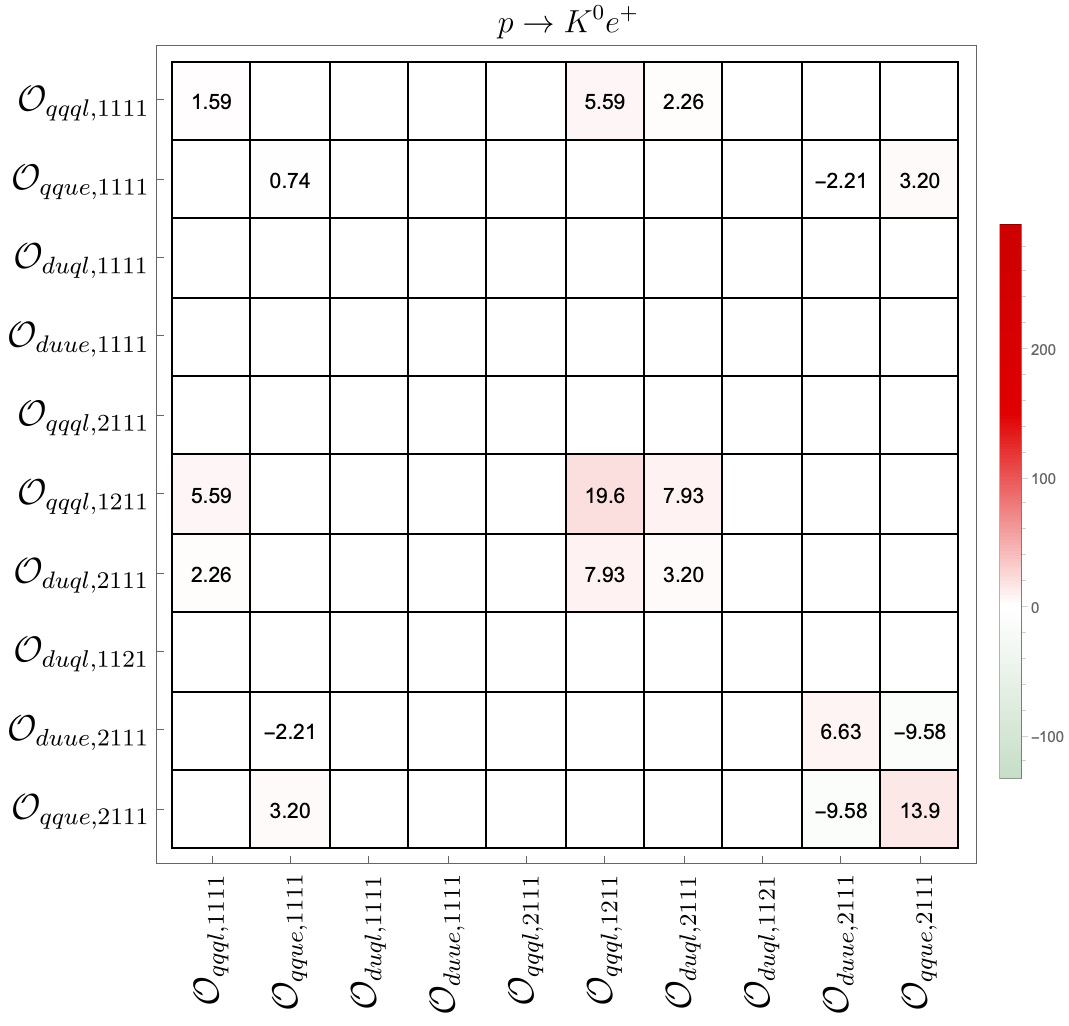}
	\end{subfigure}
	\begin{subfigure}[H]{0.55\textwidth}
		\centering
        \includegraphics[width = 1.01\textwidth]{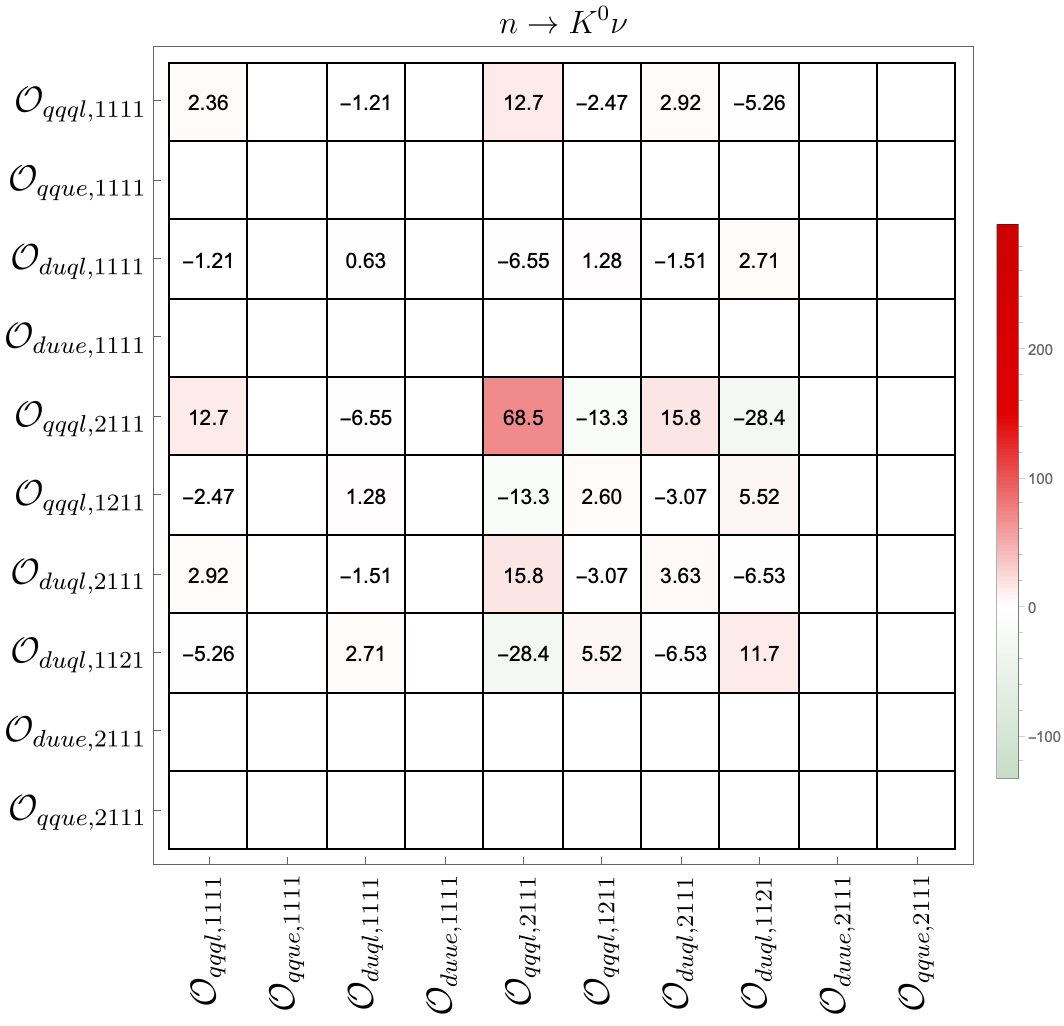}
	\end{subfigure}
 	\begin{subfigure}[H]{0.55\textwidth}
		\centering
        \includegraphics[width = 1.01\textwidth]{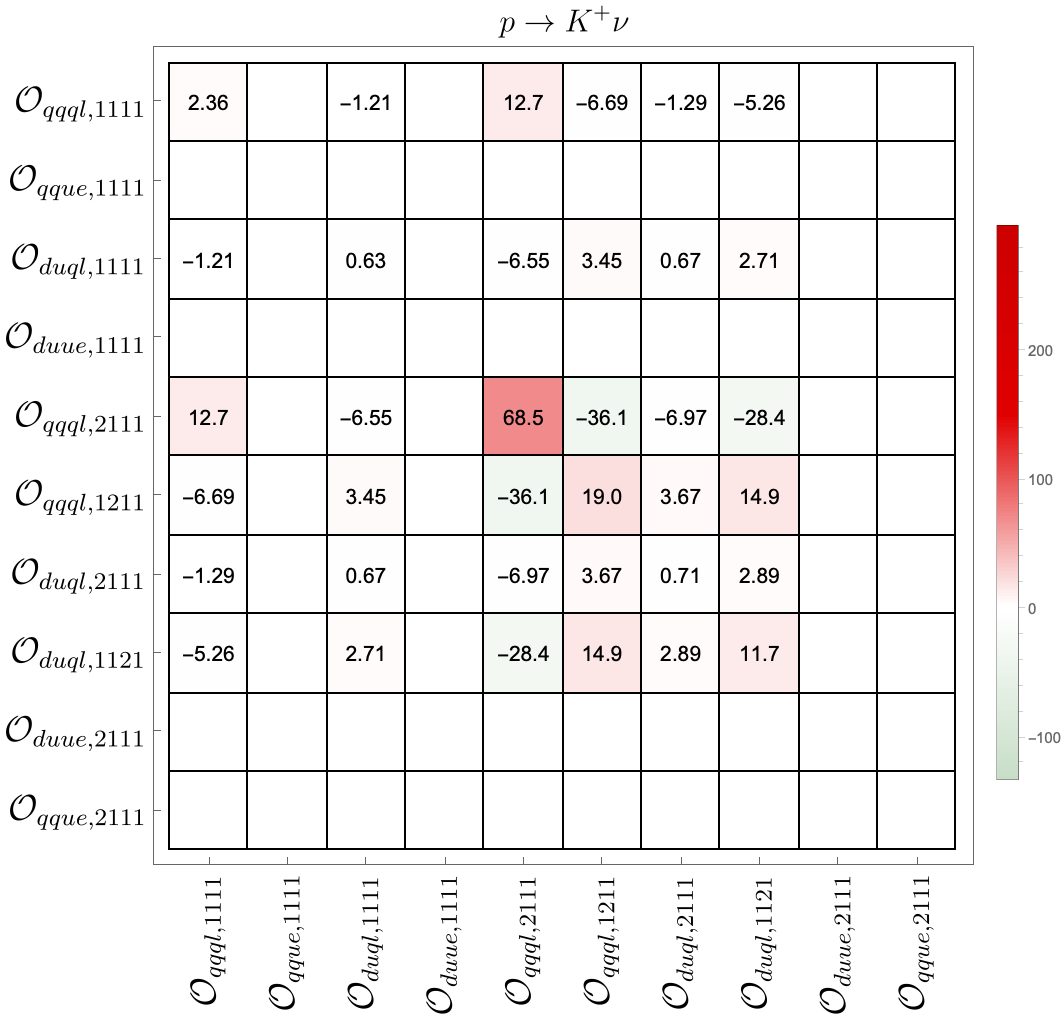}
	\end{subfigure}
\end{figure}

\begin{figure}[tbp!]
        \begin{subfigure}[H]{0.55\textwidth}
		\centering
        \includegraphics[width = 1.01\textwidth]{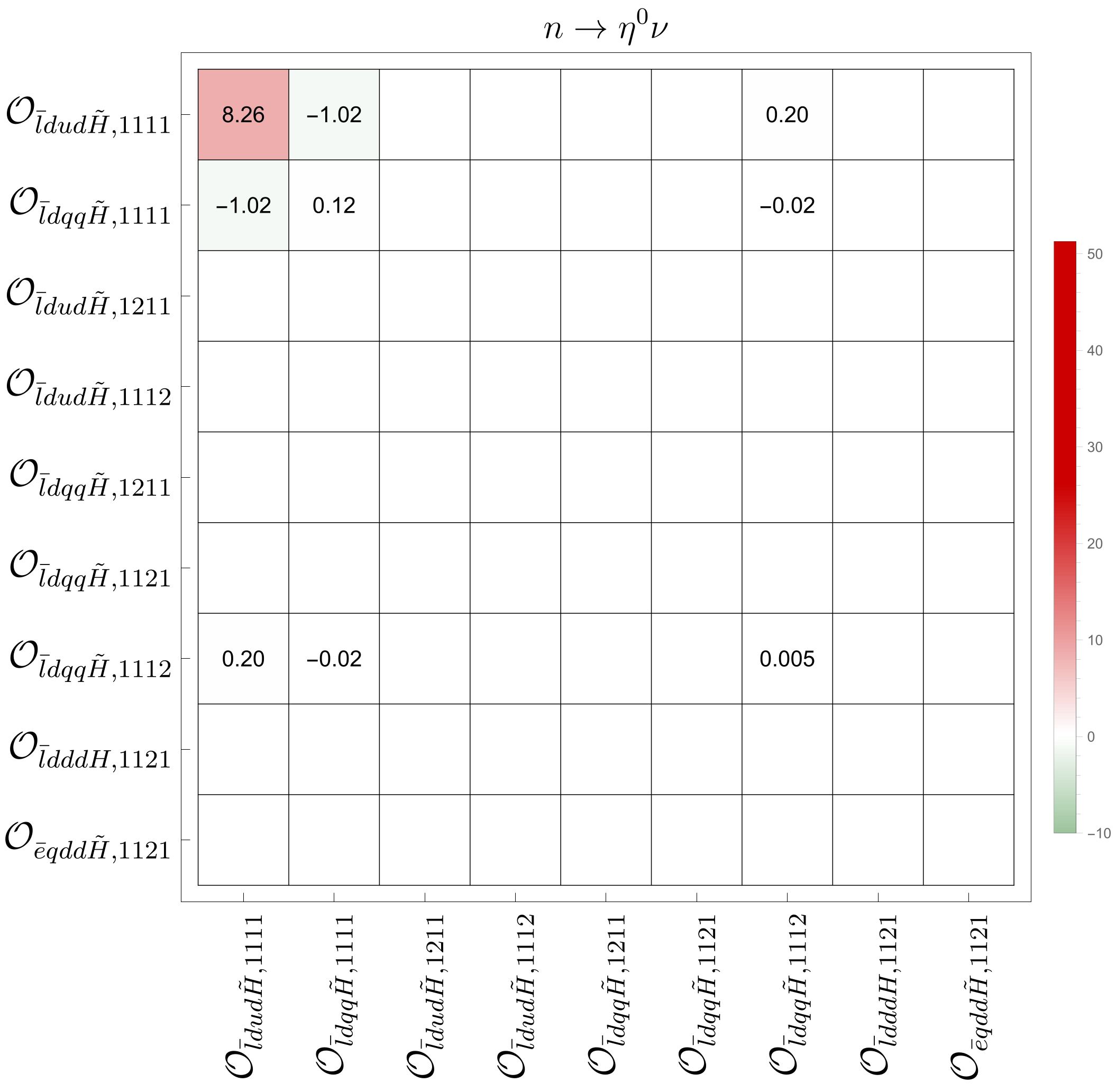}
	\end{subfigure}
	\begin{subfigure}[H]{0.55\textwidth}
		\centering
        \includegraphics[width = 1.01\textwidth]{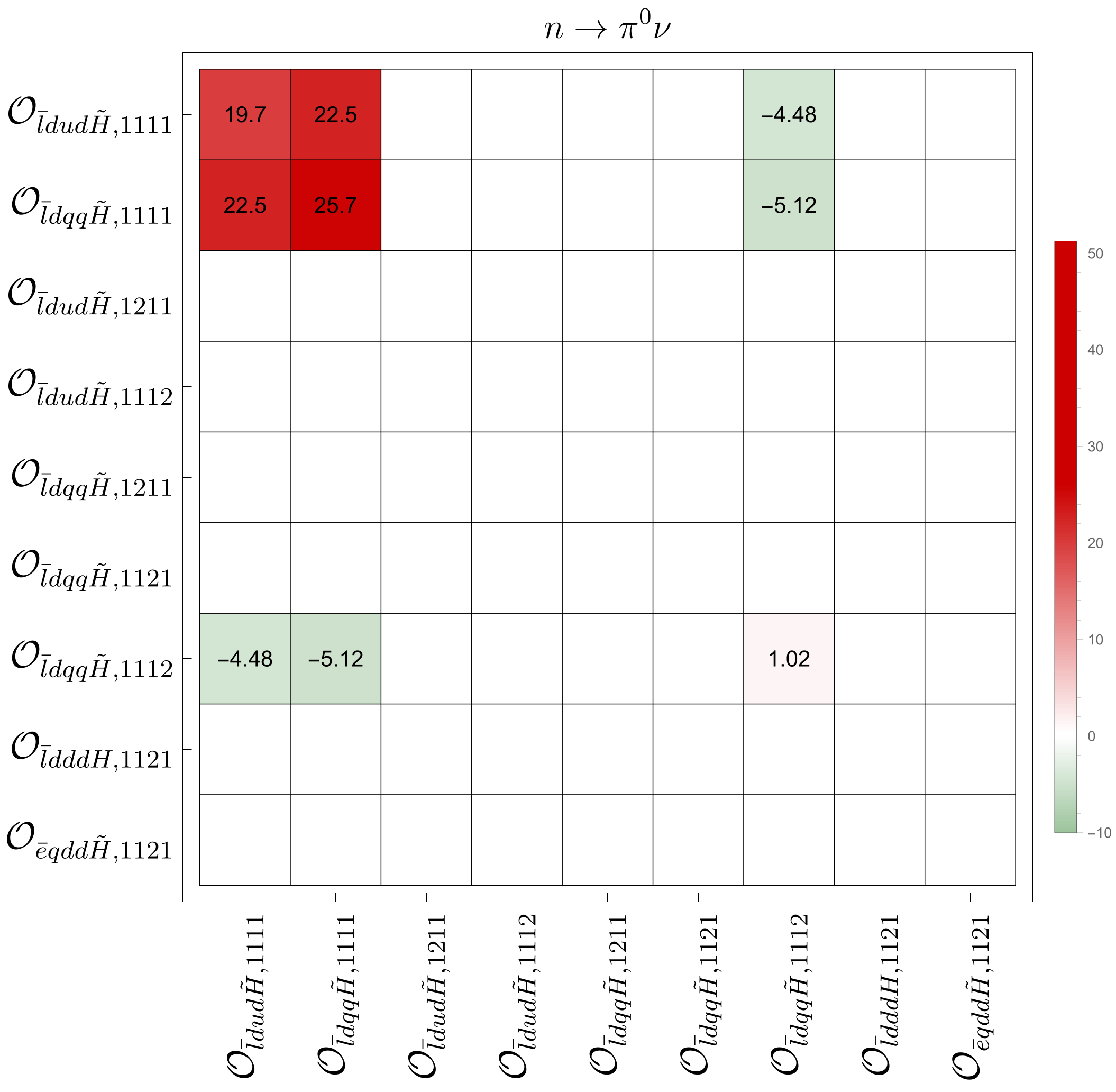}
	\end{subfigure}
 	\begin{subfigure}[H]{0.55\textwidth}
		\centering
        \includegraphics[width = 1.01\textwidth]{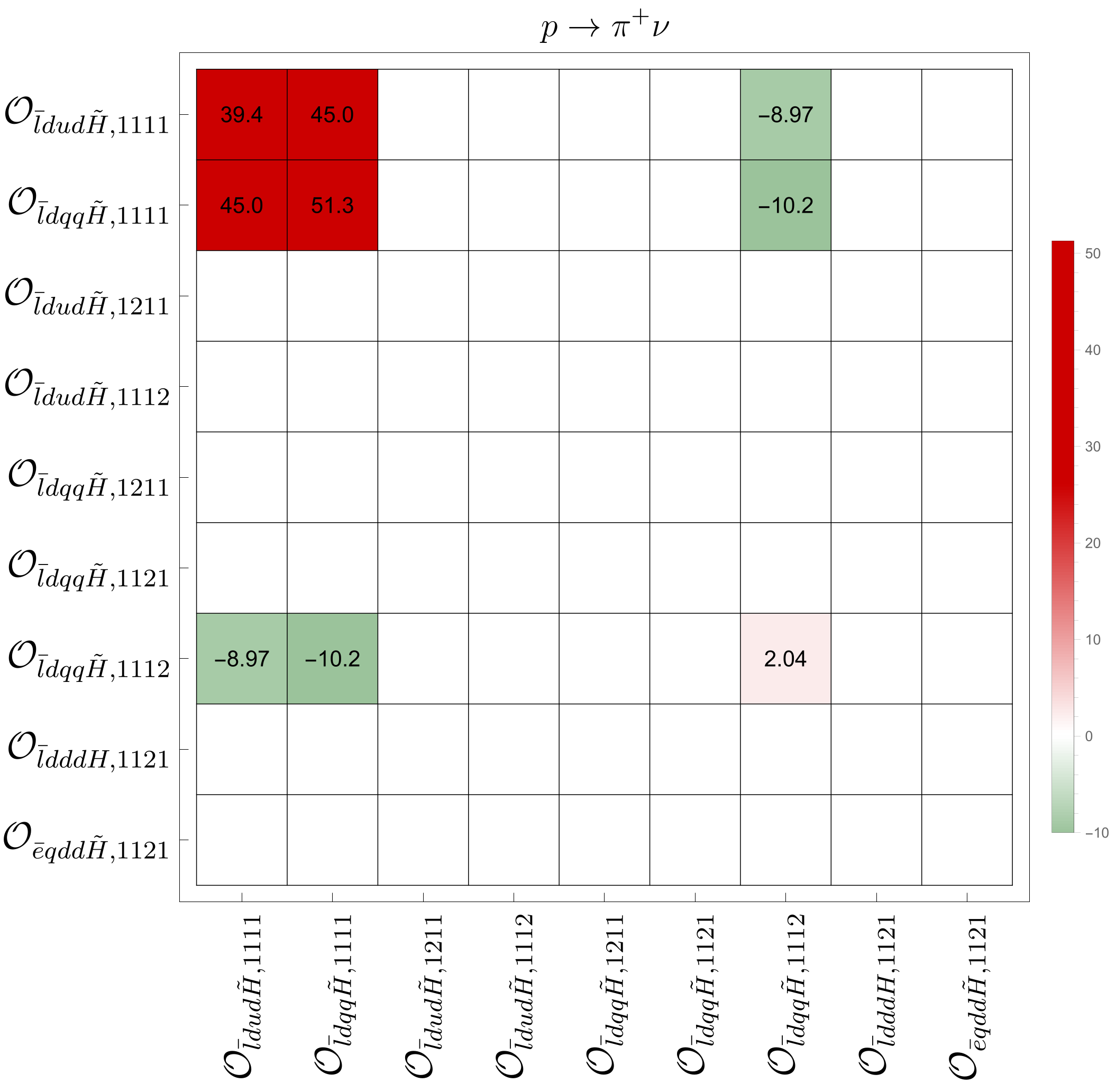}
	\end{subfigure}
        \begin{subfigure}[H]{0.55\textwidth}
		\centering
        \includegraphics[width = 1.01\textwidth]{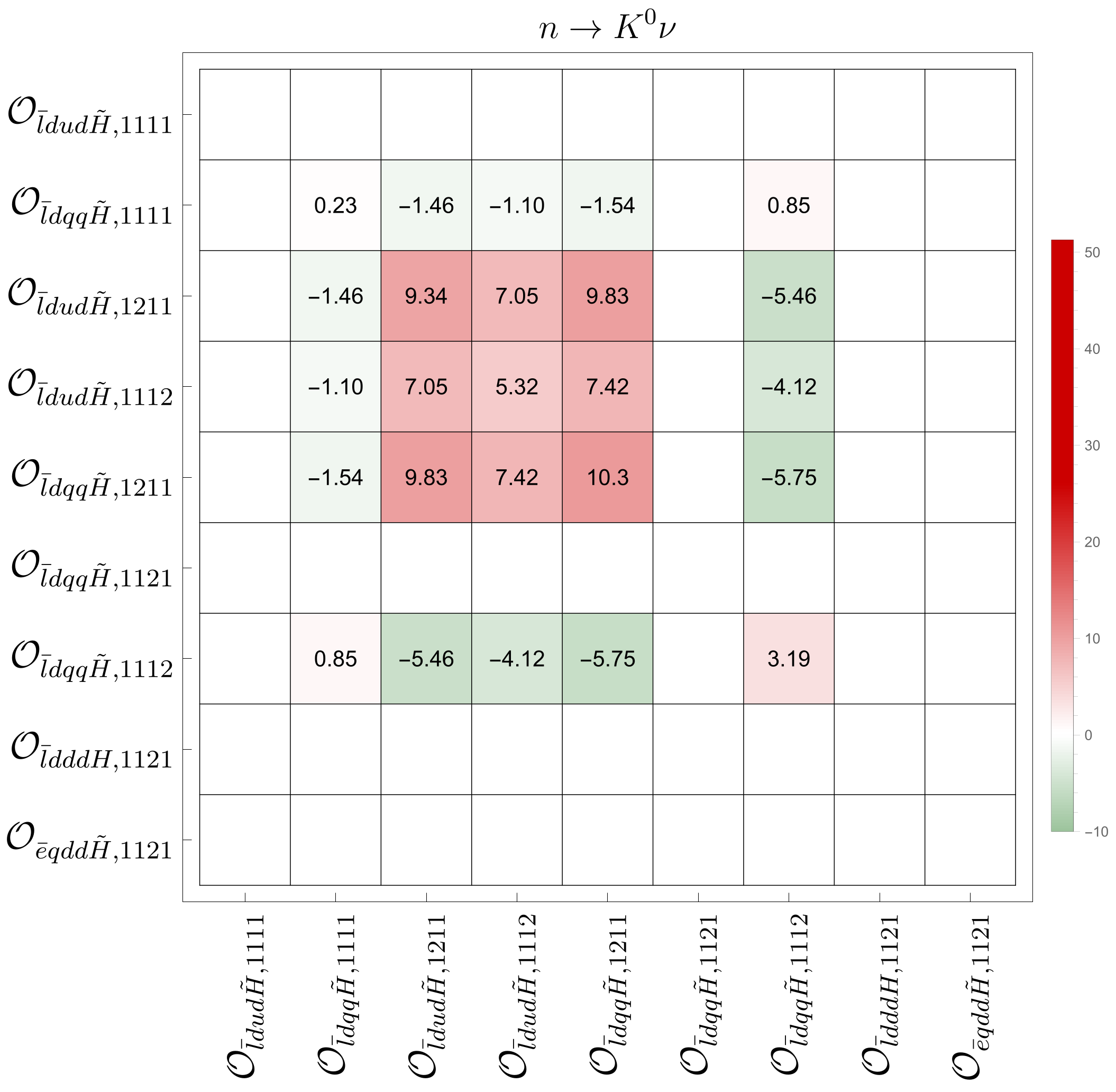}
	\end{subfigure}
        \begin{subfigure}[H]{0.55\textwidth}
		\centering
        \includegraphics[width = 1.01\textwidth]{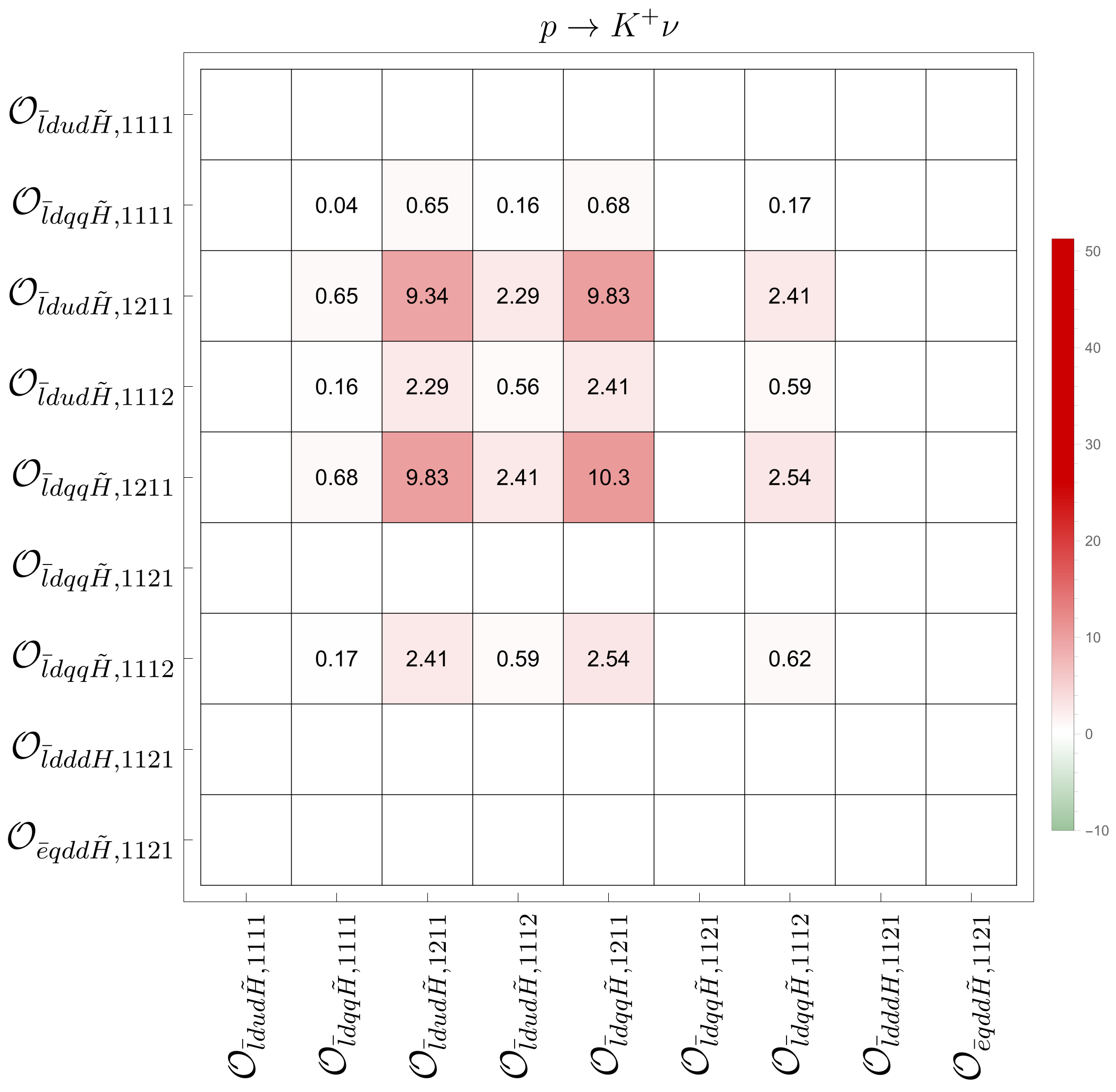}
	\end{subfigure}
        \begin{subfigure}[H]{0.55\textwidth}
		\centering
        \includegraphics[width = 1.01\textwidth]{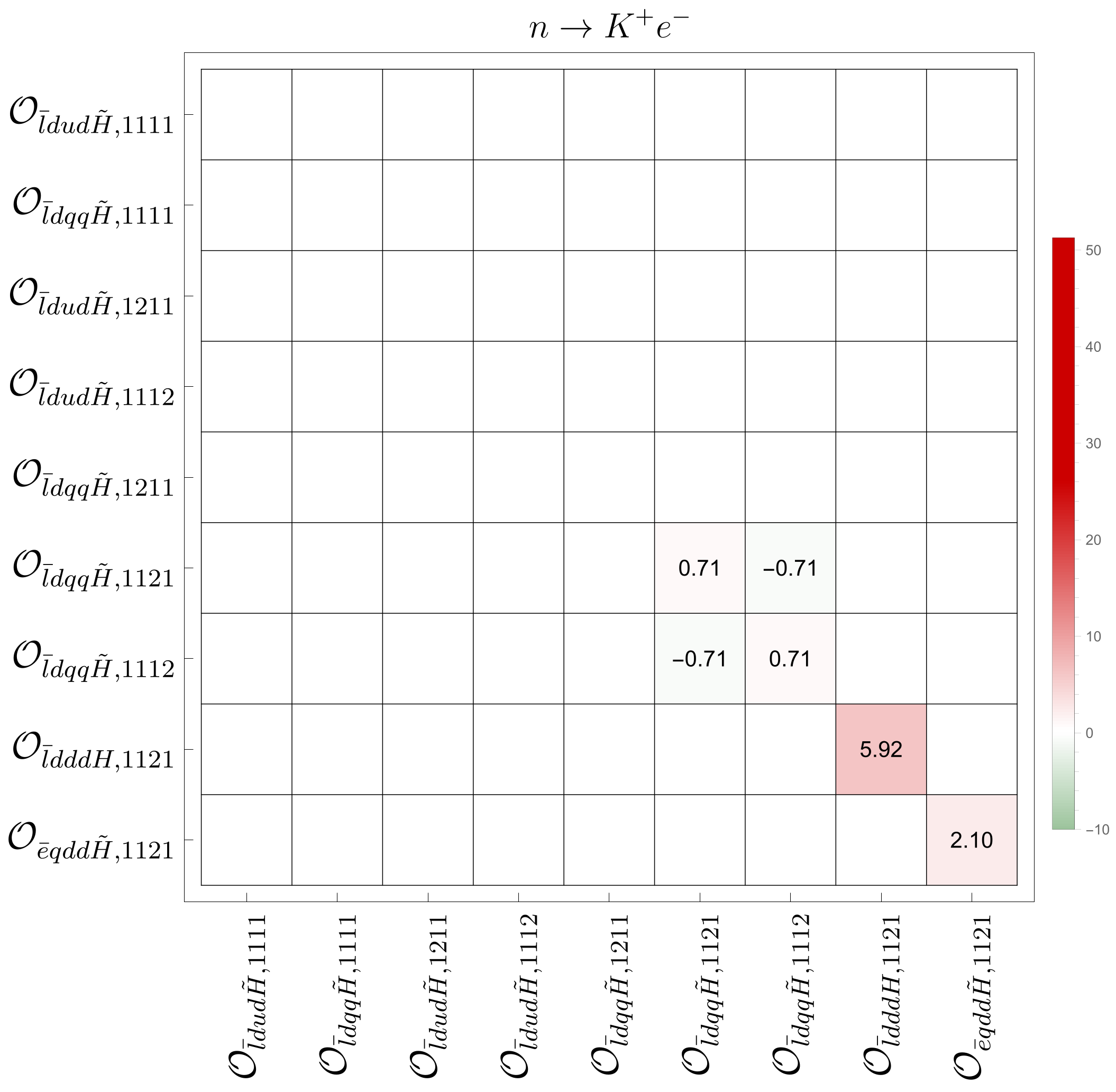}
	\end{subfigure}
\end{figure}

\clearpage

\bibliographystyle{JHEP}
\bibliography{main}

\end{document}